\documentclass[appendixfloats]{emulateapj}
\usepackage{graphicx}

\newcommand{\oii}{[\ion{O}{2}]}
\newcommand{\oiii}{[\ion{O}{3}]}
\newcommand{\hb}{H$\beta$} 
\newcommand{\ha}{H$\alpha$}
\newcommand{\lya}{Ly$\alpha$} 
\newcommand{\sii}{[\ion{S}{2}]}
\newcommand{\nii}{[\ion{N}{2}]}
\newcommand{\wlya}{W$_{Ly\alpha}$} 
\newcommand{\wha}{W$_{H\alpha}$} 
\newcommand{\kms}{km s$^{-1}$} 
\newcommand{\fesc}{$f_{esc}^{Ly\alpha}$} 

\begin{document}

\title{\lya\ emission from Green Peas:  the role of circumgalactic gas density, covering, and kinematics\altaffilmark{*}}

\author{Alaina Henry\altaffilmark{2,3}, Claudia Scarlata\altaffilmark{4}, Crystal L. Martin\altaffilmark{5} \& Dawn Erb\altaffilmark{6}} 
\altaffiltext{*}{Based on observations made with the NASA/ESA Hubble Space Telescope, which is operated by the Association of Universities for Research in Astronomy, Inc., under NASA contract NAS 5-26555.  These observations are associated with program 12928 and 11727.}
\altaffiltext{2}{Astrophysics Science Division, Goddard Space Flight Center, Code 665, Greenbelt, MD 20771; alaina.henry@nasa.gov}
\altaffiltext{3}{NASA Postdoctoral Program Fellow} 
\altaffiltext{4}{Minnesota Institute for Astrophysics, University of Minnesota, Minneapolis, MN 55455}
\altaffiltext{5}{Department of Physics, University of California, Santa Barbara, CA 93106}
\altaffiltext{6}{Department of Physics, University of Wisconsin Milwaukee, Milwaukee, WI 53211}

\begin{abstract} 
We report {\it Hubble Space Telescope}/Cosmic Origins Spectrograph observations of the \lya\ emission and interstellar absorption lines 
in a sample of ten star-forming galaxies at $z\sim0.2$.  Selected on the basis of high equivalent width
optical emission lines, the sample, dubbed ``Green Peas,'' make some of the best analogs for young galaxies in an early Universe.     We detect \lya\ emission in all ten galaxies, and 9/10 show double-peaked line profiles suggestive of low \ion{H}{1} column density.
We measure \lya/\ha\ flux ratios of 0.5-5.6, implying that 5\% to 60\% of \lya\ photons escape the galaxies. These data confirm previous findings that low-ionization metal absorption (LIS)
lines are weaker when \lya\ escape fraction and equivalent width are higher. 
However, contrary to previously favored interpretations of this trend,  increased \lya\ output cannot be the result of
a varying \ion{H}{1} covering: the Lyman absorption lines (Ly$\beta$ and higher) show a 
covering fraction near unity for gas with  $N_{H~I}\ga10^{16}~{\rm cm}^{-2}$.  Moreover, we detect no correlation between \lya\ escape and the  outflow velocity of the LIS lines, suggesting that kinematic effects do  not explain the range of \lya/\ha\ flux ratios in these galaxies.  In contrast, we detect a strong anti-correlation between the \lya\ escape fraction and the velocity separation of the \lya\ emission peaks,  driven primarily by the velocity of the blue peak.  As this velocity separation is sensitive to 
\ion{H}{1} column density,  we conclude that \lya\ escape in these Green Peas is likely regulated by the \ion{H}{1} column density rather than outflow velocity or \ion{H}{1} covering fraction. 
 \end{abstract} 

\section{Introduction}  
The \lya\ emission line is a heavily-used diagnostic in studies of high-redshift galaxies and the intergalactic medium (IGM).  
At redshifts $z\ga 2$, this feature is shifted into readily observable optical wavelengths, leading many studies to 
rely on \lya\ for  both discovery and spectroscopic redshift confirmation of galaxies.    \lya-based surveys are now extending to $z\sim6$ and beyond, where they are identifying the  building blocks of present 
day galaxies and the sources 
that are most important for metal enrichment and reionization of the IGM \citep{Martin08,Ouchi10, Hu10, Kashikawa11, Dressler11,Dressler15,Henry12, Rhoads12}.    In addition, because \lya\ photons scatter resonantly in neutral hydrogen, 
 they offer a wealth of information about this gas.  Spectroscopic line profiles can constrain gas content and kinematics \citep{Verhamme14, Erb14,Martin14}, and 
\lya\ imaging can illuminate the gaseous halos around galaxies  \citep{Zheng, Steidel11, Hayes13, Momose}.   In the epoch of reionization, 
weaker or less frequent \lya\ emission may indicate a significant neutral hydrogen fraction in the IGM   \citep{Ota10, Pentericci, Schenker,Treu12, Treu13}, an increase in the escape fraction of ionizing photons, or both \citep{Dijkstra14}.

Despite the fact that a wide range of studies rely on \lya\ emission, we do not understand how to interpret the feature. 
The \lya\ luminosity,  equivalent width, \wlya, and  \lya/\ha\ flux ratio of galaxies are seen to vary widely, and are difficult to predict from other galaxy properties like star formation rate (SFR) 
and dust extinction  \citep{Giavalisco, Scarlata09}.   Qualitatively, it is believed that the \lya\ emission line is modified by resonant scattering in the interstellar medium (ISM) and circumgalactic medium (CGM). 
This process increases the optical depth of \lya\ photons, making them more susceptible to dust absorption and sometimes turns the emission profile into absorption.  
 Alternatively, in the absence of dust, the \lya\ photons may simply diffuse to large radii and be missed by poor sensitivity to low surface brightness emission \citep{Steidel11, Hayes13}.   
 Besides dust and geometry of the HI gas, the kinematics of galaxy outflows may also play a role, since \lya\ photons may escape more easily 
when scattered by neutral hydrogen that is out of resonance with the ISM \citep{Kunth98,Wofford}.    
Nevertheless, our poor {\it quantitative} understanding of \lya\ output is cause for concern when 
 studies try to draw robust  conclusions about galaxies and the IGM from this feature.

\begin{deluxetable*} {cccccccc}[!ht]
\tablecolumns{8}
\tablecaption{Green Pea Sample \& Observations} 
\tablehead{
\colhead{ID}  &  \colhead{RA}  & \colhead{DEC}  & \colhead{$z$} &\colhead{E(B-V)$_{MW}$}  &\colhead{Rest Wavelength Coverage} & \colhead{G130M Exposure} & \colhead{G160M Exposure}  \\ 
  &   (J2000)  & (J2000) &   & (mag)   & (\AA) & (s)  & (s)
} 
\startdata
0303--0759  & 03 03 21.41  &  -07 59 23.2  &    0.164887  & 0.0877 & 975 - 1515 & 2190  & 3829  \\
1244+0216 & 12 44 23.37  &   02 15 40.4  &     0.239420  & 0.0213  &  945 - 1430 & 2042 & 6507   \\
1054+5238 & 10 53 30.80  &  52 37 52.9 &      0.252645  & 0.0132 & 910 - 1425  & 824 & 2736  \\
1137+3524  & 11 37 22.14 & 35 24 26.7 &     0.194396   & 0.0161 & 965 - 1505 & 1264 & 2340  \\
0911+1831 & 09 11 13.34 & 18 31 08.2 &    0.262236  &  0.0248 & 900 - 1435  & 2074 & 6530 \\ 
0926+4427 & 09 26 00.44  &   44 27 36.5  &    0.180698  &  0.0165 & 970 - 1505 &  5640 & 6180 \\ 
1424+4217  & 14 24 05.72 &  42 16 46.3 &     0.184800  & 0.0094 & 965 - 1220  & 1209 & 0\tablenotemark{a}  \\
1133+6514 & 11 33 03.80   &   65 13 41.4  &    0.241400   & 0.0097 & 945 - 1430  & 1232 & 4589  \\ 
1249+1234 & 12 48 34.63  & 12 34 02.9 &     0.263403  & 0.0252  & 900 - 1425  & 1644 & 6372 \\
1219+1526 & 12 19 03.98  & 15 26 08.5 &     0.195614   & 0.0239 & 965 - 1505  & 716 & 2304 
\enddata
\tablenotetext{a}{Failed Observation}
\label{sample} 
\end{deluxetable*}

Efforts to understand \lya\ escape from high redshift galaxies have generally
suffered from the limitations associated with studying faint, distant sources  (e.g.\ \citealt{Shapley03, Erb10, Erb14, Jones12, Jones13}). 
Fortunately, many of these difficulties can be overcome when studying nearby galaxies.    If low-redshift galaxies can
be found where the physical conditions are similar to high-redshift star-forming galaxies, it is possible to learn more
about the astrophysics that regulates \lya\ output.  Despite this realization, most investigations to date have struggled
to identify local galaxies that are similar to the low-mass, low-metallicity, high specific SFR galaxies 
that dominate the cosmic star-formation at early times (e.g.\ \citealt{Alavi, Stark14}). 
Instead, present samples of nearby galaxies with 
\lya\ observations are on average, more massive, dustier, metal-rich, and have lower specific SFRs
 than the bulk of high-redshift galaxies \citep{Giavalisco, Scarlata09,  Heckman11, Cowie11, Wofford, Hayes13, Hayes14, Martin14}.  
 As a possible consequence of this selection, \lya\ luminosities in low redshift samples-- when measured-- are  around an order of magnitude lower than the
  luminosities of most \lya-selected galaxies observed at $z>3$ \citep{Gronwall, Ouchi10, Hu10, Dressler11,Dressler15,  Henry12}.

The recent discovery of extremely high equivalent width emission line galaxies (the  ``Green Peas'') within the Sloan Digital Sky Survey (SDSS)  
offers a new avenue to investigate \lya\ emission from high-redshift analogs \citep{Cardamone}.     With masses reaching below $10^{9}$ M$_{\sun}$, metallicities lower than 
12 + log(O/H) $\sim 8.2$ (from direct $T_e$ measurements; \citealt{Izotov11}), and   \ha\ equivalent 
widths (\wha) exceeding hundreds of \AA,  these local galaxies may be more representative of high redshift Lyman Alpha Emitters (LAEs; \wlya $\ga 20$ \AA) and reionization epoch galaxies than other local samples.   
As a comparison, inferred emission line contamination to {\it Spitzer}/IRAC photometry provides tentative evidence for similar \wha\ in star-forming galaxies at $z\sim4-7$ \citep{Shim11,Stark13, Labbe13, Smit14}.     Ultimately, 
the {\it James Webb Space Telescope} will clarify which segment of the high-redshift populations are most analogous to the Green Peas.  At present, however, quantifying the \lya\ output from these nearby objects is a critical benchmark for future comparisons.

In this paper we present a UV spectroscopic study of Green Peas using the Cosmic Origins Spectrograph (COS) on the
{\it Hubble Space Telescope} (HST).   For a sample of ten galaxies at $0.16 < z < 0.26$, we measure the \lya\ emission that 
escapes the galaxies from within a few kpc (corresponding to the COS aperture).   At the same time, by 
observing absorption lines in both hydrogen and metals, these data allow us to quantify the role of galactic outflows, ISM, and CGM gas.   
This paper is organized as follows:  in \S 2 we describe our sample selection and the galaxy properties derived from GALEX and SDSS; in \S 3 we  
describe the COS observations and resultant data and in \S 4 we present the \lya\ line profiles and measurements;  \S 5 explores how the stellar population properties and dust of 
the Green Peas compare to other nearby samples that have been observed in \lya.   In \S 6   and \S  7 we show how the strength of the \lya\ line varies with interstellar absorption 
equivalent width and kinematics, and discuss the velocity structure of the \lya\ line. Finally, \S 8 offers a comprehensive interpretation of the data and our conclusions are summarized in \S 9.

Throughout this paper we use a \cite{Chabrier} initial mass function (IMF), and a $\Lambda$CDM cosmology with $\Omega_M = 0.3$, $\Omega_{\Lambda} = 0.7$, and $H_0 =70$ km s$^{-1}$ Mpc$^{-1}$.     We adopt the equivalent width sign convention used in literature focused on 
high-redshift galaxies: positive equivalent widths indicate emission while negative values are used to signify absorption.
All COS, GALEX and SDSS data are corrected for Milky Way extinction using attenuation measured by \cite{Schlafly}  and the \cite{Fitzpatrick} extinction law.   The foreground extinction values were obtained from the 
NASA Extragalactic Database\footnote{\url{http://ned.ipac.caltech.edu/}}, and are listed in Table \ref{sample}.

\begin{deluxetable*} {ccccccccccc}[!ht]
\tablecolumns{11}
\tablecaption{Green Pea Properties from GALEX+SDSS} 
\tablehead{ \colhead{ID}  &  \colhead{$W_{H\alpha}$}   &  \colhead{\ha/\hb}  &  \colhead{$E(B-V)_{gas}$}    & \colhead{log($L_{H\alpha}$/erg s$^{-1}$)} & \colhead{SFR} 
& \colhead{Log M/M$_{\sun}$ } & \colhead{12 + log(O/H)}  & \colhead{$M_{FUV}$} & \colhead{$\beta$  } &  \colhead{$R_{UV}$} \\   
        & \colhead{ (\AA) } &   &     \colhead{ (mag) }   &     &  \colhead{(M$_{\sun}$ yr$^{-1}$ ) } &   &   &  \colhead{ (mag) }  &  & \colhead{(kpc)}   \\
   \colhead{(1)} & \colhead{(2)} & \colhead{(3)}  & \colhead{(4)} & \colhead{(5)} & \colhead{(6)} & \colhead{(7)} & \colhead{(8)} & \colhead{(9)} & \colhead{(10)}  & 
   \colhead{(11)} 
  } 
\startdata
0303--0759  & 670  &  2.78  &    0.00 & 42.24  & 7.6  & 8.89  &  7.86  & -20.35 & -2.23  & 0.8  \\
1244+0216 &  840   & 3.10  &     0.07 & 42.78  & 26.2 & 9.39 & 8.17   & -20.32 & -1.70  & 2.6  \\
1054+5238 &  400   & 3.15  &      0.08  & 42.71 & 22.4 & 9.51 & 8.10  &  -21.31 &  -1.94 &  1.3 \\  
1137+3524  & 580   & 3.08  &      0.06  & 42.58 & 16.8  & 9.30 & 8.16  & -20.56  & -1.78 & 1.8 \\  
0911+1831 & 420  & 3.50  &       0.17  &  42.68   & 21.1  & 9.49 & 8.00  & -20.56 & -1.82  &  1.1 \\
0926+4427 &  610   & 3.20   &    0.10  &  42.49  & 13.6 & 8.52  & 8.01 & -20.58  & -1.98 & 1.0  \\ 
1424+4217  & 1100 & 3.01  &      0.04  & 42.57 & 16.5 & 8.08 & 8.04  &  -20.40  &  -1.90 & 1.0   \\ 
1133+6514 &  300  & 2.90  &     0.01   & 42.02  & 4.6  & 9.04  & 7.97  & -20.40 & -1.93 & 1.9  \\
1249+1234 &  670   & 3.09  &      0.07  & 42.50  & 13.8 & 8.79 & 8.11 & -20.25   & -1.82 & 1.8  \\
1219+1526 & 1270  & 2.87  &      0.00   & 42.43 & 11.9  & 8.09 & 7.89 & -19.94  & -1.65 & 0.7  
 \enddata
\label{properties} 
\tablecomments{ Quantities are derived from the SDSS and GALEX, or taken from \cite{Izotov11}.  Measurements errors are dominated by systematics in these high S/N data, so statistical errors are not given.  Emission line measurements include both broad and narrow components, as described in \S \ref{sample_sec}.  {\bf Column Descriptions:}  {\bf (1)} Object ID {\bf (2)} Rest-frame \ha\ equivalent width, in \AA, measured from SDSS spectrum.     {\bf (3)}  Flux ratio, F(\ha) / F(\hb)  measured from SDSS  
spectrum {\bf (4)}  Nebular gas extinction inferred from the \ha/\hb\ flux ratio, assuming the intrinsic ratio of 2.86 as described in the text.  {\bf (5)} \ha\ luminosity measured from 
the SDSS spectrum {\bf (6)} SFR estimated from $L_{H\alpha}$ using the \cite{Kennicutt} calibration, divided by 1.8 to convert from a Salpeter to \cite{Chabrier} IMF {\bf (7)} 
Stellar mass from \cite{Izotov11}, also converted from a Salpeter to \cite{Chabrier} IMF  {\bf (8)} Direct, $T_e$([\ion{O}{3}])-based metallicities taken from \cite{Izotov11}
  {\bf (9)}  FUV absolute magnitude (AB),  calculated from GALEX photometry, interpolated to provide an estimated rest-frame 1500\AA\ luminosity {\bf (10)} UV slope, defined as $F_{\lambda} \propto \lambda^{\beta}$, 
  calculated from GALEX data.  {\bf (11)} NUV Petrosian radius (see text in \S \ref{sample_sec}), calculated from the COS acquisition image.   }
\end{deluxetable*}

\section{Sample Selection and Properties} 
\label{sample_sec} 
The present sample of Green Peas were drawn from the catalog presented by \cite{Cardamone}.    
To systematically identify objects originally discovered by the Galaxy Zoo \citep{Lintott}, Cardamone et al. defined 
a color selection for objects with strong \oiii\ emission in the SDSS $r-$ band.   Following this selection, 
objects with low S/N SDSS spectra were removed and AGN were rejected on the basis of broad emission lines or 
their optical emission line ratios (\ha /\nii\ vs. \oiii / \hb; \citealt{BPT}).   The remaining sample consists of 80 galaxies with rest frame [\ion{O}{3}] $\lambda 5007$ equivalent widths of hundreds to over 1000 \AA.     

We selected Green Peas that were bright enough in the FUV to enable continuum detection with COS.  From the 
80 star-forming Green Peas in \cite{Cardamone}, we considered
 objects with GALEX photometry (GR6) and $m_{FUV} \le 20$ AB.  
 We also required $z < 0.27$, so that the FUV spectra would cover the Si IV $\lambda \lambda$ 1393, 1403 lines. 
The resulting  sample of ten galaxies is listed in Table \ref{sample}.    We verified that these galaxies have stellar masses, SFRs, metallicities, and \oiii\ equivalent widths consistent with the parent sample of Green Peas  \citep{Cardamone, Izotov11}.  
However, because of the FUV magnitude limit,  the average UV luminosity in the present sample is around 1.6 times 
higher than the average (GALEX detected) Green Peas in \cite{Cardamone}.
   
In Table \ref{properties}  we compile properties of the Green Peas, drawn from the literature or derived from GALEX and SDSS.     We describe these quantities below.    In most cases, the high S/N of these data imply 
insignificant statistical errors on the measurements.  Therefore we do not list these uncertainties.

First, we chose to re-measure the emission lines in the SDSS spectra rather than use catalog measurements.  One of the Green Peas in our sample (0303-0759) contains 
an unphysical Balmer decrement (an \ha/\hb\ flux ratio of 2.25) in the MPA-JHU DR7\footnote{\url{http://www.mpa-garching.mpg.de/SDSS/DR7/}} catalogs, which is the result of an incorrect \ha\ flux.  
This failure may be attributable to difficulty in fitting the nearly blended \ha + \nii\ lines with single Gaussians, when the lines clearly show evidence of broad  wings extending hundreds of km s$^{-1}$ 
(also noted by \citealt{Amorin12} and \citealt{Jaskot14}).    Therefore, for each of the Green Peas in this paper, we re-fit the emission line spectra with two kinematic components.  We defined a kinematic model with a narrow component 
centered at the systemic redshift ($v=0$), and a broad component which is allowed to be offset from $v=0$.  For 25 of the strongest lines in the spectrum 
(\ha\ through H9, \sii, \nii, \oiii, \oii, [\ion{Ne}{3}] and 7 \ion{He}{1} lines), the amplitudes of the broad and narrow  components were allowed to vary, but their velocity widths and centroids were required to be the same from line to line.  Hence, the resultant model has 54 free parameters (redshift, Doppler shift of the broad component, velocity widths of both the broad and narrow component, and two Gaussian amplitudes for each line). 
This model was fit to the continuum subtracted spectrum provided by the SDSS, thereby measuring redshifts (Table \ref{sample}), emission line fluxes  (for the combined broad and narrow kinematic components), and the kinematics of the ionized gas.     Equivalent widths are evaluated by taking the ratio of the emission  line flux to the median continuum flux density in a 100 \AA\ window centered on each line.   The kinematic measures derived from our fits are similar for all of the Green Peas, showing a narrow component at the systemic redshift with a FWHM around 200 \kms, and a broad component, also centered near $v\sim0$,  with a FWHM around 500 \kms.   For the isolated \hb\ line, the broad component contributes 15-35\% of the total line flux in these galaxies. 
 
 Table \ref{properties} lists the \ha\ equivalent width and the  \ha\ to \hb\ flux ratio.   From this latter quantity, we derived dust extinction assuming a \cite{Calzetti} extinction curve 
 and an intrinsic ratio of F(\ha)/F(\hb) = 2.86. This intrinsic ratio is appropriate for electron temperature $T_e = 10^4$ K and density $n_e = 10^2$ cm$^{-2}$; 
 increasing $T_e$ two-fold decreases the intrinsic Balmer decrement to 2.75 and increases the extinction, $E(B-V)_{gas}$, by approximately 0.03 magnitudes.   Dust corrected \ha\ luminosities are 
 given, and converted to SFR using the calibration given in Kennicuttt (1998; adjusted to a \citealt{Chabrier} initial mass function). 
 
 The stellar masses of the Green Peas are taken from \cite{Izotov11}.  These authors note that, for high equivalent with emission lines, 
 the corresponding bound-free continuum emission can be comparable to the stellar continuum. Therefore previous estimates of stellar mass in these
 galaxies may be too high;  to correct for this mis-estimation, Izotov et al. derived stellar masses from fits to the SDSS spectra, using models that include nebular continuum.  For consistency, we have decreased these masses by a factor of 
 1.8 to convert them from a Salpeter to \cite{Chabrier} initial mass function.   Oxygen abundances are also 
 taken from \cite{Izotov11}, where they were derived from \oiii\ electron temperature (\oiii\ $\lambda$4363).   All 10 of the Green Peas in this paper show detections in this 
 line, confirming their low metallicities. 
 
 The ultraviolet continuum luminosity and slope were derived from GALEX photometry (FUV: $\lambda \sim 1340 - 1790$ \AA, NUV: $\lambda \sim 1770 - 2830$ \AA).    The FUV fluxes were corrected for the contribution from \lya\ emission of the Green Peas (described in \S \ref{lyameas_sec}), 
 and both bands were corrected for Galactic foreground extinction (as noted in \S 1).   From the corrected FUV - NUV color, we 
 calculated the UV power law slope, $\beta$, ($F_{\lambda} \propto \lambda^{\beta}$)\footnote{The UV slope has been shown to differ systematically depending on the adopted bandpass \citep{Howell}.  Therefore, these UV slopes are not strictly comparable to $\beta$ derived from the {\it International Ultraviolet Explorer} (IUE; e.g.\citealt{Calz94}), or GALEX observations of galaxies at $z\sim0$.  We take the present measurement as a reasonable approximation; our conclusions do not rely on $\beta$.}. 
The absolute FUV magnitude was calculated by interpolating 
 between the FUV and NUV bands to obtain 
 the luminosity, $L_{\nu}$, at 1500\AA.
 
 In the final column of Table \ref{properties} we list  $R_{UV}$, the size of the galaxy measured from the NUV acquisition images. For the purposes of 
 comparison in \S \ref{haew_sec}, we choose 
 a Petrosian radius, $r_{P20}$ as adopted by \cite{Hayes14}.   This radius defines the circular isophote where the local surface brightness is 20\% of 
 the internal surface brightness.

 \begin{figure*} 
 \begin{center} 
   \includegraphics[scale=0.6, viewport=0 0 1000 350, clip]{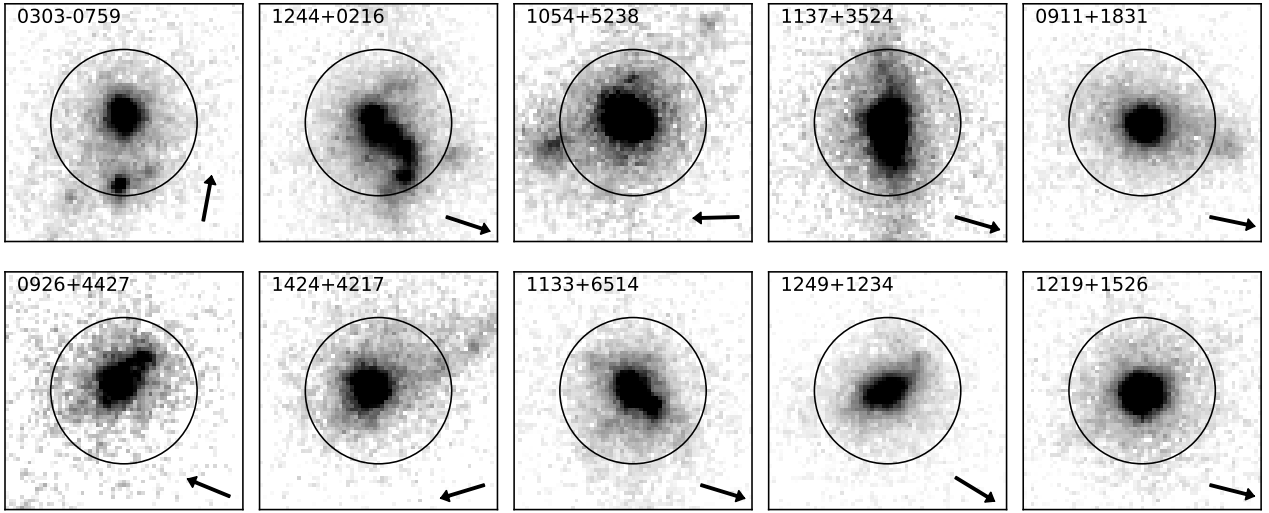}
\caption{COS/NUV acquisition images are shown for the present sample of Green Peas.  Cutouts are 1.6\arcsec\ on a side, shown oriented with 
north up and east towards the left.  The circle, with a 1\arcsec\ diameter, approximates the unvignetted portion of the COS aperture, while the full 
2.5\arcsec\ diameter aperture extends beyond the edge of the images.  The arrows indicate the dispersion direction in the spectra. The images are 
displayed in order of increasing \lya\ escape fraction (left to right, top to bottom; given in Table  \ref{lya_table} and described in \S \ref{lyameas_sec}).   \label{ta_images}  } 
\end{center}
\end{figure*}

\section{Data}

\subsection{Observations} 
The observations in this paper are part of two COS programs.  One of the galaxies that met our selection (0926+4427) was also classified as a 
Lyman Break Analog (LBA; \citealt{Heckman11}), and was previously observed with COS (GO 11727; PI T. Heckman).  The remaining 9 Green Peas were observed as part of GO 12928 (PI A.\ Henry).  

For all ten of the galaxies, the target acquisition was accomplished with NUV imaging,  configured with the Primary Science Aperture (PSA) and Mirror A.       
As detailed in the COS Instrument Handbook, an initial image is obtained and analyzed to find the peak NUV flux.  Next, {\it HST} is moved to place the peak flux
in the center of the COS aperture, and a second image is taken to verify the shift.      The target acquisition images of the Green Peas are shown in Figure \ref{ta_images}.    
These data indicate relatively compact emission, implying that  most of the UV continuum emission from these galaxies falls within the  central 
 1\arcsec\ diameter of the COS aperture (shown as circles) where vignetting is minimal.     (Likewise, if optical line emission closely follows the UV continuum, we expect negligible aperture losses in the 3\arcsec\ SDSS fiber spectroscopy.)
 
 The spectra were obtained in the FUV, using both the G130M and G160M configurations (in 9/10); with both gratings the rest-frame 
 wavelength coverage spans from approximately $950 - 1450$ \AA, or somewhat shorter/longer depending on the redshifts.  The 
 rest-frame wavelengths covered for each galaxy are listed in Table \ref{sample}, along with the exposure time for each grating.  For 1424+4217, the second orbit 
 containing the G160M observations failed; it did not qualify for a repeat observation because the program was more than 90\% complete.   In this case, the 
 observations cover the Green Pea \lya\ emission only for velocities $v\la 750$ \kms.    As recommended by the COS Instrument Handbook, all spectra were 
 taken at four FP-POS settings in each grating in order to mitigate the effects of fixed pattern noise.    
 In addition, the central wavelengths for each grating were chosen to avoid placing strong absorption lines between the A and B segments.    For one galaxy 
 (1054+5328) this choice was not possible in G130M, so two central wavelengths were used.      
 Since CALCOS does not combine data taken at different central wavelengths, we combined them by simply taking the mean (except in the gaps, where we  adopt the values from the spectrum that has coverage).  The errors were derived from taking the Poisson noise on the total counts, as we will describe below in \S \ref{noise}.

All data from GO 12928 were taken at the COS lifetime position two;  they were downloaded from MAST 2013 August 23 implying that they were processed with 
CALCOS  version 2.19.7.   The spectra from GO 11727 were taken in lifetime position one, and were downloaded on 2012 November 2  (CALCOS version 2.18.5).   

The full spectra for all ten Green Peas are shown in the Appendix. 

\subsection{Estimating Spectral Resolution}  
Because the Green Peas are spatially resolved, the spectral resolution that we achieve is reduced from  the 15 - 20 km s$^{-1}$ that is 
expected for point sources.  Nevertheless, Figure \ref{ta_images} shows that the galaxies are compact;  they do not fill the aperture, so the resolution should be
better than the 200 km s$^{-1}$ that is reported for filled aperture observations \citep{France}.   

In order to quantify the spectral resolution for the present sample, we create a model line spread function (LSF) from the NUV acquisition images and the point source
LSF\footnote{ \url{http://www.stsci.edu/hst/cos/performance/spectral\_resolution/} }.     For each galaxy,  we  create a one-dimensional profile in the dispersion direction,
 by summing the pixels in the cross dispersion direction of the acquisition image.    This profile is then convolved with the LSF to estimate how it is broadened by spatially resolved emission.  

The resultant model LSFs in both G130M and G160M have typical FWHM $\sim 12$ pixels (or 18 for the somewhat larger 1244+0216)  at 1450\AA. 
This scale is larger than the 6-8 pixel FWHM of the point-source LSF,  although both the point-source and model galaxy LSFs have wings extending beyond $\pm 20$ pixels.    
For a dispersion of 9.97 (12.23)  m\AA\  pixel$^{-1}$ in G130M (G160M),  the FWHM of the LSF corresponds to   25 (30) km s$^{-1}$ for typical Green Peas and 37 (46) km s$^{-1}$ for 1244+0216.   As will become evident in the later sections, the absorption lines are much broader and clearly spectrally resolved.   The \lya\ emission may
have somewhat poorer spectral resolution if it is more extended than the stellar component.

\subsection{Binning and Noise} 
\label{noise} 
We chose to bin the spectra to gain signal-to-noise and mitigate the potential effects of correlated noise \citep{Keeney}.  As we showed above, 
the spectral resolution for our galaxies is degraded relative to our expectation for point sources.    Binning 20 native COS pixels into one does not discard information for the galaxies in 
our sample. Therefore, we used the IDL {\tt rebin} function to calculate the mean of 20 adjacent fluxes, $f_{\lambda}$ and wavelengths, $\lambda$.   

Inspection of the error arrays for the unbinned spectra showed that CALCOS overestimated the statistical uncertainties in the nine Green Peas observed in GO 12928, which have lower S/N than the spectrum of 0926+4427\footnote{This mis-estimation is a known problem for low S/N data (Oliveira 2014, private communication).}.  We measured the noise in line-free regions of these spectra, and found that it was around three times smaller than implied by the error vector.     
Therefore, we re-calculated the error spectrum by taking the Poisson noise on the observed counts in each 20 pixel bin, 
according to \cite{Gehrels}.    The total counts per pixel were taken from the {\tt GCOUNTS} array in the one-dimensional COS spectrum, which includes both source and background counts.  
To convert the error on the counts to an error on the flux,  we divide by the exposure time and the sensitivity function.  The sensitivity function was simply estimated from the ratio of the calibrated and uncalibrated one-dimensional spectra (the {\tt FLUX} and {\tt NET} arrays provided by CALCOS).   We verified that the newly derived error vector is a sensible representation of the noise in line-free portions of the continuum.   Moreover, this method applied to the higher S/N spectrum of 0926+4427 showed good agreement with the CALCOS error spectrum. For the sake of uniformity with the rest of the sample, we adopt the re-calculated error spectrum for this object.

Significant amounts of non-Poissonian (correlated) noise have been noted in COS spectra \citep{Keeney}, so we also test for this effect in our data.   For this analysis, we repeated the spectral binning described above, creating  spectra with $N = $ 3, 5, 10, 20, 40, 100, and 200 COS pixels averaged.  We also include an unbinned spectrum to probe $N=1$.  In each spectrum, we subtracted a local continuum fit and measured the noise, $\sigma_N$,  in regions that were free from strong absorption lines.  
In the case of the Poisson limit, we expect the noise per bin to scale with the noise per single COS pixel  $\sigma_N = \sigma_1 N^{\beta}$, where $\beta = -0.5$.    In reality, we measure  $\beta = -0.43$ to $-0.48$, close to the Poisson limit.   This contribution from non-Poissonian noise is smaller 
than the contribution measured by Keeney et al;  it does not impact our conclusions so we do not correct for it.

\subsection{Velocity Precision}   
Since we are interested in kinematic features probed by our COS spectra, it is important to assess the wavelength solution and velocity precision of our data.    
We make three tests. First, we check that the geocoronal \lya\  1215.67 and \ion{O}{1} $\lambda 1302.17$  lines  lie at the correct observed 
wavelengths in the G130M 
spectra.    To make this comparison, we first remove the heliocentric velocity correction implemented by CALCOS.  
This step ensures that the geocoronal emission is in the rest-frame of the Earth.   After taking this correction into consideration, 
 the geocoronal emission features fall within -5 to +17 km s$^{-1}$ of their expected velocities.  

The second test that we make is a comparison of the Milky Way ISM absorption line velocities.  Because these lines may be Doppler shifted, they do not test the 
zero-point accuracy of the wavelength scale.  Rather, their consistency gives a measure of the precision of our absorption line velocities. 
The features that we use for this comparison are the \ion{Si}{2} $\lambda \lambda$1190.4, 1193.3,  $\lambda$1260.4, and $\lambda1526.7$, as well as \ion{C}{2} $\lambda$1334.5, and \ion{Al}{2} $\lambda 1670.8$.     
Although  higher ionization states are sometimes detected,   
 we do not include these lines becuase they can have different 
kinematics than the low-ionization lines (when observed in other galaxies, e.\ g.\ \citealt{Grimes, Steidel10}).    
From Gaussian fits to the Milky Way lines, we find velocities that are consistent within 20-40 km s$^{-1}$.

Finally, we confirm that the systemic redshift of the stars in the Green Peas are consistent with the \ion{H}{2} regions.     Redshifts measured from the photospheric  \ion{C}{3} 1175.5 line are compared to the redshifts  that we measured from the SDSS spectra  (Table \ref{sample}).  
In 8/10 galaxies, these redshifts agree to better than 41 km s$^{-1}$; for two others, the test is not possible:  the \ion{C}{3} line is undetected in 1219+1526 and it is 
contaminated by Milky Way absorption in 1137+3524.     Nevertheless, when measurable, the velocity offset between the \ion{H}{2} regions and stars are consistent 
within the uncertainties.    

In summary, these tests show that we can measure velocities in the present COS FUV spectra to better than 40 km~s$^{-1}$.   This level  of precision is confirmed for  COS observations of Ultra Luminous Infrared Galaxies \citep{Martin14}.

\section{\lya\ Emission From Green Pea Galaxies}   
\label{lyameas_sec} 

\begin{figure} 
\begin{center}
  \includegraphics[scale=0.43, viewport=1 1 800 775, clip]{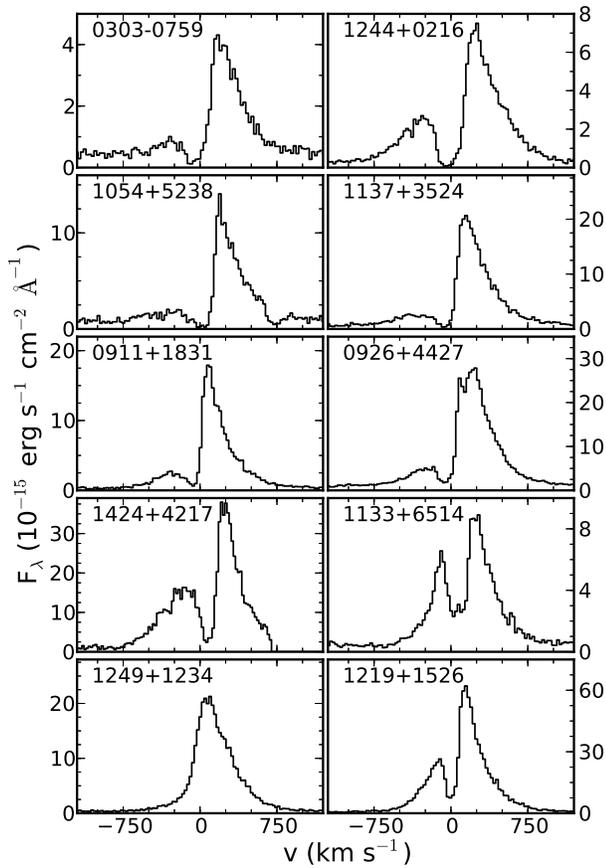} 
\caption{The 10 Green Pea galaxies observed with COS all show \lya\ in emission, even though they were not selected to exhibit this feature.   Broad wings on the lines clearly extend to  
several hundred km s$^{-1}$, and 9/10  show double-peaked emission.  The profile of 1424+4217 is truncated at the red-cutoff of the G130M spectrum; for this object, the G160M observation failed.   The spectra are displayed in order of increasing $f_{esc}^{Ly\alpha}$ (left to right, top to bottom).  \label{lya_spec} } 
\end{center}
\end{figure}

\subsection{The prevalence of strong, double peaked \lya\ emission} 
Figure \ref{lya_spec} shows the \lya\ spectra of the Green Peas in the present sample, displayed in order of increasing \lya\ escape fraction (estimated below). 
   Their appearance is remarkable, for a few reasons.  First,  all ten of the 
galaxies show \lya\ in {\it emission}.        This result is not trivial;  typical star-forming galaxies at all redshifts show a range of \lya\ strength, ranging from pure (even damped) 
absorption to P-cygni emission plus absorption, to pure emission \citep{Wofford, Pettini02, Shapley03}.    
 On the other hand, the Green Peas are not typical of nearby star-forming galaxies,  so the \lya\ emission seen in all ten galaxies is not entirely surprising.    Nearby, high equivalent width optical emission lines, like those in the Green Peas, have been  suggested to favor \lya\ emission  \citep{Cowie11}.  Furthermore, in more distant galaxies at $4 < z < 6$, the increasing incidence and strength of \lya\ emission \citep{Stark11} may go hand in hand with high equivalent width optical emission lines (inferred from contamination to broad-band photometry; \citealt{Shim11,Stark13, Labbe13}).   The detection of \lya\ emission in all ten Green Peas is 
 consistent with these claims.

The second noteworthy feature about the spectra in Figure \ref{lya_spec} is the prevalence of the double-peaked line shape.   Of the Green Peas that we observed with COS, 
9/10 share the spectral-morphology with both red-shifted and blue-shifted emission peaks.   
Again, this line shape is not typical of normal nearby star-forming galaxies.  \cite{Wofford} find only one doubled-peaked emitter in their sample of twenty \ha\ selected objects around $z\sim 0.03$.   And \cite{Martin14}, in a study of eight nearby ULIRGs, show that \lya\ can exhibit complicated kinematic profiles with broad {\it blueshifted} emission in more dusty environments. 
In contrast, at higher redshifts ($z\sim2-3$), \cite{Kulas} report that 30\% of UV-continuum selected galaxies with \lya\ emission show multiple-peaked profiles.    Comparison to the Green Peas, however, reveals the 
importance of spectral resolution in this measurement.   At the 200-500 \kms\  resolution used by \cite{Kulas}, many of the Green Pea spectra would be observed as single-peaked lines.    
Indeed, spectroscopy with 120 \kms\ resolution showed double-peaked  line profiles in 3/3 LAEs targeted by \cite{Chonis}.     Nevertheless,  the \lya\ line profiles of the Green Peas are notably different than the other low-redshift samples, where \wha\ is lower  and dust content is higher \citep{Wofford, Martin14, RT15}.    As indicated by \lya\ radiative transfer models, 
the blue peak should appear when the \ion{H}{1} column density is low, plausibly due to anisotropies in the gas distribution \citep{Verhamme14, Behrens, ZW14}.

In order to better understand how the unusual conditions in the Green Peas are influencing their \lya\ escape, 
we next provide quantitative measurements of the \lya\ lines and show how these compare to nearby samples.  
Then, in the remainder of the paper we will  use the UV absorption lines to explore how the conditions in the ISM and CGM affect the \lya\ emission. 
Radiative transfer models of the \lya\ spectral line profiles will follow in Orlitov\'a et al.\ (in prep).

\subsection{\lya\ Emission Line Measurements}

\begin{deluxetable*} {ccccccccccccc}[!ht]
\tablecolumns{13}
\tablecaption{\lya\ measurements from COS Spectroscopy of Green Pea Galaxies } 
\tablehead{
\colhead{ID}  &  \colhead{$F_{Ly\alpha}$ }    & \colhead{$L_{Ly\alpha}$}  &  \colhead{\lya/\ha} & \colhead{$f_{esc}^{Ly\alpha}$}  &  \colhead{$W_{Ly\alpha}$}   &
 $W_{Ly\alpha}^{red}$ & $W_{Ly\alpha}^{blue}$     &  &  \colhead{$v_{red}^{peak}$}  & \colhead{$v_{blue}^{peak}$} & \colhead{$v_{max}^{red}$}  
& \colhead{$v_{max}^{blue}$}  \\ 
\\
\cline{6-8}  \cline{10-13} \\
&   \colhead{(10$^{-14}$ erg s$^{-1}$ cm$^{-2}$)}     &   \colhead{(10$^{42}$ erg s$^{-1}$)} &  &  &   \multicolumn{3}{c}{(\AA)}  &  &  \multicolumn{4}{c}{(\kms)}  \\    
   } 
\startdata
0303--0759  & $1.1 \pm 0.2$     &  0.8 &    0.5 &  0.05  &  $9\pm 2$   &  $9 \pm 2$ & $ -0.1\pm 0.3$  &   & 170  & -290 &  900 & -400  \\
1244+0216 &  $2.0 \pm 0.1$    & 3.4  &     0.7 &   0.07   & $48 \pm 10$ &  $36  \pm 7$   & $12 \pm 2$ & & 250 & -280 & 1100  & -1000    \\
1054+5238 &  $1.7 \pm 0.2$   & 3.1 &     0.8  &   0.07  & $12 \pm 3$  & $12 \pm 2$   & $1 \pm 1$  & & 160  &  -250  &  700  & -700 \\  
1137+3524  & $3.8 \pm 0.2$   & 4.0 &      1.2  &  0.12   & $35 \pm 7$  & $33 \pm 7$ & $3 \pm 1$&  & 150 & -400 &  1100 & -800   \\ 
0911+1831 &  $3.3 \pm 0.1$ & 6.8  &       2.3  &  0.16  &  $59 \pm 12$  &  $48 \pm 10$ &  $8 \pm 2$   & & 90  & -280    & 1100  & -800 \\ 
0926+4427 &  $6.0 \pm 0.3$   & 5.4   &    2.3 &  0.20  &  $40 \pm 8$ &   $36 \pm 7$  & $ 5 \pm 1$&     & 160 &  -250 & 1200 & -800 \\ 
1424+4217  & $ 8.5 \pm 0.2$ & 8.0  &      2.4  &  0.25  &  $95 \pm 19$ & $ 57 \pm 12$  & $35 \pm 7 $  &  & 230 & -150  & \nodata  & -850  \\ 
1133+6514 & $2.1 \pm 0.1$   & 3.6 &     3.5   &    0.40  & $36 \pm 7$  &  $25 \pm 5$  & $11 \pm 2$ & & 230  & -100   & 1000  & -600   \\
1249+1234 &  $5.4 \pm 0.1$  & 11.3  &      4.4  &  0.41    &  $98 \pm 20$ & $74 \pm 15$  & $16 \pm 3$ & &70 &  \nodata  & 1300  & -700  \\
1219+1526 & $13.7 \pm 0.2$ & 14.7 &      5.5   &  0.62  & $164 \pm 33$ &  $118 \pm 24$  & $41 \pm 8$ &  & 140 & -100&  1300 & -950

\enddata
\label{lya_table} 
\tablecomments{\lya\ measurements from the COS spectra of the Green Peas give the fraction of \lya\ that escapes from within the few kpc probed 
by the spectroscopic aperture.      The \lya\ fluxes and equivalent widths are calculated by directly integrating the line profile to the velocity where it meets 
the continuum, and \wlya$^{red}$ and \wlya$^{blue}$ are calculated for $v >0$ and $v<0$, respectively.     The fluxes and equivalent widths assume a conservative 20\% error on the continuum flux.  
 Additionally, \lya\ kinematics give the velocities marking the red and blue peaks, as well as the maximal blue and red velocities where the emission reaches the continuum. The former velocities 
are good to better than 50 \kms\ in all cases except for the red peak of 0926+4427, which appears impacted by foreground absorption.   The maximal velocities a less certain, with typical errors 
around 200 \kms.  }
\end{deluxetable*}  
   
The \lya\ emission line measurements are presented in Table \ref{lya_table}.       The line flux is measured by directly integrating the emission line spectra out to the 
velocity where the continuum is met.  We adopt a conservative 20\% uncertainty on the continuum flux near \lya, since the broad N V 1240 \AA\ stellar wind feature makes continuum normalization difficult.  
This error makes little difference for the \lya\ flux uncertainties, but it dominates the error budget for equivalent width measurements.  
For 1424+4217, where the G160M spectrum was unavailable, we use the G130M observations even though they only cover \lya\ for $v \la 750$ km s$^{-1}$.    Since the other 
galaxies show only a small contribution at these velocities, we do not correct \lya\ measurement for missed flux.

Although the compact sizes of the Green Peas suggests minimal aperture losses in the continuum spectra, the \lya\ emission may be more extended.      At the redshifts of our sample, 
the unvignetted portion of the COS aperture shown in Figure \ref{ta_images} (1\arcsec\ diameter) corresponds to 2.8 - 4.0 kpc; the full COS aperture is 2.5 times larger. 
As a comparison, in the nearby Lyman Alpha Reference Sample (LARS), which is comprised of 14 galaxies with HST imaging of \lya, all but two objects show emission extending to at least 10 kpc \citep{Hayes14, Ostlin}.   Based on their curves of growth,  a  3-4 kpc diameter aperture would capture only one third to one half of the \lya\ flux.  
Indeed, aperture losses are confirmed for three galaxies in our sample.  One Green Pea, 0926+4427 is also identified as LARS 14, and two (1133+6514 and 1219+1526) are 
serendipitously covered by GALEX grism observations.  For the former, the COS aperture captured 40\% of the total large aperture luminosity estimated from LARS imaging 
($L_{Ly\alpha}^{total} \sim 1.4\times 10^{43}$ erg s$^{-1}$). The latter two cases show that 60 and 75\% of the GALEX grism flux is detected by COS.  Because the physical extent 
of the \lya\ emission may vary from galaxy to galaxy, and 7/10 of the Green Peas have no large-aperture measurements, we do not correct the COS measurements.
Ultimately, though, the COS measurements are interesting because they tell us about the \lya\ that is observed within the central few kpc. These quantities are important for comparing to high-redshift spectroscopic studies, where slits may subtend physical scales similar to the COS aperture. 

Besides the line fluxes, Table \ref{lya_table} lists the   \lya\ equivalent width,  luminosity, and the \lya\ escape fraction, $f_{esc}^{Ly\alpha}$.    
The latter quantity is defined as the ratio of the observed \lya\ luminosity to the intrinsic \lya\ luminosity, $L_{Ly\alpha}^{obs} / L_{Ly\alpha}^{int}$ (see also \citealt{Hayes14}).    
The intrinsic \lya\ luminosity is  inferred from the dust-corrected \ha\ luminosity times the intrinsic case-B ratio of $L_{Ly\alpha}/L_{H\alpha}  \sim 8.7.$\footnote{The intrinsic  
$L_{Ly\alpha} / L_{H\alpha}$ ratio predicted by case-B recombination theory is often reported as 8.7 and incorrectly attributed to \cite{HS87} or \cite{Brocklehurst71}.    
Instead, the canonical assumption where 2/3 of all ionizing photons lead to a \lya\ photon yields \lya\ /\hb\ = 23.1, and when \ha/\hb = 2.86 (for $T_e = 10,000K$), 
we have \lya /\ha\ = 8.1.  However, the \lya\ /\hb\ ratio is density dependent, and 2/3 of recombinations lead to \lya\ photons only in the low density limit.  
At higher densities, collisions bump electrons from 2 $^2S$ to 2 $^2P$, thereby suppressing 2-photon continuum and enhancing \lya\ emission.  
 \cite{DS03} tabulate \lya/\hb\  and \ha/\hb\ ratios for a range of temperatures and densities.     
For the present sample, the [\ion{S}{2}] $\lambda \lambda 6716, 6731$ ratio indicates electron densities, 
$n_e$, from 100 to 800 cm$^{-3}$, implying intrinsic   \lya/\ha =8.1 - 9.2.  
We adopt 8.7 as intermediate to these extremes. }   We also include measurements of the red and
blue-side $W_{Ly\alpha}$ for  the 9/10 double peaked \lya\ lines.  These quantities are calculated by directly integrating the 
emission profiles for $v>0$ and $v<0$ separately.  In two cases (0303-0759 and 1054+5238) the equivalent width of the blue peak is consistent with zero; this finding is {\it not} an indication that the 
blue peak is undetected.  Rather, the blue emission is weak, and there is net absorption around zero velocity.

Finally, Table \ref{lya_table} also lists  kinematic signatures from the \lya\ lines.   The velocities of the blue and red peaks are given, and the broad wings 
of the lines are quantified by calculating the velocity where $F_{\lambda}$ reaches the continuum.

\section{Comparison to nearby \lya\ samples}  
\label{haew_sec}

In this section, we explore the role of dust and stars in regulating the \lya\ output of the Green Peas.    In this context, it is useful to compare the Green Peas to 
nearby galaxies that have also been observed in \lya.  We choose two samples with published \lya\ measurements:   LARS \citep{Hayes13,Hayes14, Ostlin, RT15}
 and nearby galaxies identified as \lya\ emitters (LAEs; $W_{Ly\alpha} > 15-20$ \AA) from GALEX grism surveys \citep{Deharveng, Scarlata09, Cowie11}.     We acknowledge that the selection effects and aperture size likely  influence this comparison.  On one hand,  the GALEX LAEs were \lya\ selected  via slitless grism spectroscopy with a large 5\arcsec\ FWHM PSF.   Under this configuration, we expect  that most of the \lya\ emission should be included in the measurements, but the sample is biased towards rare, strong \lya\ emitting galaxies.    On the other hand, the measurements reported for the \ha-selected LARS galaxies are integrated inside Petrosian apertures that are defined using the \lya\ plus continuum images.  These apertures, which range in diameter from 2.6 - 32 kpc, do not capture all of the \lya\ emission detected at large radii in the LARS images.  For the low-mass, low-dust, high $W_{H\alpha}$ end of the LARS sample, 
the typical 4 kpc diameter aperture is not too different than the COS aperture at $z\sim 0.2$.  Indeed, for LARS 14/0926+4427, the Petrosian aperture \lya\ flux from the LARS image is comparable to (83\% of) the
the flux included in the COS spectrum.  However,  as we noted in \S \ref{lyameas_sec}, these measurements account for less than half of the total \lya\ flux in the LARS image. 
Nevertheless, we conclude that the LARS Petrosian aperture measurements and COS/Green Pea spectra are still useful, because they give a sense of how much \lya\ is escaping from the central regions of the galaxies.  
Finally, we acknowledge that  \lya\ from  nearby COS-observed galaxies has also been presented by \cite{Wofford},  
but we exclude this sample from comparison because their close proximity ($z \sim 0.03$)  implies that the COS 
aperture  subtends only 0.6 kpc (unvignetted).  Indeed, their COS NUV continuum images show much greater spatial extent 
compared to the Green Peas, suggesting that the \lya\ measurements are not easily compared to the other samples 
considered here. 


\begin{figure} 
\begin{center} 
  \includegraphics[scale=0.48, viewport= 0 0 500 400, clip]{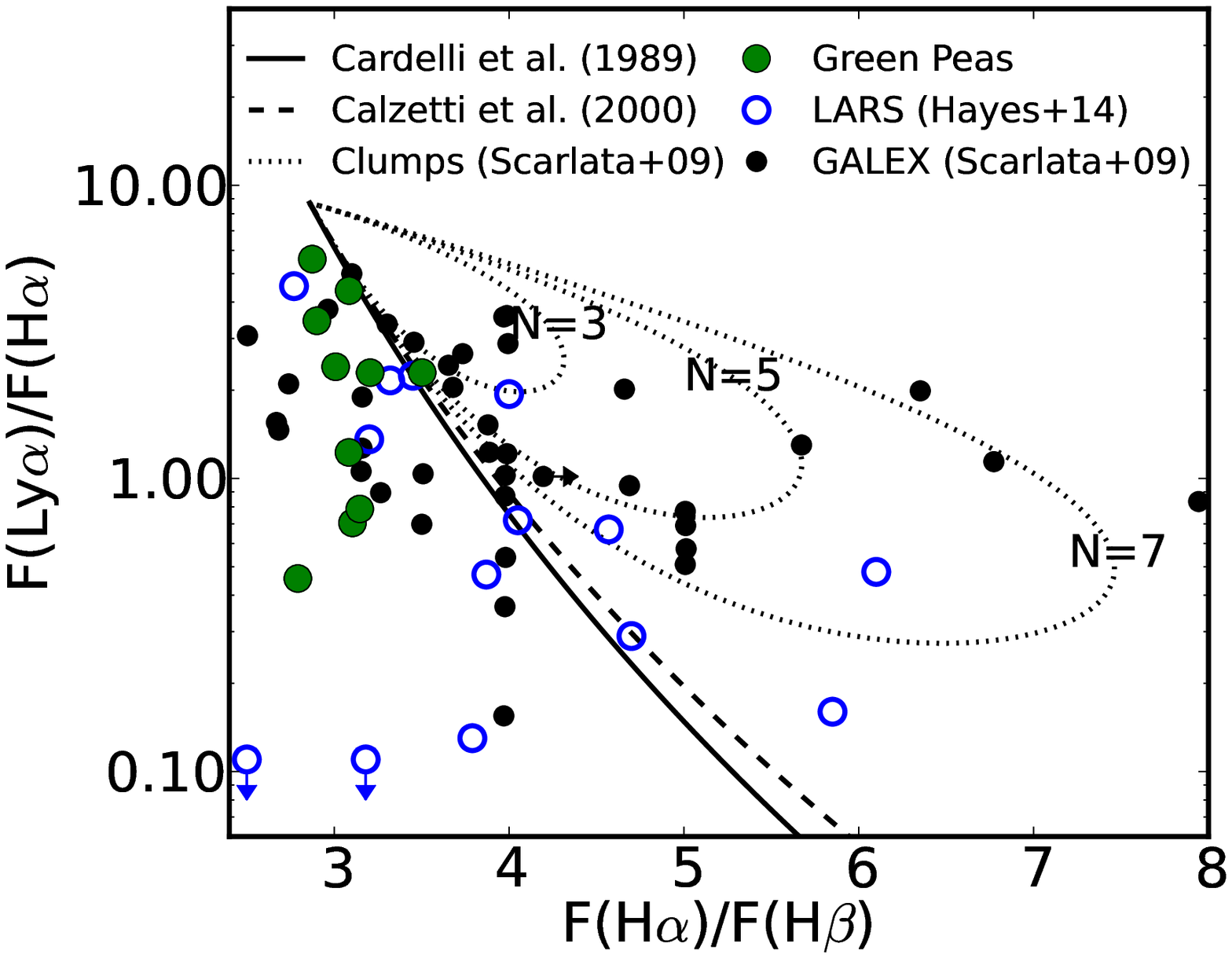} 
\caption{The low \lya\ to \ha\ flux ratios of the Green Peas cannot be explained by dust.  Solid and dashed lines show how, in the absence of resonant scattering,
 the intrinsic  \lya\ to \ha\ flux ratio would respond to dust extinction for the \cite{Calzetti} and \cite{Cardelli} extinction curves.   Additionally, clumpy dust models from \cite{Scarlata09} are
  shown for 3, 5 and 7 clumps; they form loops when high-clump optical depth returns the flux ratios to their intrinsic, dust-free limits.   Statistical errors for the Green Peas are small compared to the size of the data points,
  and are not reported for the other samples.  Comparison with the LARS galaxies and the GALEX LAEs shows that 
  the Green Peas have lower \ha/\hb\ flux ratios, indicating less dust obscuration.   GALEX LAEs identified as AGN have been excluded. 
  \label{lya_ha_hb} } 
  \end{center}
\end{figure} 

A straightforward approach for quantifying \lya\ escape is to directly compare \lya\ flux to the \ha\ and \hb\ emission that probe the nebular gas and dust. 
In Figure \ref{lya_ha_hb}, we plot the \lya\ to \ha\ flux ratio against the \ha\ to \hb\ flux ratio for 
the Green Peas, LARS galaxies, and GALEX LAEs.   The latter sample is comprised of 45 galaxies 
taken from  \cite{Scarlata09} and \cite{Cowie11}, where slit-loss corrections facilitate the comparison between \lya\ and \ha\ fluxes. Additionally, we show predictions from case-B recombination theory, assuming extinction laws from \cite{Cardelli} and \cite{Calzetti}. 
In the absence of resonant scattering, the flux ratios should follow these lines.    We also show the clumpy dust models of  \cite{Scarlata09} which form 
loops  when the limit of high clump optical depth returns the observed line ratios to the intrinsic case-B ratios. While these models are needed to explain some of the high \lya\ /\ha\ ratios in dusty galaxies, 
they do not explain line ratios of the Green Peas which have low \lya\ to \ha\ for their dust content. 
Although the Green Peas show little to no dust, their \lya\ to \ha\ flux ratios span a factor of 10.  Two of the Green Peas, 1249+1234 and 0911+1831, 
fall close to the \lya\ to \ha\ ratio that is predicted for their dust extinction, but  the remaining eight show ratios that are too low to be explained by dust extinction alone.  

\begin{figure}
\begin{center}
  \includegraphics[scale=0.4, viewport=0 0 1000 1100, clip]{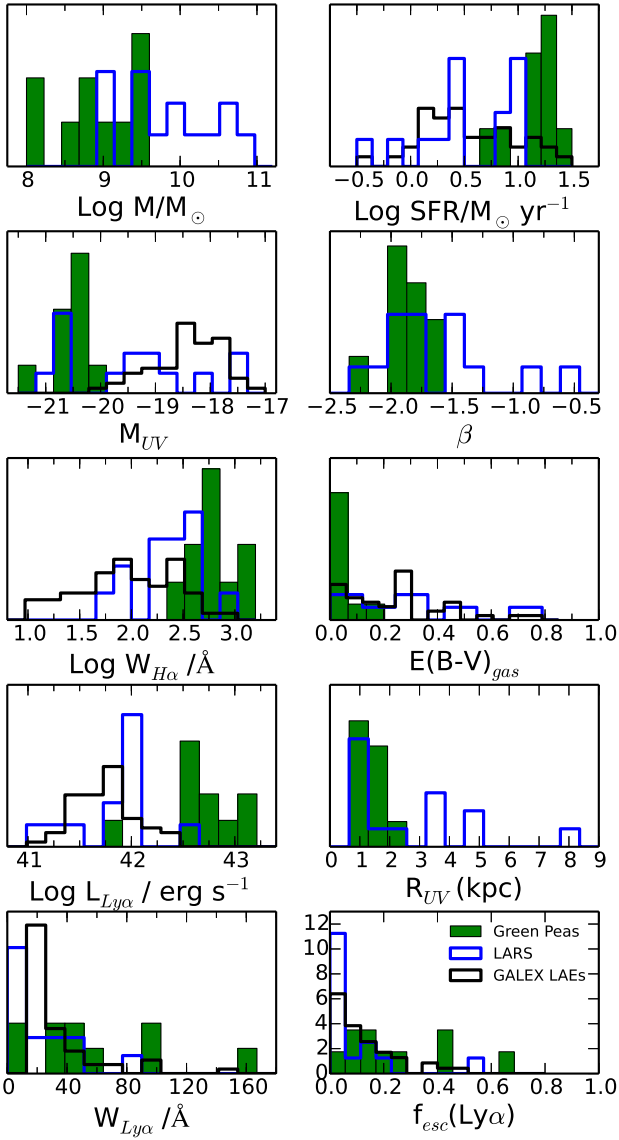} 
\caption{The Green Peas are compared to the LARS and GALEX LAE samples through their distributions of dust, stellar population properties, and \lya\ luminosity, equivalent width, and escape fraction.  
The LARS data are taken from \cite{Hayes13, Hayes14}, and the GALEX data are taken from Scarlata et al. (2009; SFRs and $E(B-V)_{gas}$) and Cowie et al. (2011; $M_{UV}$, $W_{H\alpha}$, and L$_{Ly\alpha}$ and \wlya).  
The $R_{UV}$ is the same isophotal Petrosian radius defined in \cite{Hayes13, Hayes14}, measured in the FUV for LARS and the NUV acquisition images for the Green Peas. The distributions are normalized since the samples sizes being compared are varied.  To our knowledge, stellar mass, $\beta$ slopes, and UV sizes of the GALEX LAEs have not been published.  \label{property_hist}  }
\end{center}
\end{figure}

To clarify how the Green Peas sample a different range of galaxy properties than  LARS and the GALEX LAEs, 
 Figure \ref{property_hist} shows histograms in stellar mass, SFR, $M_{UV}$, UV-slope, $\beta$, $W_{H\alpha}$, $E(B-V)_{gas}$, $L_{Ly\alpha}$, the UV isophotal Petrosian radius, $R_{UV}$, as well as  \wlya, and   $f_{esc}^{Ly\alpha}$.
     The derivation of these properties are outlined in \S \ref{sample_sec} for the Green Peas and \cite{Hayes13, Hayes14} for LARS.  For the GALEX LAEs, we take the $M_{UV}$, $W_{H\alpha}$, 
      $L_{Ly\alpha}$ and \wlya\ from Table 1 and 2 in Cowie et al.\ (2011; 44 galaxies).  The SFR and $E(B-V)_{gas}$  are calculated from the slit-loss corrected line fluxes noted above (from \citealt{Scarlata09} and \citealt{Cowie11}; 45 galaxies), assuming a \cite{Calzetti} extinction law, an intrinsic \ha/\hb\ ratio of 2.86, and the \cite{Kennicutt} SFR calibration (converted to a \citealt{Chabrier} IMF). GALEX LAEs identified as AGN have been excluded.

\begin{figure} 
 \begin{center}
\includegraphics[scale=0.43, viewport=1 1 800 460, clip]{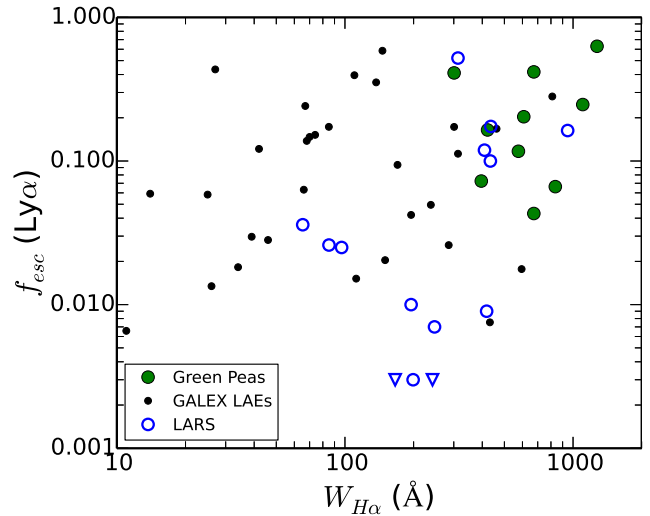} 
 \caption{The \lya\ escape fraction is strongly correlated with \wha\ in the combined sample of 
 Green Peas and LARS galaxies.      The Spearman rank correlation coefficient is 0.64, and the  probability of the null hypothesis is  $8.1\times 10^{-4}$.    The GALEX LAEs, on the other hand, show no significant correlation. The LARS measurements are taken from \cite{Hayes14}, and the quantities for the GALEX LAEs are calculated from data tabulated in \cite{Scarlata09} and \cite{Cowie11}.       }
 \label{haew_fig}  
 \end{center}
\end{figure}

Figure \ref{property_hist} shows that the Green Peas occupy a region of parameter space that is poorly sampled by other studies.     Their stellar 
masses overlap with the low-mass end  of the LARS galaxies, but extend to masses an order of magnitude smaller.   At the same time, the Green Peas' SFRs 
and UV luminosities are, on average, higher than the LARS galaxies and GALEX LAEs.   They are uniformly low in dust, with $E(B-V)_{gas} < 0.2$ and $\beta \sim -2.0$.       
Finally, Figure \ref{property_hist} shows that the sizes of the Green Peas are similar to the  more compact half of the LARS galaxies.     These different Green Pea properties seem to impact the \lya\ output:  the \lya\ luminosities are an order of magnitude larger than are observed for LARS and the GALEX LAEs.  In fact, among these three samples, only the Green Peas have \lya\ luminosities in the range of most high redshift LAEs: $L_{Ly\alpha} \ga 10^{42.5}$ erg s$^{-1}$ \citep{Ouchi10}.    Moreover, \wlya\ and $f_{esc}^{Ly\alpha}$ show a broad range of values for the Green Peas, while  the LARS galaxies are more peaked  at low values.  (The GALEX LAEs, by definition, exclude the low values of \wlya\ and $f_{esc}^{Ly\alpha}$).  In fact, Figure \ref{haew_fig} shows that in the combined set of LARS and the Green Peas, \wha\ is strongly correlated with \wlya.  The Spearman rank  correlation coefficient\footnote{The upper limits on $f_{esc}^{Ly\alpha}$ (corresponding to net \lya\ absorbers in the LARS galaxies) are set to zero in this statistical test. The Spearman correlation coefficient, using ranks, is not sensitive to upper limits in this case.}  is 0.64 and a spurious correlation is rejected with a probability of $8.1\times 10^{-4}$.      

 This observation that the low masses, high SFRs, and low dust content of the Green Peas may favor \lya\ emission is consistent with trends reported in \cite{Cowie11}.   
These authors compared GALEX LAEs to a UV-continuum selected control sample and found that the LAEs had bluer colors, more compact sizes, lower metallicities, and  higher $W_{H\alpha}$ than their non-emitting counterparts.     Similarly,  Cowie et al. find \lya\ emission to be more common in samples with higher \wha, and also reported a weak correlation between \wha\ and \wlya.   Here,  by actually comparing $f_{esc}^{Ly\alpha}$  instead of \wlya\  for the optical emission line selected samples, we see that this correlation probably originates from increased \lya\ escape, rather than young stellar populations with intrinsically high \wlya.     At the same time, however, we detect no statistically significant correlation between $f_{esc}^{Ly\alpha}$  and \wha\ for the GALEX LAEs.

Regardless, a relation between $f_{esc}^{Ly\alpha}$ and \wha\ (or any of the other properties highlighted in Figure \ref{property_hist}),   does not explain the  physical mechanism regulating \lya\ escape.   Rather, it suggests that gas properties associated with the youth of a stellar population increase \lya\ escape.  In the sections that follow, we use the ultraviolet interstellar absorption lines to further investigate the role of this gas.

\section{\lya\ emission and the absorption strength of interstellar gas}  
\begin{deluxetable*} {cccccc}[!ht]
\tablecolumns{6}
\tablecaption{Observed UV Absorption and Emission Lines} 
\tablehead{
\colhead{Ion}    & \colhead{$E_{ion}$} &  \colhead{Vacuum Wavelength} & \colhead{$f_{lu}$ }  & \colhead{$A_{ul}$} &\colhead{$E_{low} - E_{up}$}    \\ 
  & (eV) &  (\AA)  &  & (s$^{-1}$)  & (eV) 
   } 
\startdata  
\ion{H}{1} Ly$\epsilon$ & 0 & 937.80 & 0.0078 & $ 1.64 \times 10^{6}$ & 0.00   - 13.22070331 \\
\ion{H}{1} Ly$\delta$  &  &  949.74   & 0.014   & $ 4.12 \times 10^{6}$ & 0.00   -  13.0545011 \\ 
\ion{H}{1} Ly$\gamma$ & & 972.54  &   0.029   &  $1.28 \times 10^{7}$ & 0.00 - 12.7485388  \\
\ion{H}{1} Ly$\beta$  &   & 1025.72  &   0.08   & $5.57 \times 10^{7}$   & 0.00 -   0 12.0875046  \\ 
 \ion{H}{1} \lya\            &     & 1215.67  & 0.46  & $4.69 \times 10^{8}$  & 0.00  - 10.1988353 \\ 
\ion{O}{1}  & 0     & 1302.17   & 0.05    &  $3.41\times 10^{8}$  &  0.00 - 9.5213634 \\ 
\ion{O}{1}* &      &  1304.86   & 0.05   &  $2.03\times 10^{8}$ & 0.0196224 - 9.5213634 \\ 
                 &       &  1306.03   & 0.05   &  $6.76\times 10^{8}$ & 0.0281416 - 9.5213634 \\ 
\ion{Si}{2} &  8.15  & 1190.42  &  0.277   & $6.53\times 10^{8}$ & 0.00  -- 10.415200  \\
                 &         & 1193.29   & 0.575   & $2.69\times 10^{9}$ &   0.00 --10.390117\\
                 &         & 1260.42    & 1.22    & $2.57\times 10^{9}$ &  0.00 -- 9.836720  \\
                 &         &  1304.37   & 0.09  & $3.64\times 10^{8}$ & 0.00 - 9.505292 \\ 
  \ion{Si}{2}* &  & 1194.50  & 0.737  & $3.45\times 10^{9}$  &  0.035613  - 10.415200 \\
                  &    &  1197.39 &  0.150  & $1.40\times 10^{9}$  &   0.035613 -  10.390117 \\  
                &      & 1264.74  & 1.09   &   $3.04\times 10^{9}$ & 0.035613 -  9.838768 \\ 
                &      &  1309.28 & 0.08 &   $ 6.23 \times 10^{8}$ &   0.035613 - 9.505292 \\ 
\ion{C}{2} & 11.26 &  1334.53 & 0.129 &  $2.42\times 10^{8}$   &  0.00 - 9.290460    \\
\ion{C}{2}*  &           & 1335.71 & 0.115 & $2.88\times 10^{8}$ &  0.007863 -  9.290148 \\ 
\ion{Si}{3}  &  16.34  &   1206.50 &   1.67 &  $2.55\times 10^{9}$ & 0.00 - 10.276357 \\
\ion{Si}{4} &  33.49  &  1393.76  &  0.513  & $8.80\times 10^{8}$ &  0.00 - 8.895697  \\
                &              &  1402.77  &   0.255  & $8.63\times 10^{8}$ &  0.00 - 8.838528 
\enddata
\label{ions} 
\tablecomments{Atomic line data are given for the transitions considered in this paper.  Values are taken from the NIST Atomic Spectra Database.     }
\end{deluxetable*}

\label{ewsec}

 \begin{figure} 
 \begin{center}
\includegraphics[scale=0.43, viewport=1 1 800 825, clip]{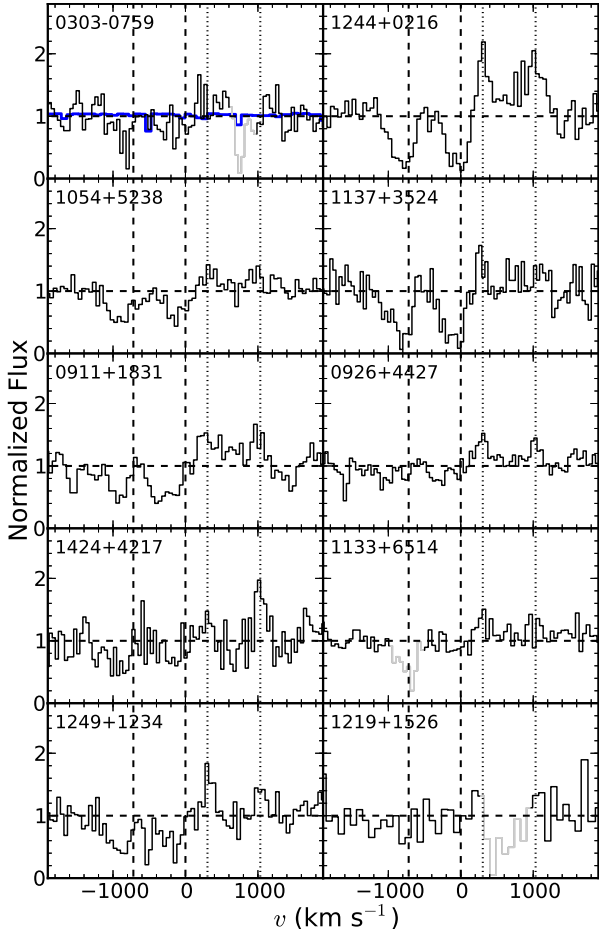} 
\caption{The COS spectra of 10 Green Peas show weak, blueshifted absorption in \ion{Si}{2} 1190.4 and 1193.3.  The velocity scale is 
appropriate for \ion{Si}{2} $\lambda$ 1193.3, with zero velocity marked by the vertical dashed line.  The vertical dashed line around -700 \kms\ shows the expected 
velocity for absorption in \ion{Si}{2} 1190.4, and the \ion{Si}{2}* emission at 1194.5 \AA\ and 1197.4 \AA\ is marked by dotted lines.    The portions of the spectra that are plotted 
 in grey are affected by Milky Way absorption or extra noise, except for the case of 1133+6514 where we suspect intervening \ion{Si}{2} $\lambda$ 1260 from an absorber at $z=0.17$.  The horizontal dashed line shows unity in the normalized spectra.   The blue spectrum, overplotted in the top left panel, is a $Z = 0.002$ Starburst99 model, which confirms that {\it stellar} \ion{Si}{2} absorption is absent from young, UV-luminous stellar populations.
 The spectra are displayed in order of increasing $f_{esc}^{Ly\alpha}$ (left to right, top to bottom).} 
 \end{center} 
\label{si21190}
\end{figure}

\begin{figure} 
\begin{center}
\includegraphics[scale=0.43, viewport=1 1 800 825, clip]{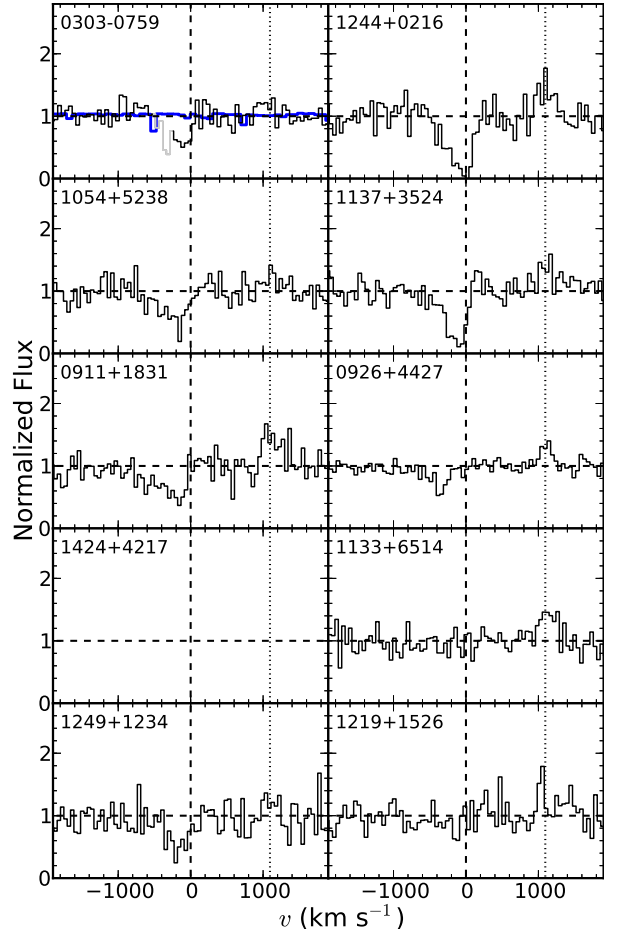} 
\caption{Same as Figure \ref{si21190}, but for \ion{Si}{2} $\lambda$ 1260.4 and \ion{Si}{2}*  $\lambda$ 1265.0.  These lines fall beyond
our wavelength coverage for 1424+4217.}
\label{si21260}
\end{center}
\end{figure} 

\begin{figure} 
 \begin{center} 
\includegraphics[scale=0.43, viewport=1 1 800 825, clip]{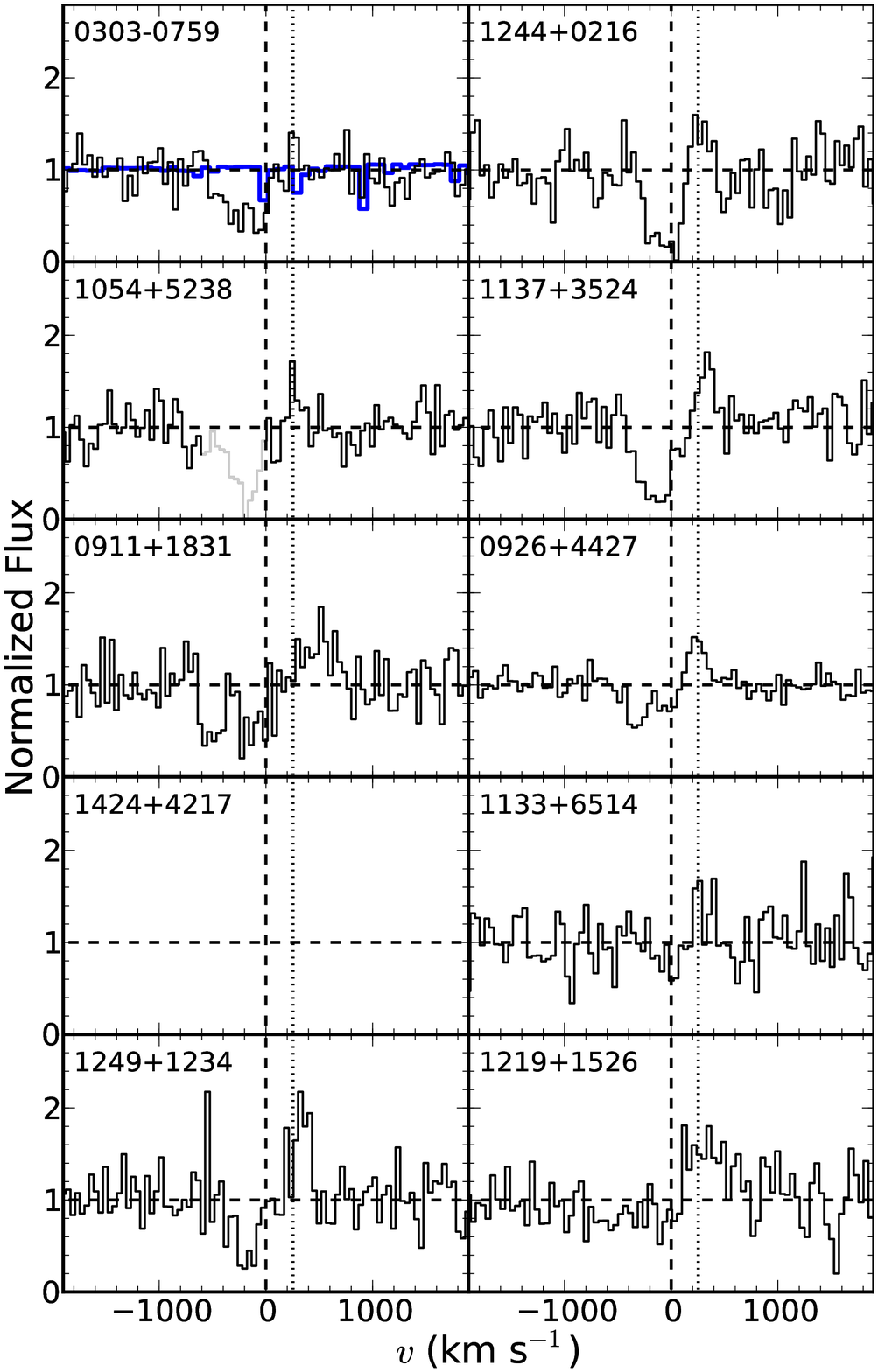} 
\caption{Same as Figures \ref{si21190} and \ref{si21260}, but for \ion{C}{2} $\lambda$ 1334.5 and \ion{C}{2}*  $\lambda$ 1335.7.   
Again, these lines fall beyond
our wavelength coverage for 1424+4217.     For 1054+5238, the \ion{C}{2} $\lambda$ 1334.5 absorption  is heavily contaminated by strong Milky Way \ion{Al}{2} $\lambda$1671.}
\label{c21334}
\end{center}
\end{figure}

 The most common approach to studying the impact of outflows on \lya\ is to observe low ionization interstellar (LIS) metal lines.    With ionization potentials less than 13.6 eV, ions such as \ion{Si}{2} and \ion{C}{2}  can trace both neutral and ionized hydrogen.   As such, these transitions are typically used to quantify the \ion{H}{1} gas that scatters \lya\ photons.   In Figures \ref{si21190}, \ref{si21260} and \ref{c21334}, we show the COS spectra covering \ion{Si}{2} $\lambda \lambda 1190, 1193$, $\lambda 1260$, and \ion{C}{2} $\lambda 1334$.    The spectra have been normalized by linear fits to the local continuum, and zero velocity is marked by a dashed vertical line.    Similarly, dotted lines mark the expected locations of fluorescent fine-structure emission lines, \ion{Si}{2}* and \ion{C}{2}*.    These lines form when electrons  
 excited by absorption in the  \ion{Si}{2} $\lambda \lambda 1190, 1193$, $\lambda 1260$, and \ion{C}{2} $\lambda 1334$ transitions subsequently decay to the excited ground state (followed by a fine structure transition to the ground state).   We exclude from our analysis the \ion{O}{1} and \ion{Si}{2} lines at $\lambda 1302$, $\lambda 1304$, because their small wavelength separation and contribution from fluorescent and resonant re-emission (\ion{O}{1}* $\lambda 1304$) complicates the interpretation (see spectra shown in the Appendix).   We list the rest-frame vacuum wavelengths, ionization potentials, absorption oscillator strengths, $f_{lu}$, and  emission coefficients, $A_{ul}$  for these lines in  Table \ref{ions}.   Additionally, we note that contamination to these ISM features by stellar absorption is negligible.   In the top left panel of  \ref{si21190}, \ref{si21260} and \ref{c21334}, we show a 50 Myr old, continuous star forming model spectrum, with $Z= 0.002$ \citep{Leitherer14}.  Similar stellar absorption is seen across a wide range of young, metal-poor stellar population properties, where UV spectra are O-star dominated.

 \begin{figure*} 
 \begin{center} 
\includegraphics[scale=0.48, viewport=1 1 850 900, clip]{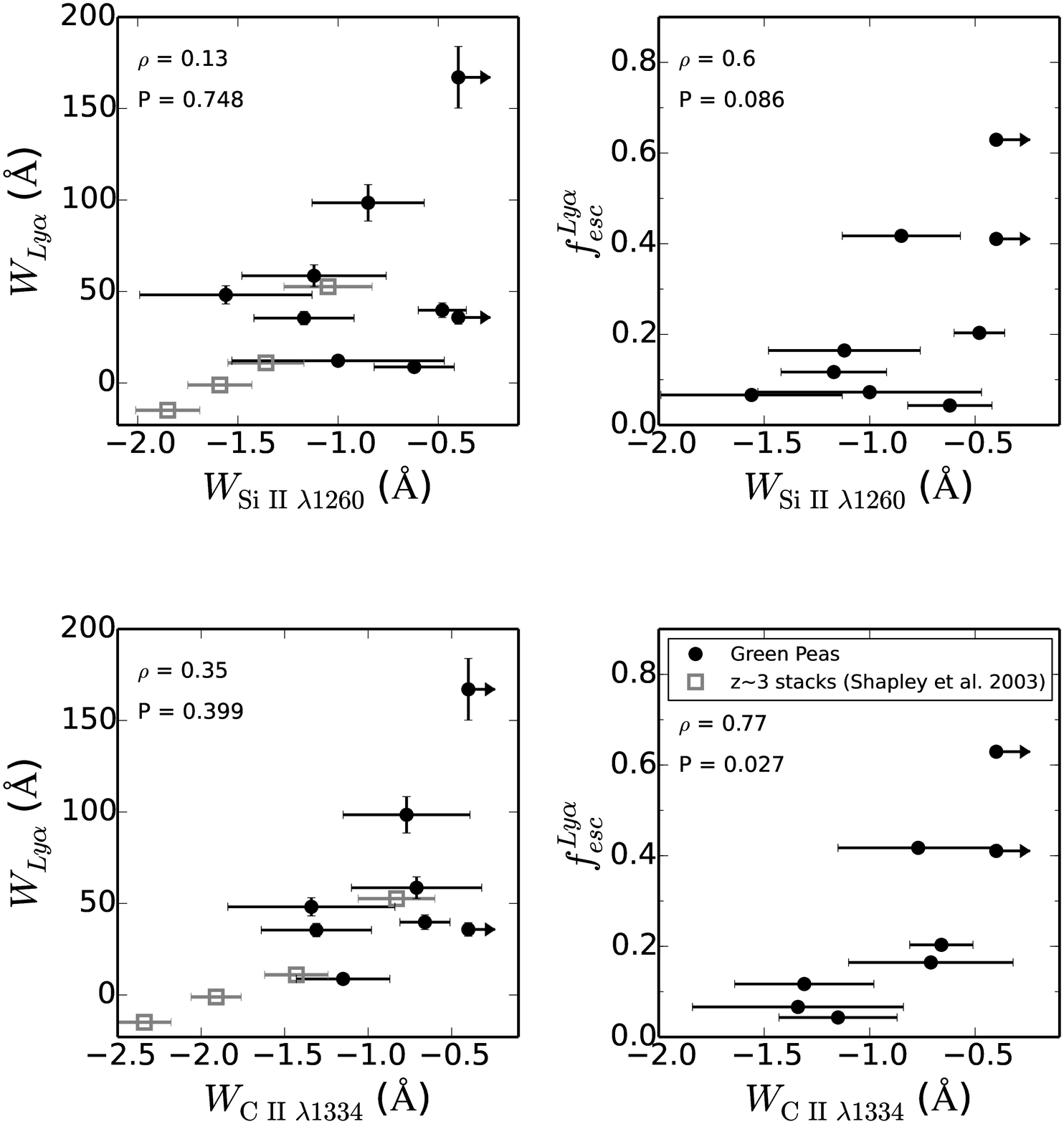} 
\caption{The rest-frame equivalent widths of \lya\ emission (left panels) and the \lya\ escape fraction, $f_{esc}^{Ly\alpha}$ (right panels)  are plotted against the absorption equivalent width of the low ionization \ion{Si}{2}  $\lambda$ 1260 and \ion{C}{2} $\lambda 1334$  lines.  The Spearman correlation coefficient, $\rho$, and probability of the null hypothesis, P, are given in the upper left of each panel. Comparison data from stacked spectra of $z\sim3$ galaxies are shown as grey squares \citep{Shapley03}.   While the Green Pea data are consistent with an extrapolation of the $z\sim3$ data, this correlation is only significant  when $f_{esc}^{Ly\alpha}$ is used to quantify \lya\ output.} 
\label{ew1260fig}
\end{center}
\end{figure*}

Weaker equivalent width of these LIS absorption lines have been associated with increased \wlya\ in stacked spectra
of galaxies at $z\sim 3-4$   \citep{Shapley03, Jones12}.   Qualitatively, the Green Peas lend some support to this scenario:  the two strongest
absorbers  1244+0216 and 1137+3524 are among the lower $f_{esc}^{Ly\alpha}$ Green Peas, and the weakest absorbers (1133+6514 and 1219+1526) are among those with the highest  $f_{esc}^{Ly\alpha}$.   To quantify this trend, we measure equivalent widths of the LIS metal absorption lines by directly integrating the normalized spectra 
out to the velocity where the absorption meets the continuum.     The equivalent width uncertainties are determined by propagating the error vector over the same velocity range,  including a systematic 
10\% uncertainty on the continuum normalization.  For undetected lines, we approximate the upper limit by taking the equivalent width of the marginally detected lines.   For the LIS lines
this threshold is around $0.5$ ($0.4$) \AA\ in the observed (rest) frame.      These measured equivalent widths are listed in Table \ref{ewtab}. 

In Figure \ref{ew1260fig}, we compare \wlya\ to the equivalent widths of  \ion{Si}{2} $\lambda 1260$  and  
\ion{C}{2} $\lambda 1334$ for both the Green Peas and the stacked $z\sim 3$ LBG sample from \cite{Shapley03}.       In this plot, the Green Peas are consistent with the  $z\sim3$ LBGs, although 
they extend to higher \wlya\ and weaker  $W_{LIS}$.    Nevertheless, we do not detect a significant correlation;  the Spearman correlation coefficient ($\rho$) and probability of the null hypothesis ($P$) are given in the upper left of each panel in Figure    \ref{ew1260fig}. 
However,  since \wlya\ is only a rough proxy for \lya\ escape,  we recast this diagram using $f_{esc}^{Ly\alpha}$  in the right panels of  Figure \ref{ew1260fig}.  
Here, the significance of the correlation is improved:  the Spearman coefficient (which, using ranks, is not sensitive to the non-detections in this case) implies
that the correlation is robust at 91\% confidence for \ion{Si}{2} and 97\% confidence for \ion{C}{2}.  Although these trends are only marginally significant, the consistency between \ion{C}{2} and \ion{Si}{2} supports a real correlation.  Moreover, these trends work in the same direction as the $z \sim 3- 4$ measurements 
from stacked spectra.     It appears that conditions which create weaker LIS absorption lines also favor greater $f_{esc}^{Ly\alpha}$.     We will return to this trend and discuss 
its possible meaning in \S \ref{interp}, where we offer a more comprehensive interpretation of the data.

\begin{figure} 
\begin{center} 
\includegraphics[scale=0.43, viewport=1 1 800 825, clip]{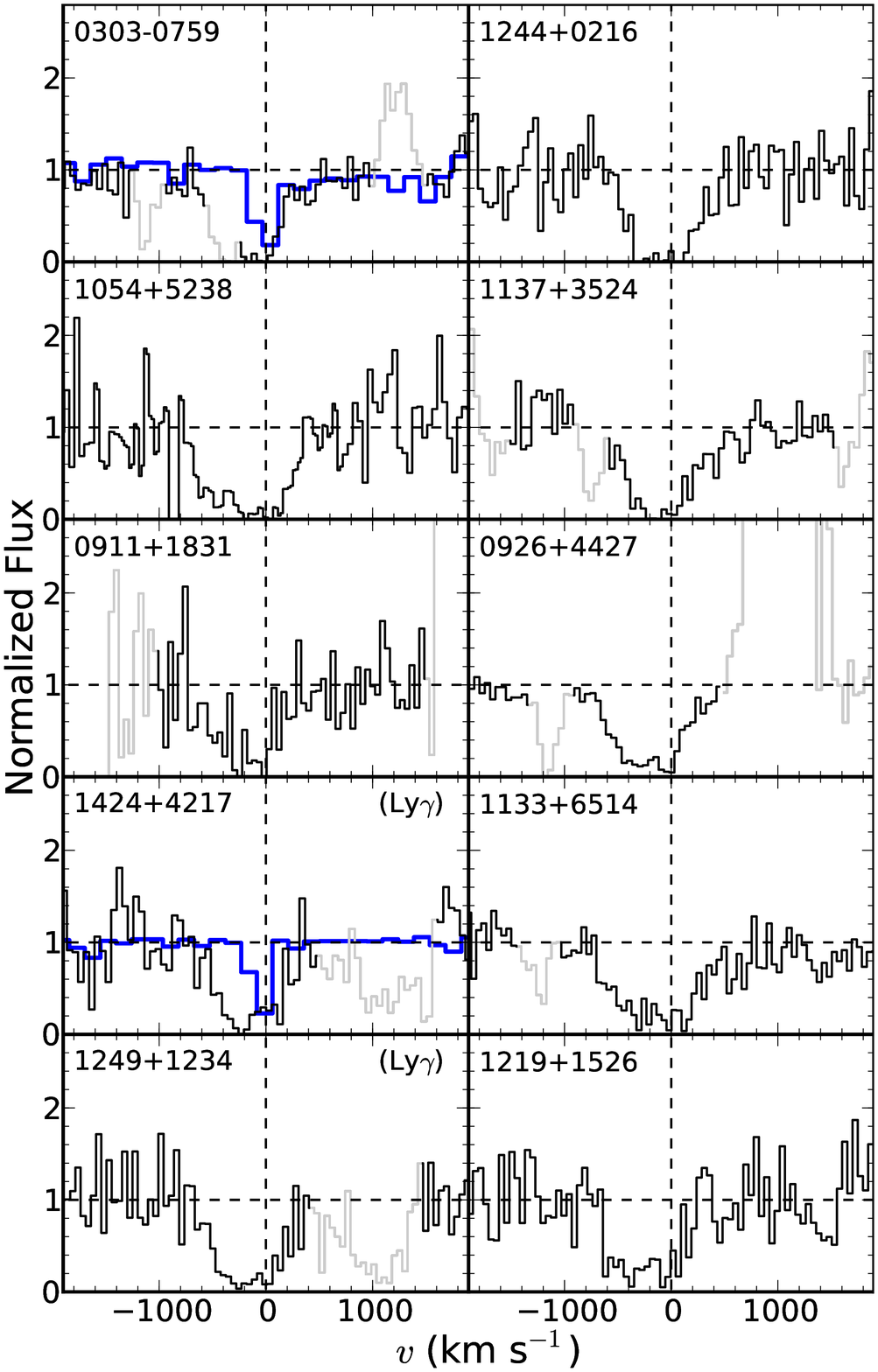} 
  \caption{Lyman series absorption line profiles directly trace hydrogen gas.       We show the Ly$\beta$ absorption line for 8/10 Green Peas, and Ly$\gamma$ for 2/10. 
  The latter have Ly$\beta$ that is contaminated by geocoronal emission (1424+4217), or fell at a noisy wavelength near the gap between the COS FUV segments (1249+1234).    
  Starburst99 model spectra,  illustrating the stellar component of the Ly$\beta$ and Ly$\gamma$ lines are shown in blue.    
  The model plotted is a 50 Myr, continuous star-forming population with $Z = 0.002$; similar profiles are present across a wide range of
 stellar population models, provided that UV spectra are O-star dominated.
 As with Figures \ref{si21190}, \ref{si21260}, and \ref{c21334}, the grey shaded regions of the spectra show contamination from Milky Way absorption, geocoronal emission, and 
  in the case of the Ly$\gamma$ profiles, the adjacent \ion{C}{3} $\lambda$977.0 line.}
  \label{lyman_fig} 
  \end{center}
  \end{figure}

Despite heavy reliance on the LIS metal lines, they remain {\it indirect} probes of the neutral hydrogen that is scattering the \lya\ emission.  
Fortunately, the COS spectra of the Green Peas allow a more direct look at the \ion{H}{1} gas, with at least one Lyman series line (besides \lya) observed in each of the 10 galaxies.  Figure \ref{lyman_fig} highlights these
features, showing either Ly$\beta$ or Ly$\gamma$ for each Green Pea.  Whenever more than one Lyman series line is observed (7/10 galaxies)
we find that their absorption profiles are consistent.   The equivalent widths of these lines (Table \ref{ewtab_h1}) are measured in the same manner as for the metal LIS lines, although we instead adopt a 20\% 
uncertainty on the continuum placement because normalization is more challenging at these wavelengths.  Additionally,   the same stellar absorption model used for the metal lines is shown for Ly$\beta$ and Ly$\gamma$.  In contrast to the metals, the stellar \ion{H}{1} absorption is significant around zero velocity.  Nevertheless, the stellar absorption does not explain the  blueshifted and highly opaque absorption,  indicating a 
significant contribution from outflowing interstellar gas.  We conclude that these lines are still useful for probing the outflowing gas at moderate to high (blueshifted) velocities.

The \ion{H}{1} absorption lines show some remarkable differences from the LIS metal lines.  
First, the equivalent widths of the Lyman series lines show no variation with \fesc\ or \wlya.  
Second, and in contrast to the spectra shown in  Figures \ref{si21190}, \ref{si21260} and \ref{c21334}, the hydrogen absorption lines show little to no residual intensity at modest blueshifted velocities (a few hundred \kms).  This finding indicates that near 100\% of the stellar light is covered by \ion{H}{1} absorbing gas at these velocities.   Moreover, the high opacity and the similarity between Ly$\beta$, Ly$\gamma$, Ly$\delta$ and Ly$\epsilon$  equivalent widths  (when more than one line is detected) indicates that these lines are saturated but not damped. Hence, the neutral hydrogen column density is poorly constrained: $N_{H I} \sim 10^{16} - 10^{20}$ cm$^{-1}$.  Remarkably, even the low end of this range leads to high \lya\ optical depth at line center, even if it is optically thin to  
hydrogen-ionizing Lyman continuum photons  ($\tau_{Ly\alpha} > 10^2 - 10^3$;  \citealt{Verhamme06, Verhamme14}).

  \begin{figure}
  \begin{center} 
\includegraphics[scale=0.43, viewport=1 1 800 825, clip]{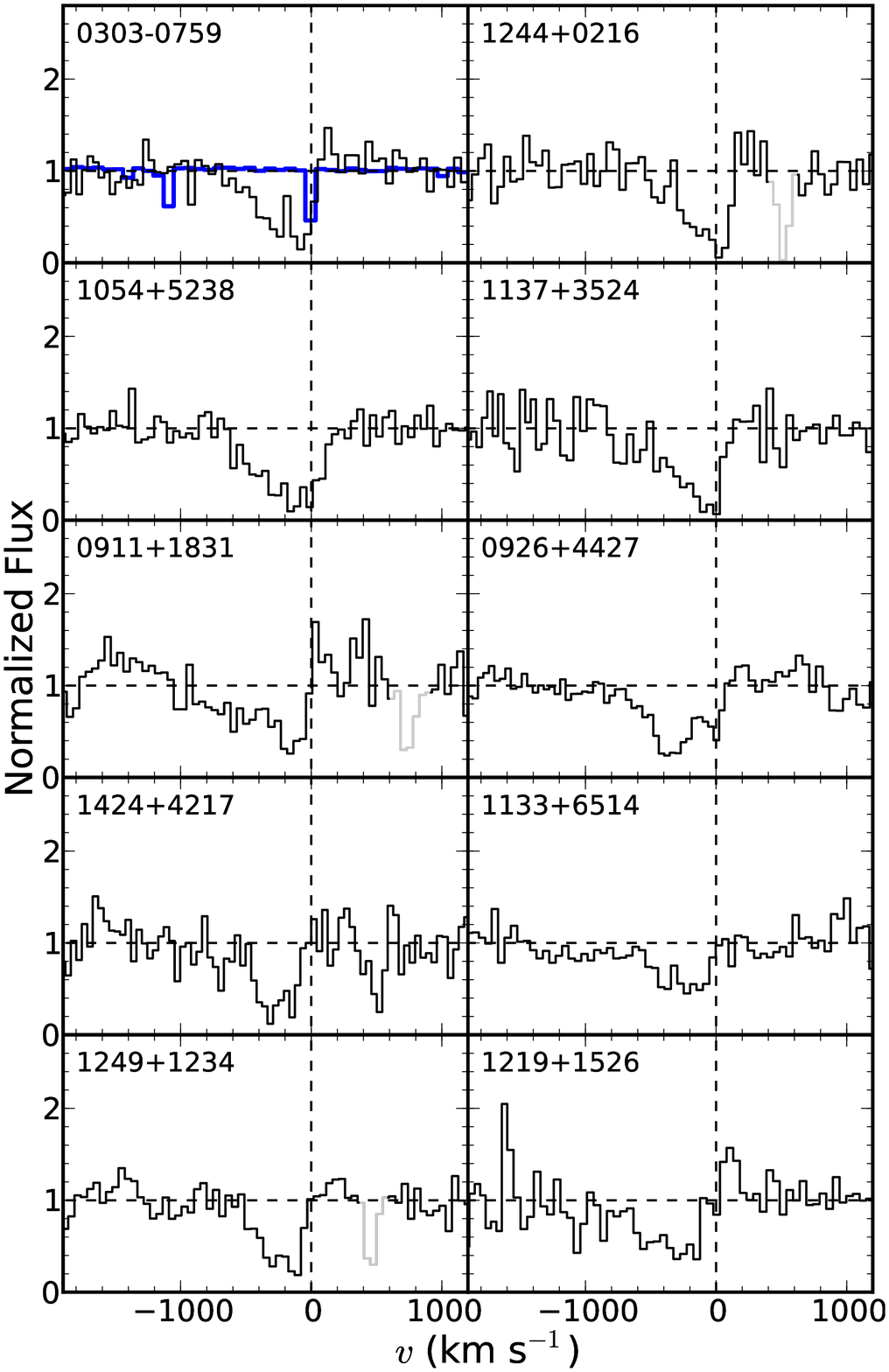} 
\caption{Same as Figures  \ref{si21190}, \ref{si21260}, and \ref{c21334}, but for \ion{Si}{3} 1206.5.}
\label{si3}
\end{center} 
\end{figure}

\begin{figure} 
\begin{center} 
\includegraphics[scale=0.43, viewport=1 1 800 825, clip]{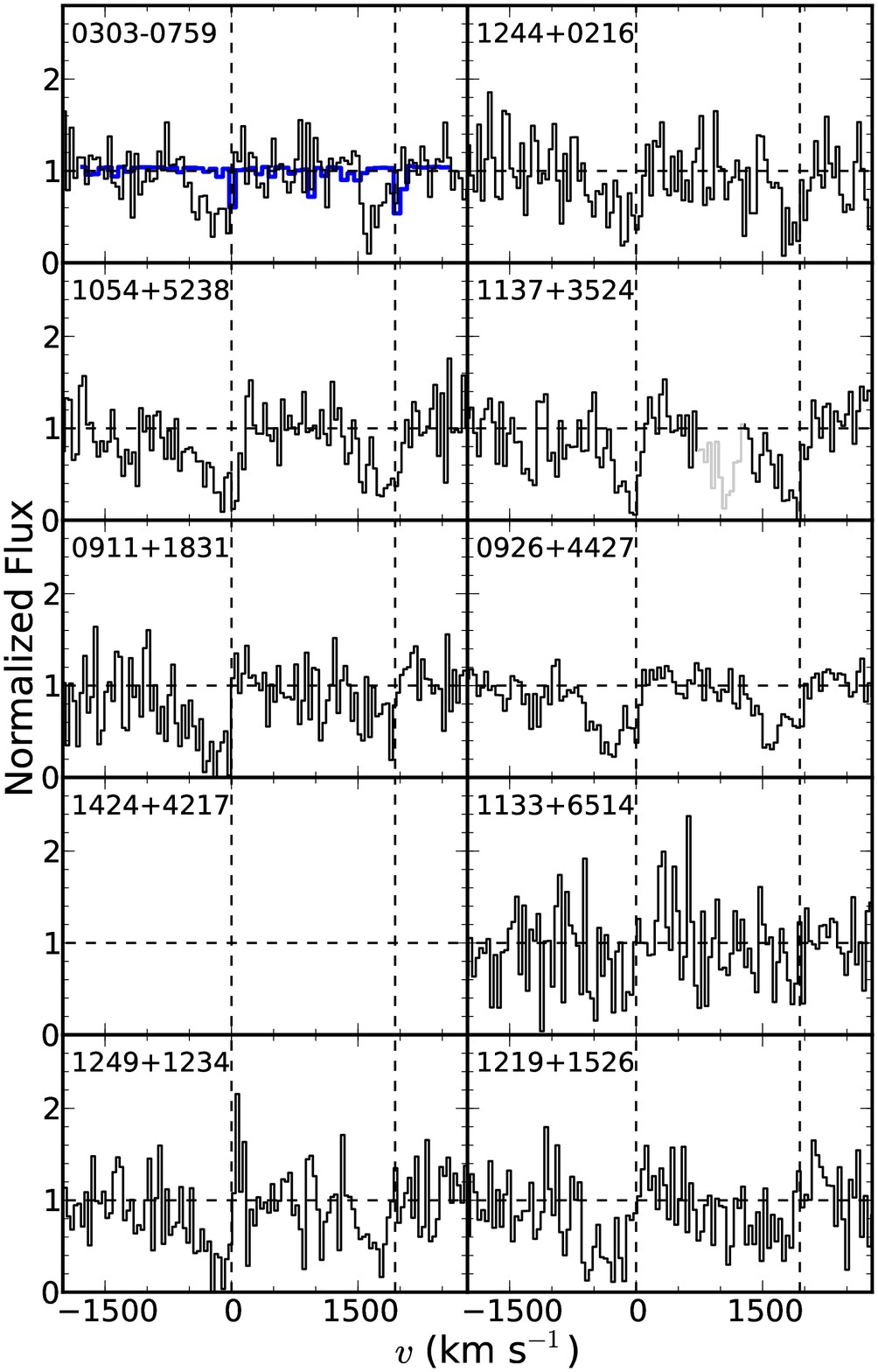} 
\caption{Same as \ref{si21190}, \ref{si21260},  \ref{c21334}, and \ref{si3}, but for \ion{Si}{4} $\lambda \lambda 1393.8, 1402.8$.  The velocity
scale is appropriate for \ion{Si}{4} $\lambda 1393.8$, and the dashed line around 1900 \kms\  marks zero velocity for  \ion{Si}{4} $\lambda 1402.8$.   }  
\label{si4} 
\end{center}
\end{figure}

Finally,  we show 
 higher ionization  \ion{Si}{3} $\lambda 1206.5$ and \ion{Si}{4} $\lambda 1394, \lambda 1403$ in Figures \ref{si3} and \ref{si4}.   
The equivalent widths of these lines are also listed in Table \ref{ewtab}.   Unlike the LIS metal lines,   
the  equivalent widths of these higher ionization lines do not appear to change across the sample.    This conclusion is most  apparent 
for the \ion{Si}{3} lines,  all of which have good S/N and $W \sim -1.3$ \AA.     In this aspect, the Green Peas are also consistent 
with the $z\sim3$ LBGs:  \cite{Shapley03} report that, despite the relation between \wlya\ and $W_{LIS}$,  $W_{Si~IV}$ is the same
in each of their four \wlya-defined stacks.

 \section{Do Outflows Help the \lya\ photons escape?}   
 
 \begin{figure*}
\begin{center} 
\includegraphics[scale=0.50, viewport=1 1 900 400, clip]{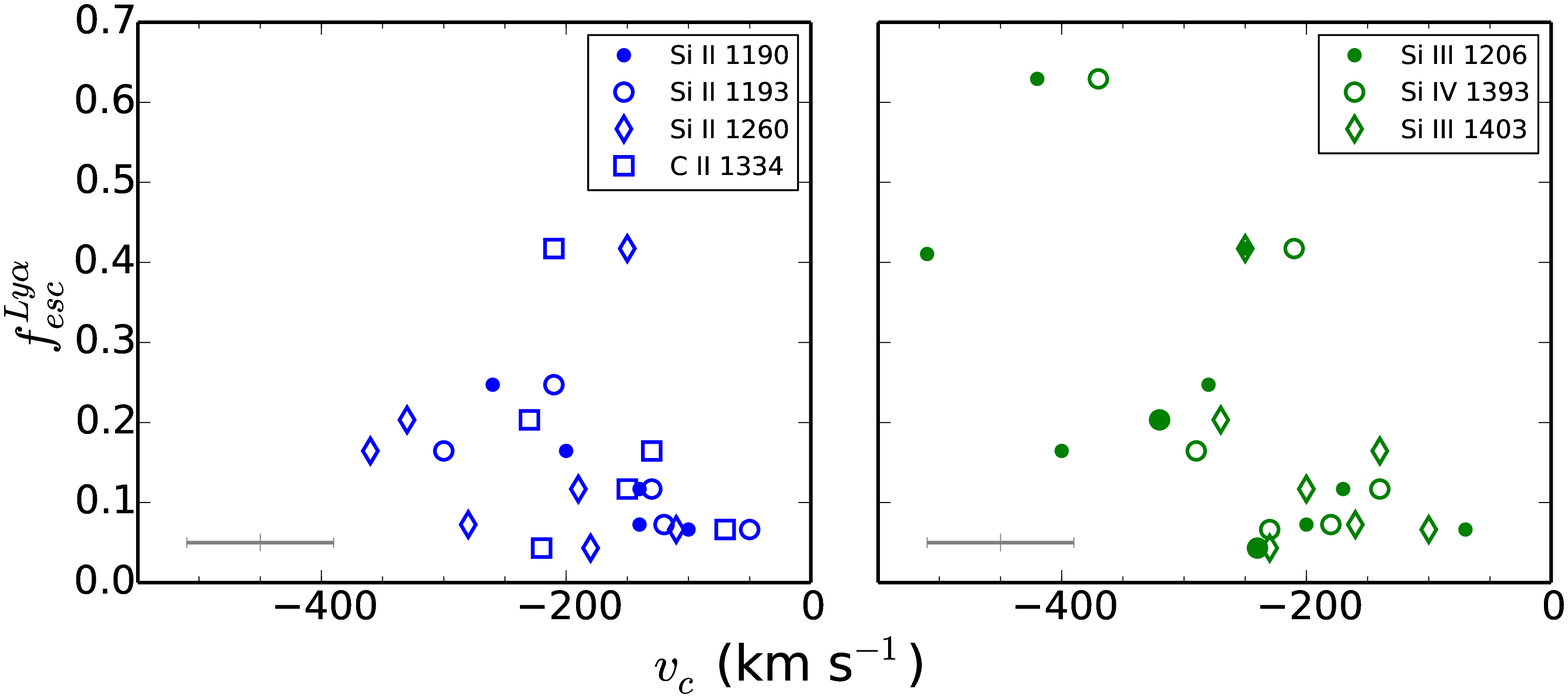}  \\
\includegraphics[scale=0.50, viewport=1  1 900 400, clip]{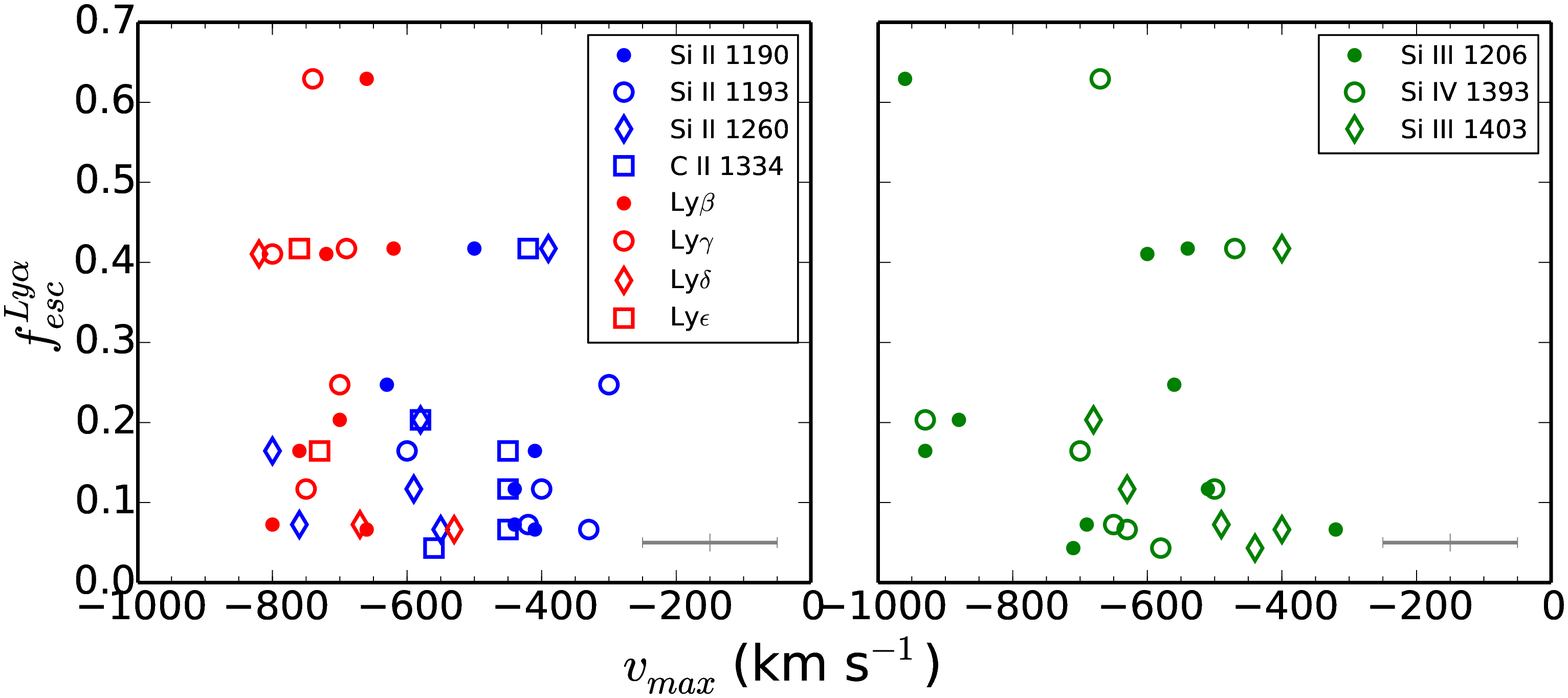}
\caption{The \lya\ escape fraction shows no convincing evidence for a correlation with outflow velocity in the low-ionization metals  or \ion{H}{1} (left panels), but hints at a possible 
trend with the velocity of more highly ionized gas (right panels).    The top panels show the equivalent width weighted characteristic outflow velocities of different ions (Table \ref{vcent_table}), with
a typical 60 \kms\ error bar plotted in grey. The bottom panels show the maximal outflow velocity (Tables \ref{vmax_lis} and \ref{vmax_ly}), measured from the velocity where the absorption profile meets the continuum.  The typical errors in this case are around 100 \kms.   The bottom left panel includes the \ion{H}{1} absorption (red points) which occupy a consistently narrow range of maximal velocities ($\sim -600$ to $-800$ \kms), and are systematically bluer than the metal line tracers that are often use to probe \ion{H}{1}. } 
\label{fesc_vc}
\end{center}
\end{figure*}

The kinematics of the CGM may play an important role in regulating \lya\ escape from galaxies.  
When the photons scatter in outflowing gas, they can Doppler shift out of resonance with the bulk of the ISM 
and escape more easily.  Hence, we may expect that a large velocity gradient in the CGM promotes strong \lya\ emission. 
Early studies find evidence for this scenario.  In a sample of eight local starburst galaxies,   \cite{Kunth98}  find that half of their
sample with damped \lya\ absorption showed no evidence of outflowing gas in their LIS lines.  Yet the other half of their sample that showed \lya\ emission 
exhibited outflows with velocities around 200 km s$^{-1}$.  Similarly, \cite{Wofford} report similar findings from COS observations of 20 nearby galaxies.   In their sample,  seven \lya\ emitting galaxies have $v_{LIS} \sim -100$ \kms, whereas ten galaxies with damped \lya\ profiles have $v_{LIS} \sim -20$ \kms.   And most recently, \cite{Martin14} show that \lya\ escape from ULIRGs is enhanced when the wings of the \lya\ and \oiii\ lines reach greater blueshifted velocities.

 In this section, we will test whether the \lya\ escape is aided by outflowing gas in the Green Peas.  To quantify the outflow kinematics, we will use metal and hydrogen absorption lines, and also explore the kinematic measures from the \lya\ emission.

\subsection{Gas Kinematics in Absorption} 
\label{kin_neutral_abs} 
Figures  \ref{si21190}, \ref{si21260} and \ref{c21334} illustrate the  kinematics of the LIS ions.  When the lines are detected, they are blueshifted with wings extending several hundred km s$^{-1}$.    To measure the  gas kinematics, we estimate 
characteristic outflow velocities, $v_c$.  Since many of the lines have non-Gaussian shapes,  we avoid
fitting centroid velocities, and instead  calculate the equivalent width weighted velocity: 
\begin{equation} 
{v_c } = {{\int v (1 - f_{norm}) d \lambda}  \over {\int (1-f_{norm}) d \lambda} }, 
\end{equation} 
where $f_{norm}$ is the normalized flux, $v$ represents velocity at each wavelength, and the denominator is easily recognized as the equivalent width. 
The resultant characteristic velocities are  listed in Table \ref{vcent_table}.    The errors on these velocities are calculated using 1000 Monte Carlo realizations where
the spectrum is perturbed according to its error vector and a 10\% continuum normalization uncertainty.      
We also quantify the maximum outflow velocity by determining where the absorption trough reaches the continuum.    These velocities are 
listed in Table \ref{vmax_lis}; the uncertainties are calculated using a Monte Carlo simulation in the same manner as for the characteristic outflow velocities.    For completeness, and to obtain some kinematic information for 1133+6514 and 1219+1526 which have no LIS lines, we also list the characteristic and maximum outflow velocities for the higher ionization 
 \ion{Si}{3} and \ion{Si}{4} lines.   And likewise, maximum outflow velocities for the Lyman series lines are listed in Table \ref{vmax_ly} (where we have again included a 20\% continuum normalization uncertainty).    However, because of the stellar component and the breadth of the \ion{H}{1} absorption lines, we do not measure characteristic outflow velocities for these features.

\begin{figure*} 
\begin{center} 
  \includegraphics[scale=0.56, viewport=1 1 900 550,clip] {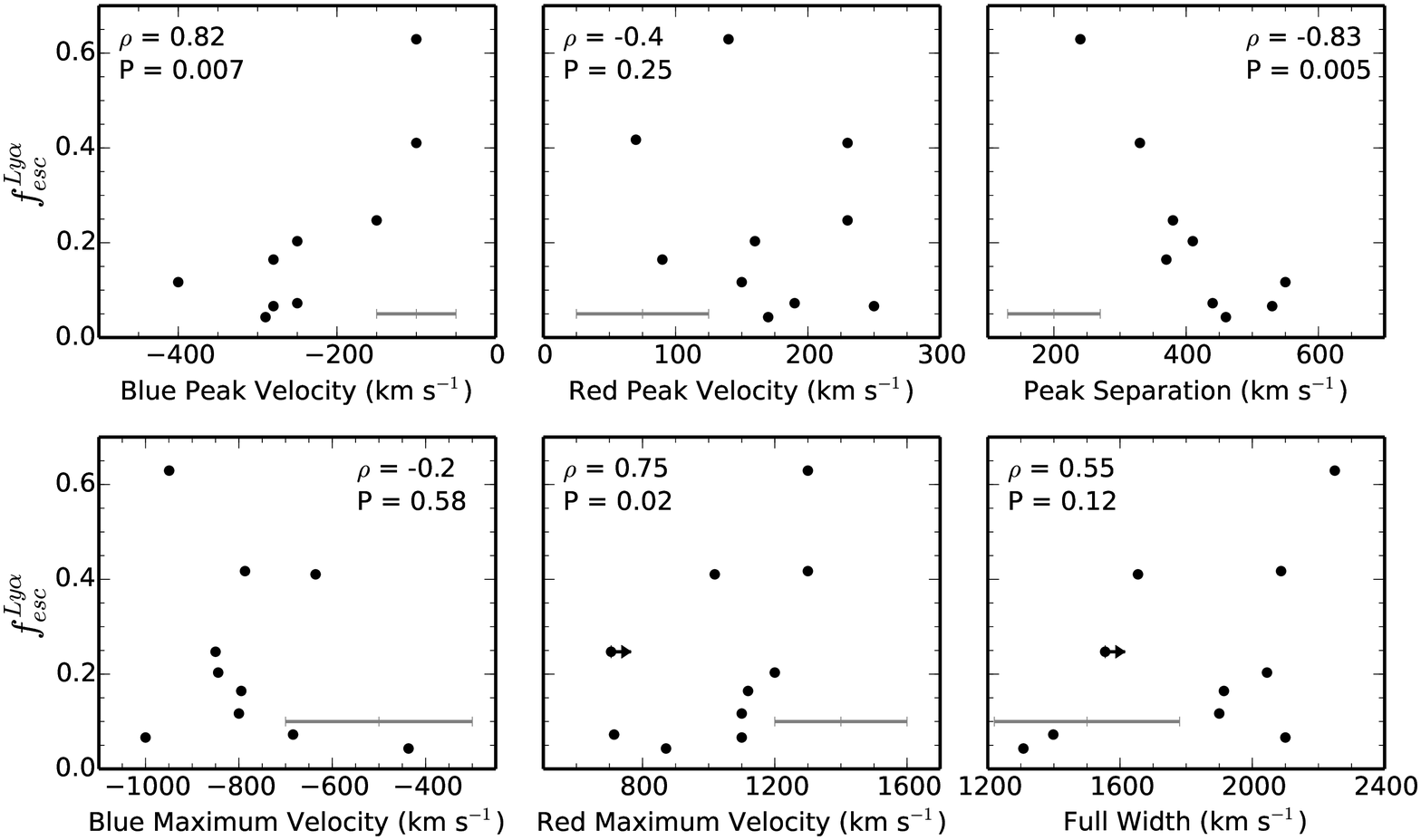} 
  \end{center}
\caption{The velocity of the blue \lya\ peak, as well as the velocity separation between the blue and red peaks are strongly correlated 
with the escape of \lya\ photons.  Other measures, including the maximal velocities, and the velocity of the red peak are only weakly 
related to \lya\ escape.  Each panel is labelled with the Spearman correlation coefficient, $\rho$, and $P$, the probability that the 
correlation arose by chance.  The grey error bars show the typical uncertainty in each measurement. } 
\label{fesc_lyakin} 
\end{figure*}

Figure \ref{fesc_vc} shows how $f_{esc}^{Ly\alpha}$ depends on these kinematic features for the \ion{H}{1}, LIS metals, and higher
ionization lines.    The top two panels focus on $v_c$, with different symbols indicating  different transitions.   The upper left panel shows that we detect no significant trend between
$f_{esc}^{Ly\alpha}$  and outflow velocity measured from the LIS metals.     However, we must note that two of the  three highest $f_{esc}^{Ly\alpha}$ galaxies cannot be included in this plot and correlation test, because they are undetected in all LIS absorption lines (1133+6514 and 1219+1526).    Their addition in the upper right panel of Figure \ref{fesc_vc}, where we show the higher ionization states of Si, hints at a correlation.     For   \ion{Si}{3}, which is detected at good S/N in all ten Green Peas, the Spearman correlation coefficient rules 
out a spurious relation at 98\% confidence.  But this trend is only significant with the addition of the  two high   $f_{esc}^{Ly\alpha}$ Green Peas.   In the bottom panels, where we show the maximal velocities, the metals tell a similar story:  the Green Pea 1219+1526 (the highest  $f_{esc}^{Ly\alpha}$ object)  has greater maximal outflow velocity than most when \ion{Si}{3} is considered.  But otherwise, we detect no trend.       Finally, in the bottom left panel of Figure \ref{fesc_vc} we  add maximal outflow velocities reached by \ion{H}{1} Lyman series lines (red points).   These lines consistently reach 
$-700$ to $-800$ \kms, and show no trend with $f_{esc}^{Ly\alpha}$.     In summary, since the only possible correlation detected in Figure \ref{fesc_vc} relates $f_{esc}^{Ly\alpha}$ 
with higher ionization lines that do not trace \ion{H}{1}, we conclude that these data give no compelling evidence for a scenario where  \lya\ escape is enhanced by scattering in outflowing \ion{H}{1} gas.  

\begin{deluxetable*} {ccccccccc}[!ht]
\tablecolumns{8}
\tabletypesize{\footnotesize}
\tablecaption{Equivalent widths of ISM  metal absorption lines}  
\tablehead{ \colhead{Object ID} & \colhead{\ion{Si}{2} $\lambda$1190}  & \colhead{\ion{Si}{2} $\lambda$1193} & \colhead{\ion{Si}{2} $\lambda$1260} & \colhead{\ion{C}{2} $\lambda$1334} 
&  \colhead{\ion{Si}{3} $\lambda$1206} & \colhead{\ion{Si}{4} $\lambda 1393$}  & \colhead{\ion{Si}{4} $\lambda 1403$}  \\ 
\colhead{} & \multicolumn{7}{c}{ (\AA) } }  
\startdata
0303--0759 &  $-0.4 \pm 0.3$   & $-0.4 \pm 0.3$             & $-0.6\pm 0.2$          &  $-1.2 \pm 0.3$ & $-1.3\pm 0.2$  &  $-1.1\pm 0.4$    & $-0.9 \pm 0.4$        \\ 
1244+0216 &   $-1.1 \pm 0.3$    & $-1.0 \pm 0.3$          & $-1.6 \pm 0.4$         &  $-1.3 \pm 0.5$ &  $-1.1 \pm 0.3$  & $-1.3 \pm 0.8$    &  $-1.0 \pm 0.8$    \\ 
1054+5238 &  $-0.8 \pm 0.3$   &  $-0.5 \pm 0.3$          & $-1.0  \pm 0.5$      &   \nodata            &    $-1.8 \pm 0.5$ & $-1.9 \pm 1.2$    & $-1.3 \pm 1.2$    \\  
1137+3524 &  $-1.0  \pm 0.3$     & $-1.1 \pm 0.3$        &   $-1.2 \pm 0.3$         & $-1.3 \pm 0.3$   &  $-1.4 \pm 0.2$  & $-1.3 \pm 0.3$   & $-1.5 \pm 0.5$      \\ 
 0911+1831&  $-0.6 \pm 0.2$     & $-0.9 \pm 0.2$          &  $-1.1 \pm 0.4$        &  $-0.7 \pm 0.4$    &   $-1.3 \pm 0.4$   & $-1.9 \pm 0.7$ & $>-0.6$                  \\ 
0926+4427 &  $-0.4 \pm 0.1$     & $-0.3 \pm 0.1$         &  $-0.5 \pm 0.1$      &   $-0.7 \pm 0.2$   &  $-1.5 \pm 0.2$  &  $-1.7 \pm 0.3$   & $-1.3 \pm 0.2$       \\  
1424+4217 & $-0.7\pm 0.4$     & $-0.3 \pm 0.3$            & \nodata                 &   \nodata           &    $-1.0\pm 0.3$  &  \nodata             &   \nodata                 \\  
1133+6514 &   \nodata              & $> - 0.4$                      & $>-0.4 $                 &   $>-0.4$           &   $-1.3 \pm 0.4$  & $>-0.6$               &  $-0.6$              \\ 
1249+1234 & \multicolumn{2}{c}{$-1.8 \pm 0.4$\tablenotemark{a} }        &  $-0.9\pm 0.3$          & $-0.8 \pm 0.4$    &   $-1.1 \pm 0.2$  & $-1.3 \pm 0.5$   &   $-0.9 \pm 0.6$     \\  
1219+1526 &   $> -0.4$                & $> -0.4$                  &  $>-0.4$                   &  $>-0.4$              &    $-1.2\pm 0.3$  & $-1.6 \pm 0.6$   &  $-0.7 \pm 0.4$  
\enddata
\label{ewtab} 
\tablecomments{Rest-frame equivalent widths of metal absorption lines discussed in this paper.  }
\tablenotetext{a}{This equivalent width is the total for the blended \ion{Si}{2} $\lambda1190$, $\lambda1193$ lines.}
\end{deluxetable*}

\begin{deluxetable*} {ccccc}[!ht]
\tablecolumns{5}
\tabletypesize{\footnotesize}
\tablecaption{Equivalent widths of  \ion{H}{1} absorption lines}  
\tablehead{ \colhead{Object ID} & \colhead{Ly$\beta$}  & \colhead{Ly$\gamma$} & \colhead{Ly$\delta$}  & \colhead{Ly$\epsilon$}  \\ 
\colhead{} & \multicolumn{4}{c}{ (\AA) } }  
\startdata
0303--0759 &  $-2.6\pm 0.6$    &  \nodata        & \nodata            &  \nodata      \\
1244+0216  & $-2.6 \pm 0.7$   & \nodata        & $-2.1 \pm 0.6$   &   \nodata    \\
1054+5238 & $-2.8 \pm 0.7$   & \nodata        & $-2.5 \pm 0.6$    & $1.9 \pm 0.5$   \\  
1137+3524 &  $-2.5 \pm 0.8$ &   $-2.5 \pm 0.6$   & \nodata  & \nodata  \\ 
 0911+1831 & $-1.8 \pm 0.7$ & \nodata     & \nodata   &      $1.2 \pm 0.5$   \\
0926+4427 & $-2.4 \pm 0.5$  & \nodata        &  \nodata            & \nodata     \\ 
1424+4217 & \nodata             & $-1.9 \pm 0.5$  & \nodata        &  \nodata  \\ 
1133+6514 &  $-2.9 \pm 0.7$  & \nodata        &  $-2.0 \pm 0.7$  &  \nodata    \\   
1249+1234&  $-2.2 \pm 0.7$  & $-2.1 \pm 0.6$  &  \nodata        &  $2.2 \pm 0.8$ \\
1219+1526 &  $-2.0 \pm 0.7$  &  $-2.6 \pm 0.7$ & \nodata    &    \nodata  
\enddata
\label{ewtab_h1} 
\tablecomments{Rest-frame equivalent widths of \ion{H}{1} Lyman series absorption lines discussed in this paper.  Lyman series equivalent widths contain a contribution of  around 0.6 to 0.8 \AA\ from stellar absorption.  }
\end{deluxetable*}

  \begin{deluxetable*} {cccccccc}[!ht]
\tablecolumns{7}
\tablecaption{Characteristic outflow velocities from ISM absorption lines} 
\tablehead{ \colhead{Object ID} & \colhead{$v_{1190}$  }  & \colhead{$v_{1193}$ } & \colhead{$v_{1260}$  } & \colhead{$v_{1334}$} &  \colhead{$v_{1206}$} &   
\colhead{$v_{1393}$}  & \colhead{$v_{1403}$}  } 
\startdata
0303--0759 &  \nodata\tablenotemark{a}   & \nodata\tablenotemark{a}   & $-180 \pm 60$                     & $-220 \pm 60$                 & $-240\pm50$                                  & $-240 \pm 60$                 & $-230 \pm 80$                    \\
1244+0216 &  $ -100 \pm 20$                  & $-50 \pm 20$                       & $-110\pm 20$                       & $-80 \pm 20$                  & $-70 \pm 20$                                  &  $-230 \pm 70$                 & $-100 \pm 50$                    \\
1054+5238 &  $-140 \pm 100$               & $-120 \pm 80$                      & $-280 \pm 70$                     & \nodata\tablenotemark{b}   & $-200\pm40$                               &  $-180 \pm 80$                 & $-160 \pm 70$                    \\  
1137+3524  & $-140 \pm 30$                & $-130 \pm 30$                         & $-190 \pm 30$                   & $-150 \pm 40$                   &  $-170 \pm 40$                       &  $-140 \pm 30 $                        &  $-200 \pm 40$                  \\ 
 0911+1831 &  $-200 \pm 80$              & $-300 \pm 90$                          & $-360 \pm 70$                   & $-130 \pm 90$                  & $-400 \pm 90$                        & $-290 \pm 50$                           & $-140 \pm 100$                 \\
0926+4427 & \nodata\tablenotemark{a}  &  \nodata\tablenotemark{a}   & $-330 \pm 170$                     & $-230 \pm 100$                &$-320 \pm 60$                               & $-320 \pm 70$                  & $-270 \pm 70$                   \\ 
1424+4217  & $-260 \pm 80$                  & $-210 \pm 90$                     & \nodata\tablenotemark{c}    & \nodata\tablenotemark{c}  & $-280 \pm 60$                              & \nodata\tablenotemark{c}  & \nodata\tablenotemark{c}     \\ 
1133+6514 &  \nodata\tablenotemark{a}   & \nodata\tablenotemark{a}   & \nodata\tablenotemark{a}    & \nodata\tablenotemark{a}  & $-270 \pm 50$\tablenotemark{e} & \nodata\tablenotemark{a} & \nodata\tablenotemark{a}  \\
1249+1234 & \nodata \tablenotemark{d}  & \nodata \tablenotemark{d}   & $-150 \pm 60$                   & $-210 \pm 60$                   &$-250\pm60$                               &  $-210 \pm 50$                    & $-250 \pm 80$                    \\
1219+1526 & \nodata\tablenotemark{a}   &  \nodata\tablenotemark{a}   & \nodata\tablenotemark{a}   &\nodata\tablenotemark{a}   & $-420 \pm 90$                            &  $-370 \pm 80 $                   &  \nodata\tablenotemark{a}   

\enddata
\label{vcent_table} 
\tablecomments{Equivalent width weighted velocities are given for \ion{Si}{2} $\lambda \lambda 1190.4, 1193.3$, $\lambda$ 1260.4, \ion{C}{2} $\lambda$ 1334.5, \ion{Si}{3} $\lambda$ 1206.5, and \ion{Si}{4} $\lambda \lambda$ 1393.8, 1402.7. } 
\tablenotetext{a}{Line is marginally detected or completely undetected, so reliable kinematic information is unavailable.} 
\tablenotetext{b}{Line is contaminated by Milky Way absorption.}  
\tablenotetext{c}{Line is not covered by the present spectra.} 
\tablenotetext{d}{\ion{Si}{2} $\lambda \lambda$ 1190.4, 1193.3 lines are blended.}  
\tablenotetext{e}{The blue wing of the line may extend more than $-1200$ km s$^{-1}$, but uncertainties in continuum normalization make this extent unclear (see Figure \ref{si3}).   We report the velocity from a Gaussian fit 
to the main absorption component of this line.}
\end{deluxetable*}

 Remarkably, Figure \ref{fesc_vc} shows that   the maximal outflow velocities measured by the \ion{H}{1} Lyman series lines 
 are systematically bluer than the maximal velocities of the LIS metal lines.    
This result demonstrates that the  \ion{H}{1} absorption is markedly more sensitive to low density, high velocity gas that cannot 
  be detected in the LIS metal absorption lines.    Yet this \ion{H}{1} likely plays an important role in
 scattering \lya\ photons;  we will explore this topic more in \S \ref{lyakin_sec} where we compare
 the Lyman series absorption and \lya\ emission kinematics.

 \begin{deluxetable*} {cccccccc}[!ht]
\tablecolumns{7}
\tablecaption{Maximum Velocities of Metal Absorption Lines  } 
\tablehead{ \colhead{Object ID} & \colhead{$v_{1190}^{max}$  }  & \colhead{$v_{1193}^{max}$ } & \colhead{$v_{1260}$  } & \colhead{$v_{1334}^{max}$} &  \colhead{$v_{1206}^{max}$} &   
\colhead{$v_{1393}^{max}$}  & \colhead{$v_{1403}^{max}$}  } 
\startdata
0303--0759 &  \nodata\tablenotemark{a}               & \nodata\tablenotemark{a}            & \nodata\tablenotemark{b}               & $-560 \pm 50$  & $-710 \pm 100$  & $-580 \pm 120$   & $-440 \pm 30$       \\
1244+0216 &   $ -410 \pm 60$  & $-330 \pm 70$   & $-550 \pm 90$   &  $-350 \pm 60$  & $-320 \pm 80 $  & $-630\pm 90  $    & $-400 \pm 70$    \\
1054+5238 &  $-440 \pm 60$   & $-420 \pm 100$  & $-760  \pm 150$ &   \nodata\tablenotemark{b}           & $-690 \pm 100 $ &  $-650 \pm 110$  & $-490 \pm 70 $  \\  
1137+3524 & $-440 \pm 90$    & $-400 \pm 60$   & $-590 \pm 130$    & $-450 \pm 40$  &  $-510 \pm 110$  &  $-500 \pm 100 $ &  $-630 \pm 160$     \\ 
 0911+1831&  $-410 \pm 110$  & $-600  \pm 50$   & $-800 \pm 200$   & $-600 \la v \la -300$\tablenotemark{f}      & $-930 \pm 240$    & $-700 \pm 160$  &  \nodata    \\ 
 0926+4427 &  \nodata\tablenotemark{a}              & \nodata\tablenotemark{a}                & $-580 \pm 130$  & $-580 \pm 90$  & $-880 \pm 100$  & $-930 \pm 90$    & $-680  \pm 60$     \\ 
1424+4217 & $-630 \pm 130$  & $-300 \pm 70$    & \nodata\tablenotemark{c}                &   \nodata\tablenotemark{c}          & $-560 \pm 70$    & \nodata\tablenotemark{c}                & \nodata\tablenotemark{c}                     \\ 
1133+6514 &  \nodata\tablenotemark{a}               & \nodata\tablenotemark{a}             & \nodata\tablenotemark{a}              &   \nodata\tablenotemark{a}           &   $\sim -600$\tablenotemark{d}      & \nodata\tablenotemark{a}                & \nodata\tablenotemark{a}                \\
1249+1234 &  $-500 \pm 50$   & \nodata\tablenotemark{e}            & $-390  \pm 70$    & $-420 \pm 90$ & $-540\pm50$      & $-470 \pm 120$    & $-400 \pm 110$     \\
1219+1526 & \nodata\tablenotemark{a}              &  \nodata\tablenotemark{a}              & \nodata\tablenotemark{a}             &  \nodata\tablenotemark{a}            & $-960 \pm 160$   &  $-670 \pm 220 $ &  \nodata\tablenotemark{a}             
\enddata
\label{vmax_lis} 
\tablecomments{Maximum velocities (in km s$^{-1}$) are given for the same lines as in Table \ref{vcent_table}. }
\tablenotetext{a}{Line is marginally detected or completely undetected, so reliable kinematic information is unavailable.} 
\tablenotetext{b}{Line is contaminated by Milky Way absorption.}  
\tablenotetext{c}{Line is not covered by the present spectra.} 
\tablenotetext{d}{The blue wing of the line may extend more than $-1200$ km s$^{-1}$, but uncertainties in continuum normalization make this extent unclear (see Figure \ref{si3}).   Most of the absorption, however, 
lies within $-600$ km s$^{-1}$. }  
\tablenotetext{e}{The continuum is not reached before the absorption from neighboring \ion{Si}{2} $\lambda 1190.4$ is reached. Therefore the maximum outflow velocity from this line is unreliable.}  
\tablenotetext{f}{This absorption profile appears divided into two parts, possibly from a noise spike that crosses unity in the normalized spectrum around -300 km s$^{-1}$.  The maximum velocity in the
more blueshifted portion of the profile is around -600 km s$^{-1}$.}  
\end{deluxetable*}

\begin{deluxetable*} {ccccccccc}[!ht]
\tablecolumns{6}
\tablecaption{Maximum Velocities of Lyman Series Absorption Lines} 
\tablehead{ \colhead{Object ID} & \colhead{$v_{Ly\beta}^{max}$}  & \colhead{$v_{Ly\gamma}^{max}$} & \colhead{$v_{Ly\delta}^{max}$ } & \colhead{$v_{Ly\epsilon}^{max}$ }   } 
\startdata
 0303--0759 &  $-800 \la v \la -200$\tablenotemark{a}   & \nodata\             & \nodata                      & \nodata      \\
1244+0216 &   $-660 \pm 70$   & \nodata               & $-530 \pm 40$          &  \nodata      \\
1054+5238 &  $-800 \pm 60$      & \nodata                & $-670 \pm 50$      &   \nodata            \\  
 1137+3524 &  \nodata\                 & $-750 \pm 90$            &    \nodata         &  \nodata       \\ 
 0911+1831&  $-760 \pm 180$     & $-1200  \pm 270$   &  \nodata              &  $-730 \pm 130$       \\
0926+4427 &  $\sim -700 \pm 100$\tablenotemark{b}          & \nodata              &  \nodata                    & \nodata\              \\ 
1424+4217 & \nodata                  & $-700 \pm 110$       & \nodata                &   \nodata                    \\ 
1133+6514 &   $-720 \pm 60$    & $-800 \pm 170$  & $-820 \pm 160$      &   \nodata\              \\
1249+1234 &  $-620\pm 100$     & $-690 \pm 90$              &  \nodata         & $-760 \pm 170$      \\
1219+1526 & $-660 \pm 70 $        & $-740 \pm 30$            &  \nodata          &  \nodata            
\enddata
\label{vmax_ly} 
\tablecomments{Maximum velocities, (in \kms) are given for the Lyman series lines.    Lines which are not listed are either contaminated by Milky Way absorption, fall beyond the blue-wavelength cutoff
of our COS spectrum, or near the gap between the FUV segments.   Unlike the metal lines, we do not fail to detect absorption from \ion{H}{1} gas. }
\tablenotetext{a}{Blue wing of line impacted by Milly Way absorption.} 
\tablenotetext{b}{The normalized continuum does not reach unity before nearby Milky Way absorption sets in (see Figure \ref{lyman_fig}).  Visual inspection of the spectrum suggests an approximate 
maximum velocity of around $-700 \pm 100$ \kms.}
\end{deluxetable*}

\subsection{The velocity structure of \lya\ emission}   
\label{lyakin_sec}

The \lya\ emission line profiles give another probe of the outflowing gas kinematics, as the broad emission is (at least in part) generated by resonant  scattering 
in the gas around the galaxy.  Table \ref{lya_table}  gives kinematic measures of the \lya\ lines, including the velocities of the red and blue peaks, and the 
maximal velocities of the red and blue wings of the lines.   Figure \ref{fesc_lyakin}  shows how 
 $f_{esc}^{Ly\alpha}$ changes with these \lya\ kinematic measures.   Statistically significant correlations appear in some, but not all, of these quantities. 
 Each panel in Figure \ref{fesc_lyakin} lists the Spearman correlation coefficient, $\rho$, and the probability, $P$, that the correlation arises by chance.  For this analysis, we exclude the Green Pea with single-peaked \lya\ emission (1249+1234) from the statistics and plots involving the blue peak and the peak separation, but it is included in the other diagnostics.   Similarly, the maximal red velocity for 1424+4217 is a lower limit (the spectrum was truncated due to a failed observation), and this galaxy is excluded from the relevant statistics.

Figure \ref{fesc_lyakin} shows that the separation between the \lya\ emission peaks becomes smaller  when $f_{esc}^{Ly\alpha}$  is larger (top right panel), and
that this trend is driven primarily by a shift in the blue emission peak (top left panel).   At the same time, the galaxies with higher  $f_{esc}^{Ly\alpha}$ may show somewhat lower velocities for their red \lya\ peaks, but the correlation is not significant.   Compared to the  blue emission peaks, the red \lya\ emission peaks also inhabit a smaller range of velocities.  These findings are consistent with studies of high-redshift galaxies ($z\sim 2-3$), where the kinematics of the red \lya\ emission peak has been studied by multiple groups.  Notably, \cite{Hashimoto} and \cite{Shibuya} measure a red peak velocity, $\Delta v_{Ly\alpha} \sim 200 $ \kms\ for 10-20 LAEs with \wlya $>50$ \AA, compared with 450 \kms\ measured for UV-continuum selected Lyman Break Galaxies (LBGs; \citealt{Steidel10}).  In this sense, the Green Peas are more similar to the high-redshift LAEs than with the LBGs.   Indeed, for a 
larger sample of 158 galaxies, \cite{Erb14} report a significant anti-correlation between \wlya\ and $\Delta v_{Ly\alpha}$, such that \lya\ is stronger
when it emerges closer to the systemic velocity.    Our trend of increasing $f_{esc}^{Ly\alpha}$  with peak separation fits naturally with this scenario; increased emission near 
the systemic velocity serves to shift the emission peaks closer together.     A visual inspection of  Figure \ref{lya_spec} supports this idea;  where the \lya\ emission is strong, there is net emission around $v\sim0$,  whereas net absorption appears around the systemic velocity when $f_{esc}^{Ly\alpha}$  is low.    Here, for the first time, we show that the kinematics of the blue \lya\ emission peak show even greater variation than the red peak.

This tight correlation between $f_{esc}^{Ly\alpha}$ and  peak separation suggests that 
\lya\ escape is determined by the neutral hydrogen column density.    In short, when the column density is low, 
the \lya\ photons can escape nearer to the systemic velocity,  with less scattering in the expanding circumgalactic envelope.    
Variations in dust, on the other hand, preserve the shape of the profiles while decreasing the strength of the emission  \citep{Verhamme06, Verhamme08, Verhamme14, Behrens}.  
Indeed, the  \ion{H}{1} density has already been used to explain the trends with the red \lya\ peak velocity, $\Delta v_{Ly\alpha}$, seen in 
$z\sim 2-3$ galaxies (discussed above;  \citealt{Chonis, Hashimoto, Shibuya, Erb14}).
Moreover, recent radiative transfer models from \cite{Verhamme14} show that  small separations between peaks, $\Delta v < 300$ \kms, imply $N_{HI} \la 10^{18}$ cm$^{-2}$, whereas higher separations $300 <  \Delta v < 600$ \kms, arise in galaxies with $10^{18} \la N_{HI} \la 10^{20}$ cm$^{-2}$.   

However, the \lya\ radiative transfer models models of \cite{Verhamme14} do not produce
all the features in our data.  First, unlike the Green Peas, increasing $N_{HI}$ in the models shifts the red peak to higher velocities while holding the blue peak at a nearly fixed velocity.  Second, in modeled profiles, the high-velocity wings of the \lya\ reach only to $\pm$ a few hundred \kms\ in contrast to the several hundred \kms\ observed in the Green Peas. This discrepancy may be attributable to an assumed 
{\it intrinsic} \lya\ profile which neglects the broad wings observed in the optical emission lines.   Indeed, \cite{Martin14} successfully produce broad \lya\ wings  on eight ULIRGs by modeling the \lya\ emission as a superposition of intrinsic (broad + narrow) and scattered components. 
  And the third difference between the data and models of Verhamme et al.: the double-peaked emission profiles appear mostly when the shell expansion velocity is low, $v<100$ \kms.  Yet the Green Peas show $-300 \la v_c \la -100$ in 
their LIS lines.  Nevertheless, \lya\ radiative transfer modeling offers some intriguing lines of investigation.  As \cite{Verhamme14} already noted,  the galaxies with small peak separations may have low enough column densities to be optically thin to the hydrogen-ionizing Lyman continuum (LyC) photons.  The Green Peas with the most closely spaced peaks, 1219+1526 and 1133+6514, make excellent candidates for followup observations aimed at direct detection of the LyC.    Remarkably, among the present sample of Green Peas, 1133+6514 shows lower \wha\ and \wlya\ despite its high $f_{esc}^{Ly\alpha}$. These conditions are also consistent with LyC leakage.

In addition to the peaks of the \lya\ emission, we explore the kinematics probed in the wings of the lines. 
The bottom panels of Figure \ref{fesc_lyakin} show how the maximal velocity reached in the \lya\ lines relates to $f_{esc}^{Ly\alpha}$.   In these diagnostics, the typical 200 \kms\ uncertainty on the maximal velocities makes it difficult to draw robust conclusions.   Although only the center panel (illustrating the maximal velocity in the red wing of the line) shows a significant relation, these plots hint at a scenario where broader \lya\ emission wings may be associated with increased  \lya\ escape.    This finding is qualitatively consistent with
the sample of eight ULIRGs studied by  \cite{Martin14}, where  increased \lya\ escape was associated with greater blueshifted \lya\ and  \oiii\ velocities.   In this case, the close association of the blueshifted \lya\  and [\ion{O}{3}] $\lambda 5007$ kinematics suggested an in-situ production of \lya\ photons in the high velocity gas (rather than scattering).  As demonstrated by \cite{Martin14}, this emission would result from cooling of the hot galactic wind.     However, with the resolution of the SDSS spectra,  it remains difficult to test whether 
the \lya\ kinematics in the wings correspond closely with the  \oiii\ and \ha\ kinematics.   Ultimately, more work is needed to obtain
higher spectral resolution observations of the nebular gas, so that we may compare the line profiles in greater detail.  
We defer this analysis to a future study.

\begin{figure} 
\begin{center}
  \includegraphics[scale=0.40, viewport=1 1 600 850,clip] {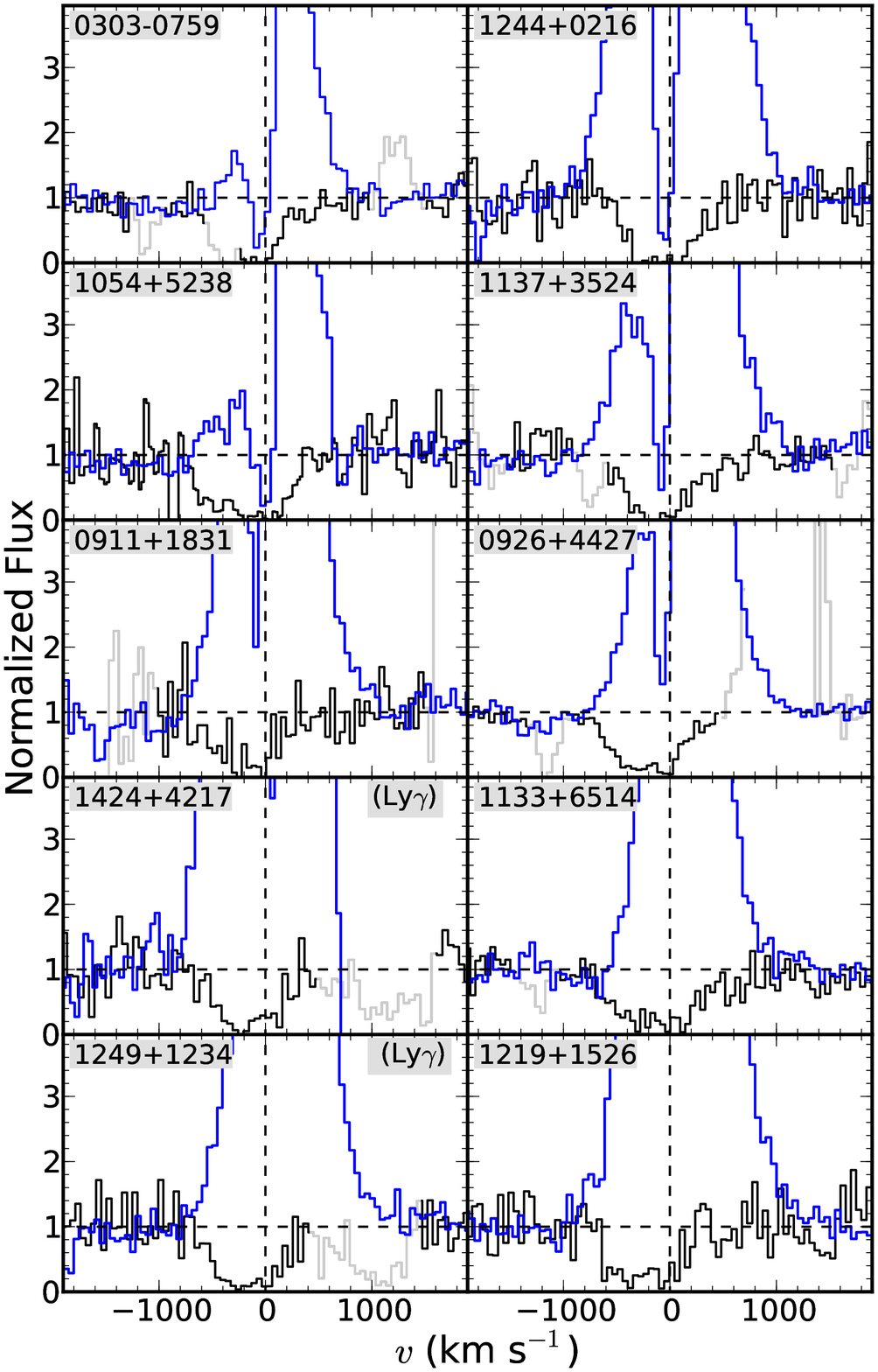} 
\caption{The wings and peaks of the \lya\ line profiles are compared to the absorption line profiles.    The \lya\ line (blue, with peaks not shown), exhibits blueshifted emission which extends to the same maximal velocity as the Ly$\beta$ (or Ly$\gamma$) lines (shown in black). } 
\end{center} 
\label{lya_velocity_compare} 
\end{figure} 

 \begin{figure*} 
 \begin{center}
\includegraphics[scale=0.8, viewport=0 0 575 400,clip] {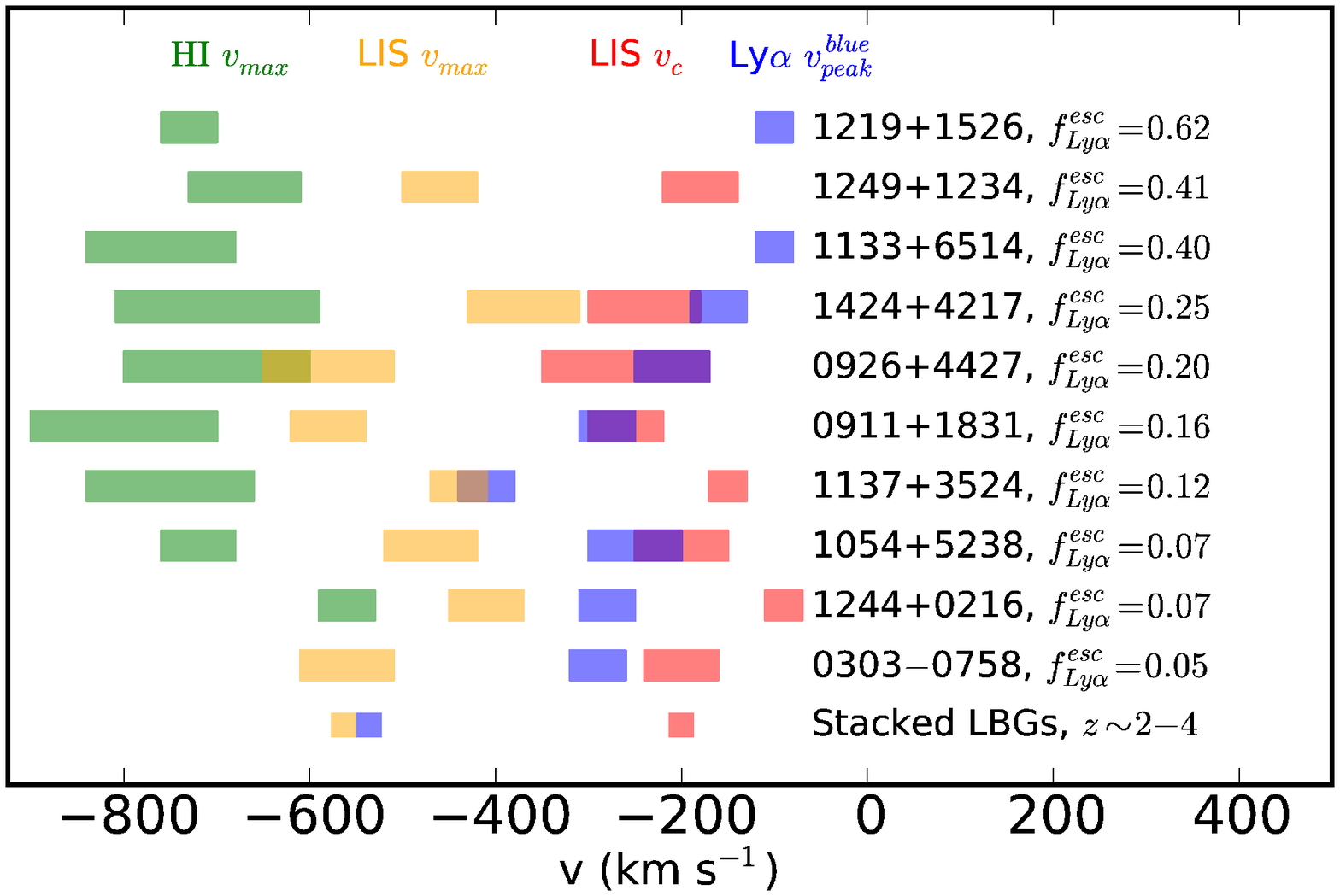}
\end{center} 
\caption{A summary of the kinematic features in the Green Pea data is compared to results from stacked spectra 
at $z \sim 2-4$ \citep{Steidel10, Jones12}.  For the Green Peas, multiple transitions are combined to give the range of velocities probed by \ion{H}{1} absorption (green) or LIS metals (\ion{Si}{2} and \ion{C}{2}; orange and red).  In these cases, bars represent the error weighted mean and its uncertainty.   Additionally, the Green Pea data are sorted in order of increasing $f_{esc}^{Ly\alpha}$ (bottom to top).  In contrast to the results from high-redshift composite spectra (shown at bottom), most of the Green Peas show \lya\ emission emerging much closer to the systemic velocity, near $v_c$ rather than the LIS $v_{max}$ . }
\label{velocity_summary}   
\end{figure*} 

In order to more fully understand what the \lya\ kinematics are measuring, we also compare the spectra to   \ion{H}{1} absorption profiles measured by the Ly$\beta$ or Ly$\gamma$ lines in Figure  \ref{lyman_fig}.  This comparison shows that the maximal velocities of the \lya\ and \ion{H}{1} absorption are well matched (except for possibly 1244+0216), extending to around
 $-700$ \kms\ in most cases.   
The presence of \ion{H}{1} absorption over the same range of velocities as the blueshifted \lya\ emission is noteworthy; it demonstrates clearly that the gas which scatters  \lya\ photons
exists in an envelope spanning a range of velocities.    Shell models, where the cool gas exists at a single velocity, cannot describe absorption spanning several hundred \kms, even though
they often provide satisfactory fits to \lya\ emission profiles \citep{Verhamme08}.   Previous studies (e.\ g.\ \citealt{Pettini02, Kulas}) have already noted this shortcoming of the shell model, drawing a similar conclusions from the velocity gradient seen in metal lines.  Here, for the first time, we are able to demonstrate the need for an envelope of \ion{H}{1} gas, without the need for metals 
as a proxy.  This added constraint, it turns out, is important since the LIS and Lyman series lines show somewhat different kinematics in their maximal outflow velocities.     As we showed in Figure \ref{fesc_vc}, the LIS metal lines reach maximal velocities around $-400$ to $-600$ \kms, whereas the Lyman series lines extend to between $-600$ and $-800$ \kms.      We interpret this difference as an indication that the LIS metal absorption lines are  insensitive to the gas that constitutes the wings of the Lyman series lines. Nevertheless, the excellent correspondence between the blueshifted \lya\ emission and \ion{H}{1} absorption velocities suggests that this high velocity, low density gas is still important for scattering \lya\ photons and creating the profiles that we observe.

Stacking analyses of UV-luminous galaxies at $z\sim 2-4$ have also explored the origin of the blue peak emission by comparing it to LIS lines \citep{Steidel10, Jones12}.  These 
studies have found that the velocity of the blue peak most closely corresponds with the velocities in the blue
wing of the LIS absorption lines, around   $-500$ to $-600$ \kms.  In other words, the blue \lya\ emission is strongest not where the apparent 
optical depth of the LIS lines is greatest, but instead where it is decreasing.   The Green Peas show 
that these characteristics of stacked spectra is not uniform among galaxies.       In Figure \ref{velocity_summary},  
we summarize the kinematic measurements from \ion{H}{1} absorption, LIS metal absorption (\ion{Si}{2} and \ion{C}{2}), and blueshifted  \lya\ emission.    We show the velocity range probed by each set of features, by computing 
the error weighted mean and its uncertainty.     The Green Peas are sorted by $f_{esc}^{Ly\alpha}$, with a comparison from stacked spectra of LBGs at $z\sim 2-4$ shown at bottom (from \citealt{Steidel10} and \citealt{Jones12}). 
While the LIS absorption velocities are similar between the  high redshift samples and Green Peas,
the latter show blueshifted \lya\ emission emerging closer to the systemic velocity and $v_c$.  In fact, while the lower $f_{esc}^{Ly\alpha}$ Green Peas show the blue peak emission falling between the LIS $v_c$ and $v_{max}$, this trend may break down as  $f_{esc}^{Ly\alpha}$ increases.     Both 0926+4427 and 1424+4217 show that the blueshifted \lya\ peak probably emerges at lower blueshifted velocities than $v_c$.  Moreover, if the LIS metals which are too weak to detect in 1133+6514 and 1219+1526 still follow the kinematics of the more highly ionized gas, we might expect this trend to be stronger.  These galaxies show $v_c$ (\ion{Si}{3}) of $-270$ and $-420$ \kms\ respectively, compared to their \lya\ $v_{peak}^{blue}$ around -100 \kms.   

In summary, we conclude that while the wings of the \lya\ emission lines emerge over the same range of velocities probed by 
\ion{H}{1} absorption, the peak emission shows little relation to any of the outflow kinematic measures.   This finding is 
consistent with the conclusions from previous high-redshift studies, where it is argued that the emergent \lya\ emission is more
strongly affected by the optical depth of gas near the systemic velocity than by outflow kinematics \citep{Steidel10, Law12, Chonis, Erb14}.

\section{Summary \& Discussion}  
\label{interp}  
In our analysis so far, we have explored the role of ISM and CGM gas in regulating the amount of \lya\ emission that 
we observe from the Green Peas.    In this section we will discuss the physical implications of our measurements.    
First, however, we summarize the findings from the previous sections: 

\begin{enumerate} 
\item  The  ``Green Pea''  classification identifies objects with prominent \lya\ emission and little to no \lya\ absorption.  Nine of the ten
galaxies in our sample have \lya\ profiles that show both blue and red peaks, and unlike other low redshift samples, their \lya\ luminosity and 
\wlya\ reach the range probed by high redshift \lya\ surveys (e.g.\ \citealt{Ouchi10, Henry10, Henry12}).   

\item Comparison to other nearby populations with published \lya\ measurements \citep{Scarlata09, Cowie11, Hayes14, Ostlin, Martin14}, showed that the Green Peas have (on average) lower masses, higher SFRs and specific SFRs, brighter $M_{UV}$, bluer UV slopes, higher \wha, lower dust extinction, and smaller sizes.  

\item Despite similarly low dust extinction in the Green Pea sample, we measure \lya\ escape fractions (within the COS aperture) that span a factor of ten: $f_{esc}^{Ly\alpha} = 0.05 -0.62$.   

\item We detect LIS absorption in 8/10 Green Peas, and confirm previous findings that  
weaker  $W_{LIS}$ is associated with stronger \lya\ emission (e.\ g.\ \citealt{Shapley03}), 
albeit, extending prior measurements to higher \wlya\ and weaker $W_{LIS}$.
 Although the correlation between $W_{LIS}$ and \wlya\ is not significant in our data, a tentative correlation is found 
when we use $f_{esc}^{Ly\alpha}$  directly, instead of \wlya\ as a proxy. 
Furthermore, the robustness of this relation is also supported by detection in two independent measures: \ion{Si}{2} 1260\AA\ and \ion{C}{2} 1334\AA.   At the same time, we show  that the equivalent width of \ion{Si}{3} $\lambda$1206 does not vary among the present sample.

\item Two of the ten Green Peas show no LIS metal detections, and these are found among those with  the highest \lya\ escape fraction, $f_{esc}^{Ly\alpha} \ge 0.40$, within the COS aperture.

\item Absorption in the \ion{H}{1} Lyman series  ($\beta$, $\gamma$, $\delta$ and/or $\epsilon$) is detected in all 
ten Green Peas.  Unlike the LIS metal lines, the \ion{H}{1} absorption appears uniform in kinematics and equivalent width across the sample.       Furthermore, and in contrast to the LIS metal lines, the Lyman series lines are unambiguously saturated, with little to no residual intensity at line center.

\item Kinematic measurements from  \ion{Si}{2} and \ion{C}{2} show no correlation between $f_{esc}^{Ly\alpha}$ and outflow velocities ($v_c$ or $v_{max}$) in the LIS metal lines, although two of the three highest $f_{esc}^{Ly\alpha}$  Green Peas cannot be included due to the absence of detectable LIS absorption.     The data do show a marginal correlation between  $f_{esc}^{Ly\alpha}$ 
and the outflow velocities ($v_c$ and $v_{max}$) for \ion{Si}{3}, but because this line does not trace 
neutral hydrogen, this trend is not compelling evidence that scattering in the cool, neutral phase of the outflow is aiding  \lya\ escape.

\item The \lya\ escape fraction, $f_{esc}^{Ly\alpha}$,   shows a significant anti-correlation with the velocity separation between the blue and red peaks, which is driven by  shifts in the blue peak velocity.     This trend is consistent with a changing neutral hydrogen column density, where lower \ion{H}{1} density allows more \lya\ to emerge near the systemic velocity.  

\item The velocity of the blue \lya\ emission spans the same range of velocities as the \ion{H}{1} absorption lines, demonstrating (for the first time with \ion{H}{1}) that the \lya\ emission 
arises from an envelope of gas spanning several hundred \kms.   At the same time, the maximal outflow velocities reached by the Lyman series lines are consistently bluer than the 
maximum outflow velocities of the  LIS metal lines; the good agreement between the velocities probed by blue \lya\ emission and Ly$\beta$/Ly$\gamma$ absorption 
implies that compared to LIS metals, these lines are a better proxy for the gas that scatters \lya\ photons. 

\end{enumerate}

These observations allow us to take a closer look at how the interstellar and circumgalactic 
gas regulates the escape of \lya\ photons from the central few kpc of the Green Peas (as probed
by the COS aperture).   In past studies, a few different \lya\ escape  mechanisms have been considered as possibly important.   In  the remainder of this section  we take a comprehensive look at the present Green Pea data, and explore which \lya\ escape models are consistent with our results.

\paragraph{Galaxy Outflow Kinematics}  
Galactic outflows are one mechanism for increasing the transmission of \lya\ photons.  In this case,  the \lya\ photons can scatter in \ion{H}{1} gas, which is Doppler shifted with respect to the ISM.   The scattered photons are then out of resonance with the ISM and escape more easily.  Previous studies have noted this effect: \cite{Kunth98}, \cite{Wofford}, and \cite{RT15} show             that   \lya\ emitting galaxies have outflow velocity $v\sim -100$ to $-200$ \kms, while \lya\ absorbing galaxies show no outflows (LIS absorption is near $v\sim0$).  On one hand, the Green Peas are consistent with these previous measurements; all LIS lines are measured with $v_c \sim -100$ to $-200$ \kms, and all the Green Peas show \lya\ emission.  On the other hand, despite similar outflow velocities, the Green Peas show a factor of eight difference in the \lya\ escape fraction ($0.05 \le f_{esc}^{Ly\alpha} \le 0.41$).  Taken together with the  previous studies, these data suggest that outflows may be necessary to permit \lya\ escape, but they are insufficient to explain the wide range of \lya\ emission strengths.   Ultimately, more work is needed to clarify the role of outflows, and confirm that the relation between outflow velocity and \lya\ escape is not secondary to some other physical characteristic like the ionization state or geometry of outflowing gas.

\paragraph{Neutral Hydrogen Covering} 
Studies using stacked, composite spectra of high-redshift galaxies ($3 \la z \la 4$) have suggested
that \lya\ escape is regulated by the covering fraction in neutral hydrogen, $f_c$ \citep{Shapley03, Jones12, Jones13}.       The motivation for this model is the correlation between W$_{LIS}$ and \wlya.    Since the LIS  metal lines appear saturated in composite spectra, but the absorption is not black, the interpretation is that this absorbing gas does not cover the entire galaxy.  Weaker W$_{LIS}$ suggests  lower  $f_c$ in \ion{H}{1} gas, resulting in greater $f_{esc}^{Ly\alpha}$.

While our data confirm the correlation between W$_{LIS}$ and \wlya\ (see \S \ref{ewsec}), we
cannot attribute this relation  to non-uniform covering in the \ion{H}{1}.   By directly probing the 
\ion{H}{1} Lyman absorption series, rather than using metals as a proxy, we showed that the lines
are saturated and opaque (or nearly so) in all ten Green Peas (see Figure \ref{lyman_fig}).   Hence, {\it we conclude that low covering of \ion{H}{1} cannot explain the variations in $f_{esc}^{Ly\alpha}$ seen in our sample. }

Besides affecting our interpretation of \lya\ escape, 
the addition of the Lyman series lines requires that we re-evaluate  
our picture of how the CGM is structured around high-redshift galaxies.  
Previous analyses  of UV spectra support a model where saturated LIS metal absorption
arises from cool, dense clouds of gas entrained in a more highly ionized
outflow \citep{Shapley03, Heckman11}.   
For the Green Peas, the LIS metal lines are 
difficult to interpret;  because of their low S/N, 
and the probable contribution from emission filling  
\citep{Prochaska11, Scarlata14}, we cannot determine 
whether the lines are optically thin or optically thick.    
On one hand, if the LIS metals from the Green Peas are
optically thick, then the data require a density dependent 
covering fraction, $f_c$.  This scenario requires a 
pervasive low density, low ionization component surrounding
higher density clumps; only the latter give rise to detectable metal absorption. 
On the other hand, the data could  also be described by a homogeneous 
low density envelope of gas, which gives rise to optically thin LIS metal absorption. 
 The \ion{Si}{2} lines that we observe become optically thick at line center
  for $N_{Si~II}  \sim 0.9 - 3.8\times 10^{13}$ cm$^{-2}$; assuming solar abundance [Si/H] = -4.5 \citep{Asplund}, $Z = 0.1 Z_{\sun}$, and
  100\% of the Si exists in \ion{Si}{2}, the implied \ion{H}{1} columns would be around $10^{18} - 10^{19}$ cm$^{2}$.  
  These are consistent with our observations of the Lyman absorption lines, which we inferred were on the flat part of the curve of growth.

A difference between the spatial distribution of detected metals and \ion{H}{1} has been noted before.  
\cite{Quider10} inferred the existence of a pervasive \ion{H}{1} component in the lensed ``Cosmic Eye'' at $z\sim 3$. 
Similar to the Green Peas, the spectrum of this galaxy showed partial covering 
  in metal lines, yet complete covering in damped \lya\ absorption.   The present COS data demonstrate that the ``Cosmic Eye'' is not alone in this characteristic.   These observations suggest that a widespread, low density \ion{H}{1} component may be common. 
  
 \paragraph{Spatially Extended \lya\ Emission} 
 In \S \ref{lyameas_sec} we noted that extended \lya\ emission, falling outside the COS aperture, may be significant for this sample.    Three galaxies 
 have serendipitous \lya\ measurements that should approximate a ``total'' flux.  The LARS imaging showed that the COS aperture captured 40\% of the flux of 0926+4427 \citep{Hayes14}, and GALEX grism spectra show that we detected 60 and 75\% of the flux for 1133+6514 and 1219+1526.    Intriguingly, for this small subsample, the fraction of \lya\ flux detected by COS increases with $f_{esc}^{Ly\alpha}$.    However, an aperture correction alone still does not account for all of the \lya\ photons produced in \ion{H}{2} regions; the small but significant amount of dust  
 measured in \S \ref{sample_sec} is probably still important, especially when coupled with resonant scattering.    Systematic uncertainties on the absolute dust correction for \lya\ preclude a full and accurate accounting 
 of the flux. 
 
 The emission and absorption line profiles in our UV spectra give another means for testing the significance of aperture effects. 
First, in a qualitative sense, aperture losses can mimic some of the features observed in our \lya\ spectral profiles.     For a spherically expanding shell, the outermost regions are expanding in the plane if the sky, with projected velocity $v\sim0$ (see Figure 2 in \citealt{Scarlata14}).    
 Consequently, including a greater contribution from these regions would increase the contribution from \lya\ emission near the systemic velocity, while also increasing  $f_{esc}^{Ly\alpha}$.   At the same time, a greater contribution from spatially extended, scattered emission would also increase the amount emission from metals.  This emission would both ``fill in'' the resonant Si and C absorption lines in our data, and would also produce fluorescent \ion{Si}{2}* and \ion{C}{2}* emission.    In the absence of an aperture, and for a spherical geometry, we expect the equivalent 
 widths of emission and absorption to sum to zero:  $W_{abs} + W_{em} = 0$ \citep{Prochaska11}.   But if the finite COS aperture misses some of the 
 scattering CGM gas, $W_{em}$ will be decreased \citep{Scarlata14}.   We tested whether \ion{C}{2} and \ion{Si}{2} transitions in the present sample
 were consistent with $W_{abs} + W_{em} = 0$, but found that this measurement is extremely sensitive to the precise placement of the continuum.   
 Because of the 10\% normalization uncertainty appropriate for our data,  we cannot determine whether scattered \ion{Si}{2} and \ion{C}{2} emission is missing from the COS aperture. 
 
 Ultimately, direct detection of extended \lya\ emission, similar to the approach adopted for the LARS galaxies \citep{Hayes13, Hayes14, Ostlin}, is needed.    Such data would clarify  how much total \lya\ emission emerges from the Green Peas, while determining its spatial extent and distribution, and whether any regions show \lya\ absorption.

\paragraph{Neutral Hydrogen Density} 
Another galaxy property that should  impact \lya\  output is the neutral hydrogen
column density.   With a greater density of \ion{H}{1}, the \lya\ photons will undergo 
increased scattering.    In this case, they are more susceptible to absorption by dust, and 
must diffuse further in frequency from the line core before escaping.   

Variations in the \ion{H}{1} density offer the most promising explanation for the range of 
$f_{esc}^{Ly\alpha}$ that we detect in the Green Peas.     In \S \ref{lyakin_sec} (see Figure \ref{fesc_lyakin}), we showed  that the $f_{esc}^{Ly\alpha}$ correlates tightly with the  
velocity separation between the blue and red \lya\ emission peaks.     This correlation 
is driven mostly by a shift in the blue emission peak, which appears closer to the systemic velocity 
when $f_{esc}^{Ly\alpha}$  is high.  Since we see no correlation between $f_{esc}^{Ly\alpha}$ and 
the outflow kinematics discussed in \S \ref{kin_neutral_abs},  the kinematic trends seen in the \lya\ lines more probably arise from  a sequence in  \ion{H}{1} density.  In fact, \cite{Verhamme14} have already attributed small peak separations to low \ion{H}{1} column densities.  In this model, the  
peaks are closer together when  low densities allow the \lya\ photons to escape close to the systemic velocity.   Hence, in the present sample, we attribute the increase  in $f_{esc}^{Ly\alpha}$ to a decrease in the neutral hydrogen density.   
 It is plausible that decreased dust contents correspond with lower  \ion{H}{1} column density, but more sensitive dust measurements are needed to test this effect.   Finally, we note that  variations in \ion{H}{1} density may explain the correlation between $f_{esc}^{Ly\alpha}$ and $W_{LIS}$ which we showed in \S \ref{ewsec};  if the metal absorption 
 arises from optically thin gas in a homogeneous envelope, the lower columns of neutral \ion{H}{1}  would also imply  weaker LIS absorption.

\section{Conclusions}

 In this paper, we have presented a COS/FUV spectroscopic study of ten Green Pea galaxies at $z\sim0.2$.   We have focused on understanding how \lya\ photons escape from the ISM and CGM.  
 The \lya\ emission line is frequently used in high-redshift studies, not only to discover objects, but also to infer crude constraints on the properties of galaxies and the IGM.  Yet our limited
understanding of how \lya\ photons escape from galaxies hampers our ability to fully interpret high-redshift observations \citep{Dijkstra14, Dressler15}.   The detection of \lya\ emisison in all ten of the Green Peas suggests  that in the lowest mass,  least dusty, highest SFR galaxies  this feature may appear in the majority of cases. 

Beyond simple detection of \lya,  the UV absorption line data provides important constraints  on 
the ISM and CGM gas that regulate \lya\ output.  These data show that while cool outflows traced by LIS metal absorption may play a role in permitting \lya\ escape,  they do not explain the widely varying \lya\ strengths observed in the Green Peas.   Moreover, aided by the low redshifts of these galaxies and the absence of \lya\ forest absorption,  for the first time we are able to directly analyze \ion{H}{1} absorption in the Lyman series.     These transitions show clear differences from the LIS metal lines which are usually used as a proxy for \ion{H}{1}; most importantly,  the \ion{H}{1} covering fraction, $f_c$, is uniformly high for gas with $N_{H}{1} > 10^{16}$ cm$^{-2}$.  The \lya\ photons do not escape the Green Peas through holes which are completely devoid of CGM/ISM gas.   Instead, the kinematic variations in the \lya\ emission profiles are strongly suggestive that  \ion{H}{1} column density is the primary characteristic  regulating \lya\ escape in these galaxies. 
The ubiquity of double-peaked profiles, and the small velocity separation between the peaks strongly suggests that a  low column density of \ion{H}{1} gas is a typical characteristic among Green Pea galaxies.

This study demonstrates that nearby, high-redshift analog galaxies are useful local laboratories, where a wealth of high fidelity data 
can provide challenges for frequently adopted models.   By observing the ``Green Pea'' galaxies, we have provided a fresh look at
 objects which, to our knowledge, are excellent analogs for young galaxies in an early Universe.   
 Yet much work remains before we can
 be certain that the conclusions drawn here are applicable at moderate to high-redshifts.   Certainly, spectroscopy with {\it James Webb Space Telescope}  will clarify the overlap between high-z galaxies and their nearby counterparts.  Ultimately, this study serves as a strong motivation for future investigations.  An even greater understanding of these galaxies could be achieved with maps of the 
 spatially extended \lya\ emission, as well as sensitive measures of the small amount of dust 
 present in these systems.

  \acknowledgements    
  We acknowledge Marc Rafelski, Sanchayeeta Borthakur, Tucker Jones,  Amber Straughn, and Jonathan Gardner   for helpful discussions. 
  AH is supported by HST GO 12928 and an appointment to the NASA Postdoctoral Program at the Goddard Space Flight Center, administered by Oak Ridge Associated Universities through a contract with NASA.  CLM acknowledges partial support from NSF AST-1109288.
 This research has made use of the NASA/IPAC Extragalactic Database (NED) which is operated by the Jet Propulsion Laboratory, California Institute of Technology, under contract with the National Aeronautics and Space Administration.  AH and CS also acknowledge travel support and gracious hosting from the Nordic Institute for Theoretical Physics during their program, {\it \lya\ as an Astrophysical Tool}.

 \appendix 
  \section{COS spectra of the Green Peas} 
  \begin{figure}[!h]
 \begin{center}
 \includegraphics[scale=0.40, viewport=10 10 1000 430,clip]{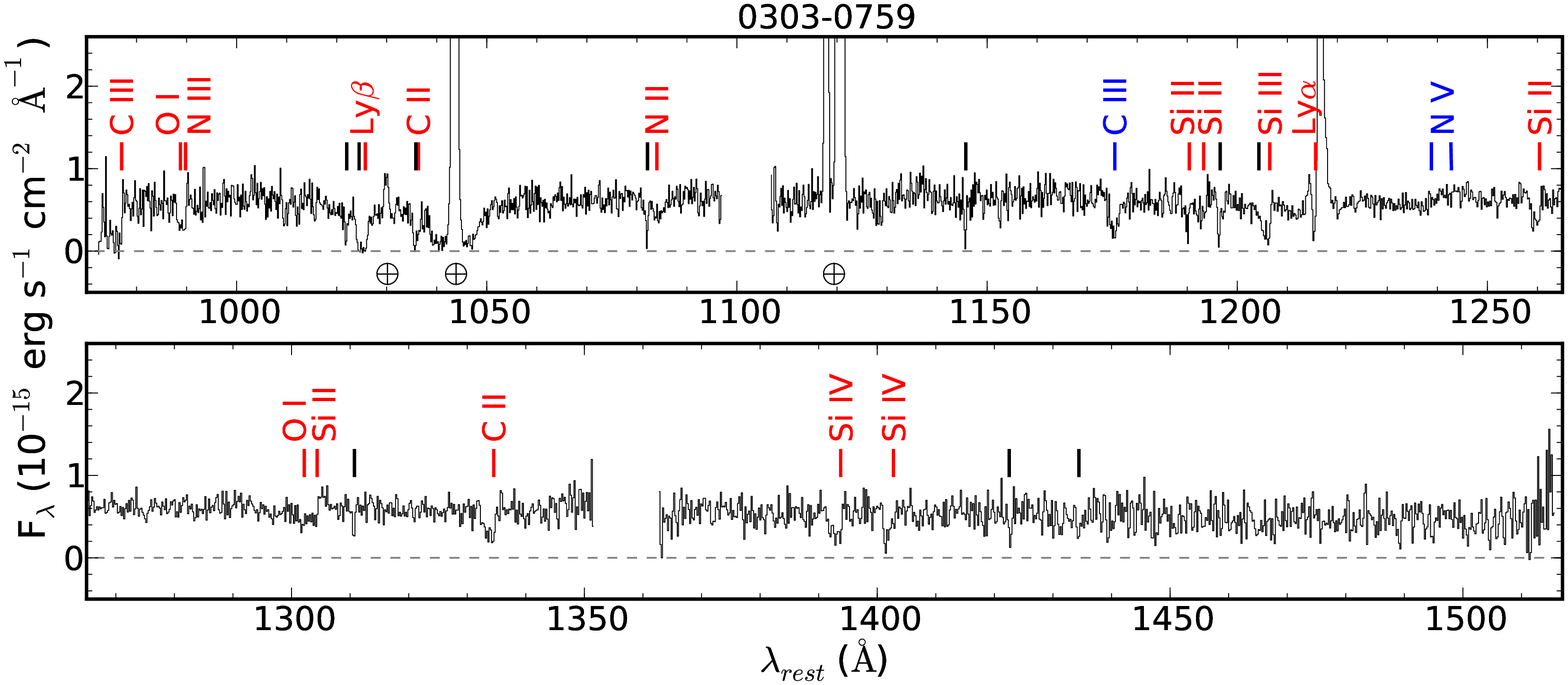}
 \caption{COS G130M + G160M spectrum of the Green Pea galaxy 0303-0759.  Gaps in the spectrum demarcate the wavelengths between the COS FUV A and B segments.  Red markers indicate the expected locations of ISM features in the Green Pea and blue labels show stellar features.  Black vertical lines indicate Milky Way features, whereas $\earth$ symbols mark geocoronal emission.  This sight line contains damped Milky Way \lya\ absorption, apparent around $\lambda_{rest} \sim 1042$ \AA.} 
 \label{0303_fullspec} 
 \end{center}
 \end{figure}  
 
 \begin{figure}
 \begin{center}
 \includegraphics[scale=0.40, viewport=10 10 1000 430,clip]{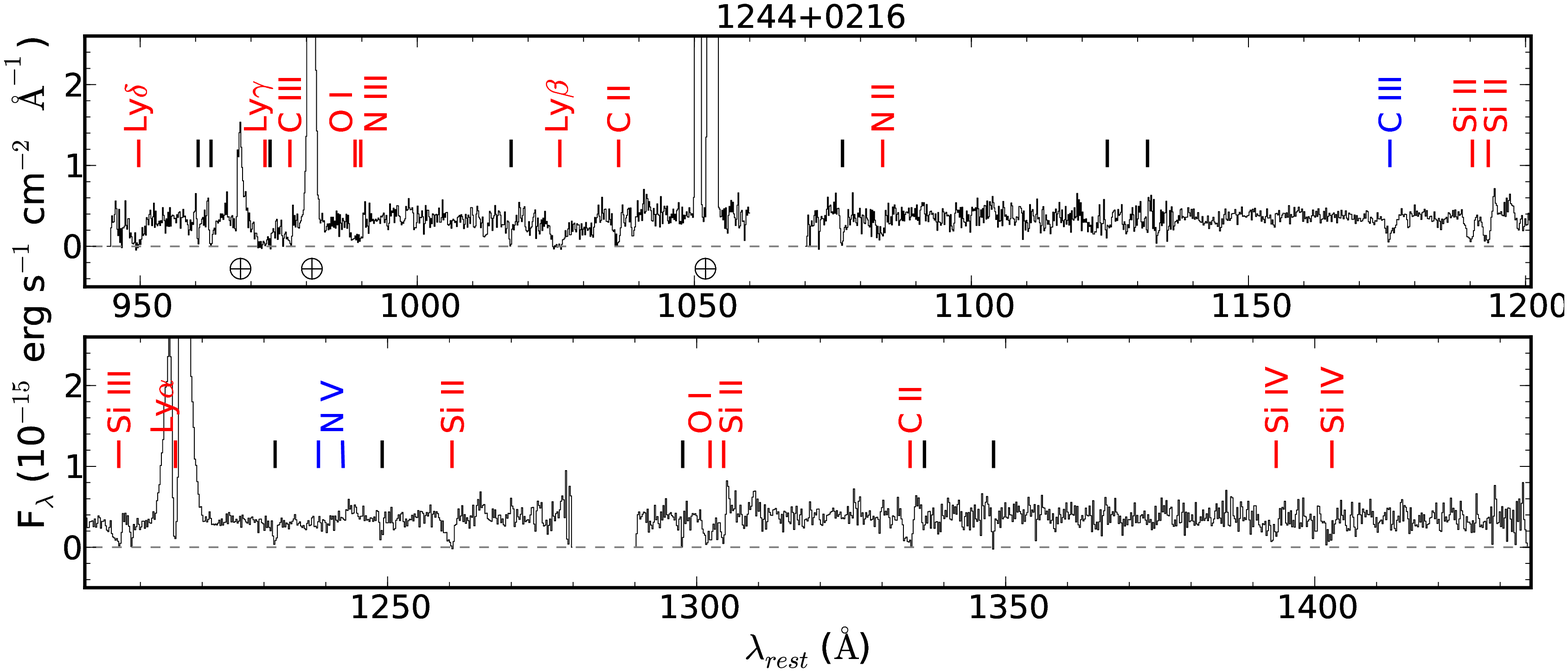}
 \caption{Same as Figure \ref{0303_fullspec}, but for 1244+0216.  This sight line does not contain damped Milky Way \lya\ absorption. } 
 \end{center}
\end{figure}

\begin{figure}
\begin{center} 
 \includegraphics[scale=0.40, viewport=10 10 1000 430,clip]{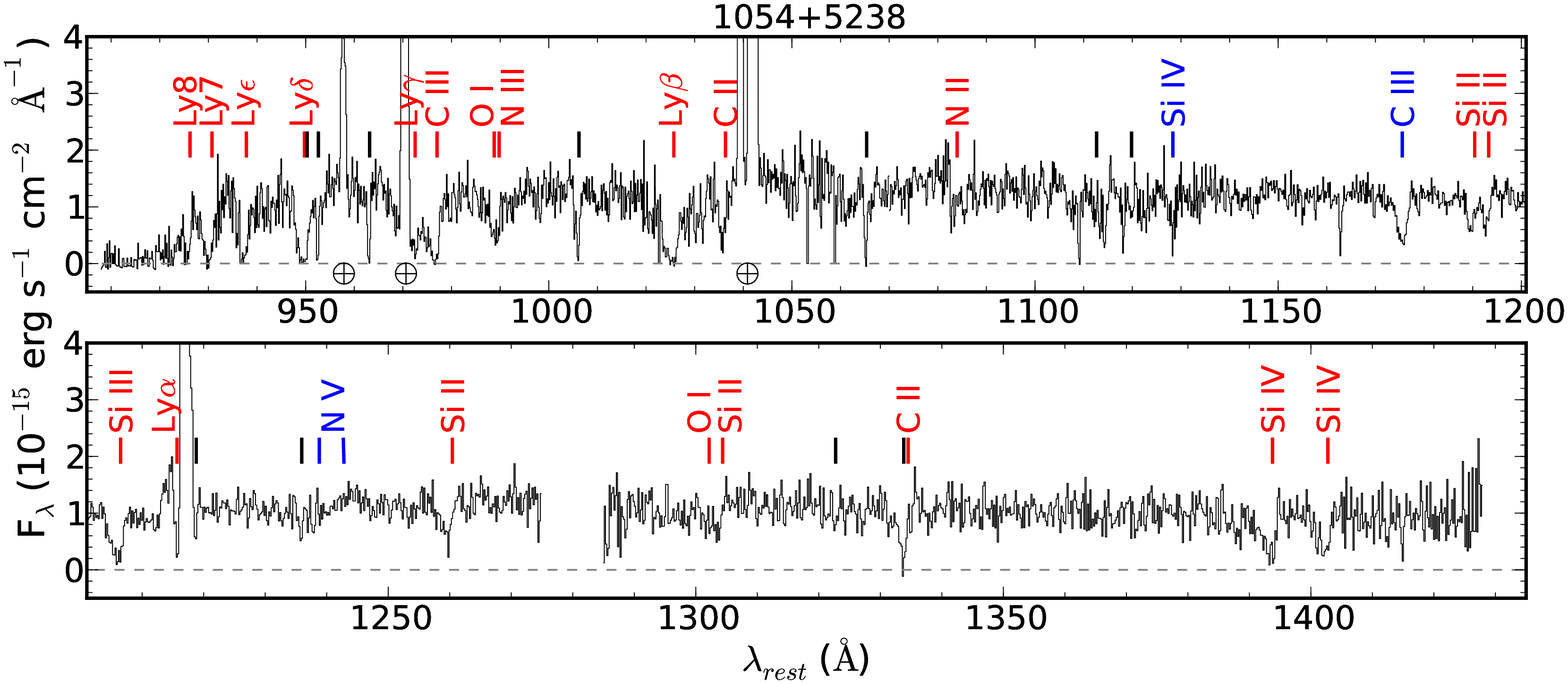} 
 \end{center}  
 \caption{Same as Figure \ref{0303_fullspec}, but for 1054+5238.    } 
\end{figure}

\begin{figure}
\begin{center} 
  \includegraphics[scale=0.40, viewport=10 10 1000 430,clip] {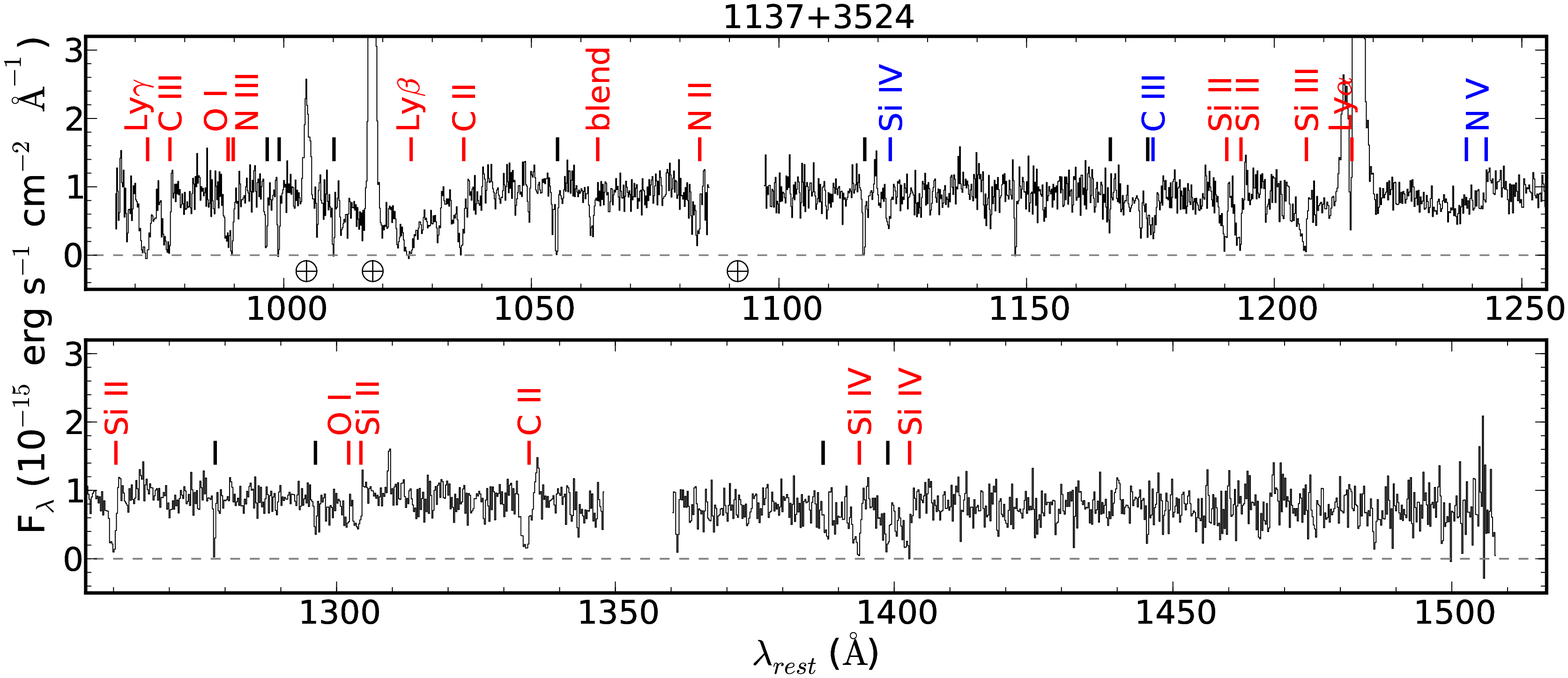} 
 \caption{Same as Figure \ref{0303_fullspec}, but for 1137+3524.}  
 \end{center} 
\end{figure}

\begin{figure}
\begin{center}
  \includegraphics[scale=0.40, viewport=10 10 1000 430,clip]{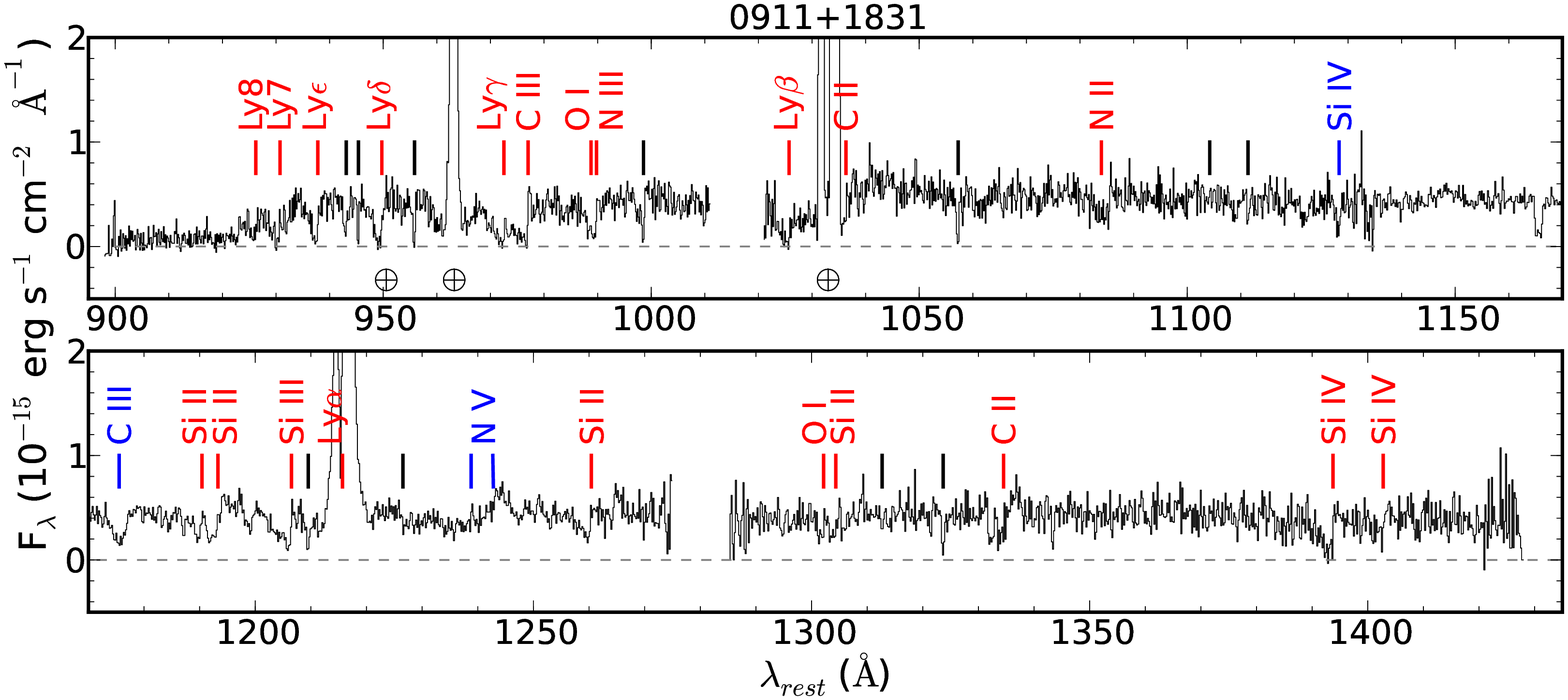} 
 \caption{Same as Figure \ref{0303_fullspec}, but for 0911+1831.  }  
 \end{center}
\end{figure}

\begin{figure}
\begin{center}
   \includegraphics[scale=0.40, viewport=10 10 1000 430,clip]{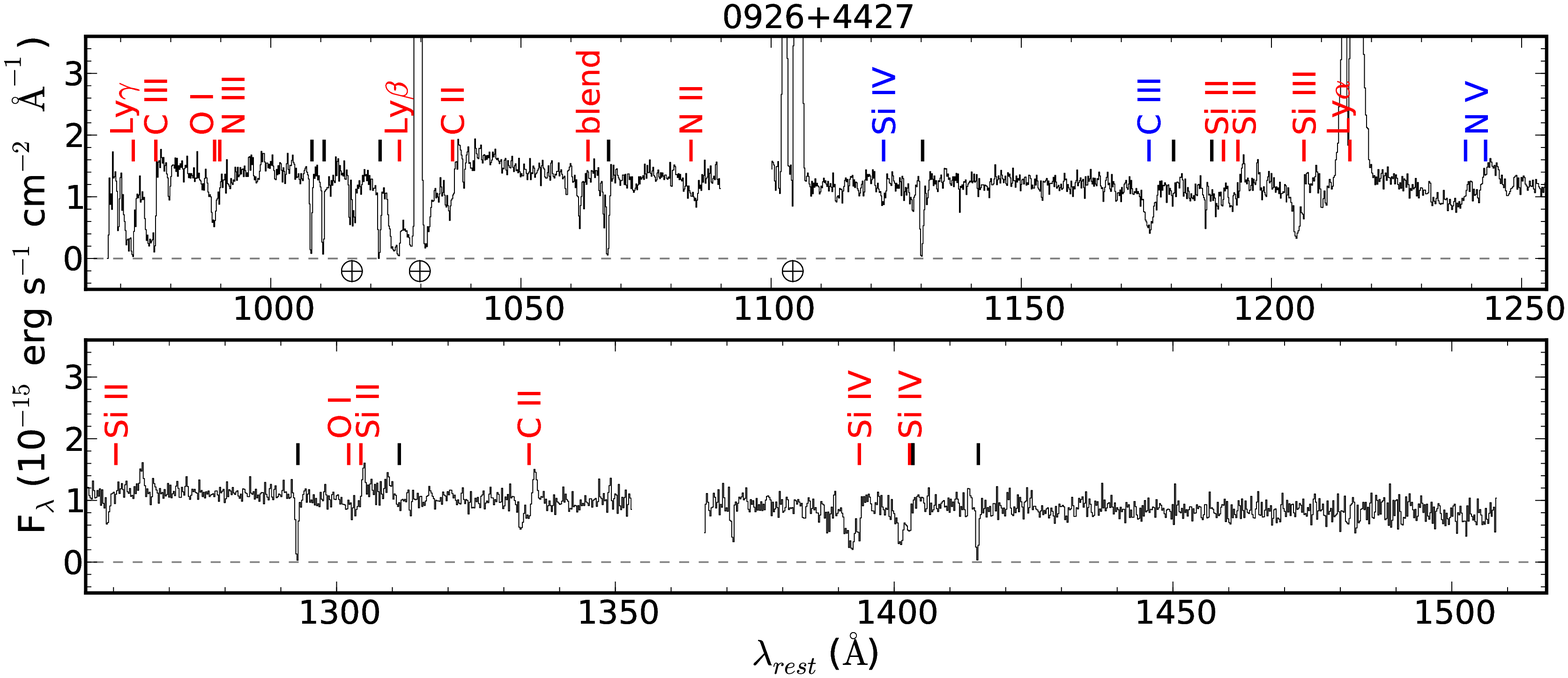} 
 \caption{Same as Figure \ref{0303_fullspec}, but for 0926+4427.  This galaxy was observed as part of the Lyman Break Analog sample presented in \cite{Heckman11}; the data are from GO 11727.     } 
 \end{center}
 \end{figure}

\begin{figure}
\begin{center} 
   \includegraphics[scale=0.40, viewport=10 10 1000 300,clip] {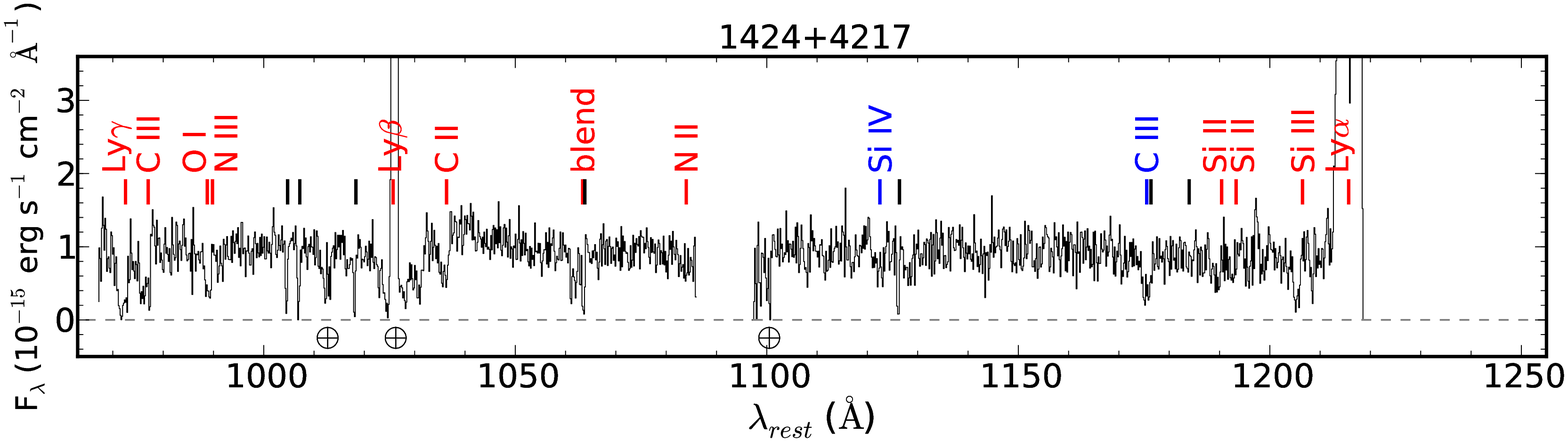} 
 \caption{Same as Figure \ref{0303_fullspec}, but for 1424+4427 .  This spectrum has shorter wavelength coverage because the G160M observation failed and our program did 
 not qualify for a repeat.   } 
 \end{center}
\end{figure}

\begin{figure}
\begin{center}
  \includegraphics[scale=0.40, viewport=10 10 1000 430,clip] {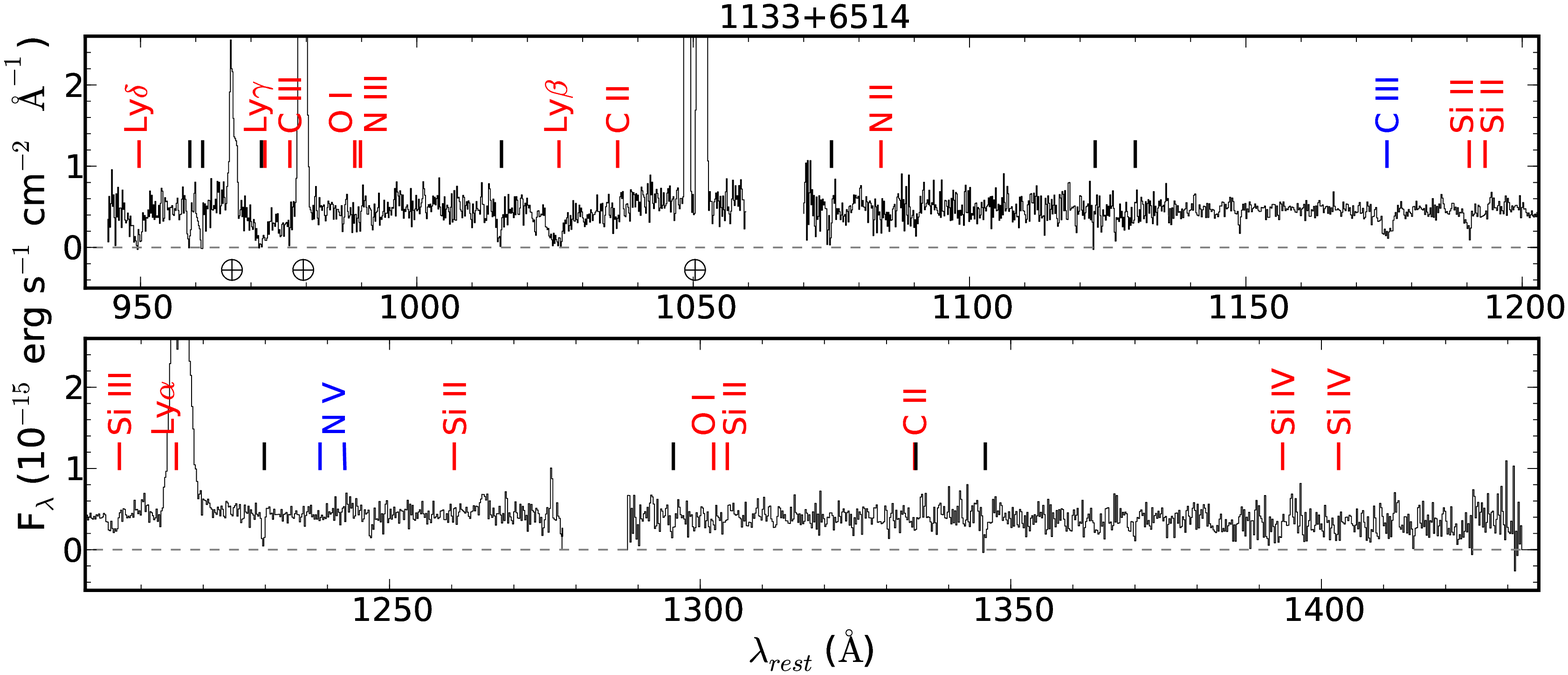} 
 \caption{Same as Figure \ref{0303_fullspec}, but for 1133+6514.   } 
 \end{center}
\end{figure}

\begin{figure} 
\begin{center} 
   \includegraphics[scale=0.40, viewport=10 10 1000 430,clip] {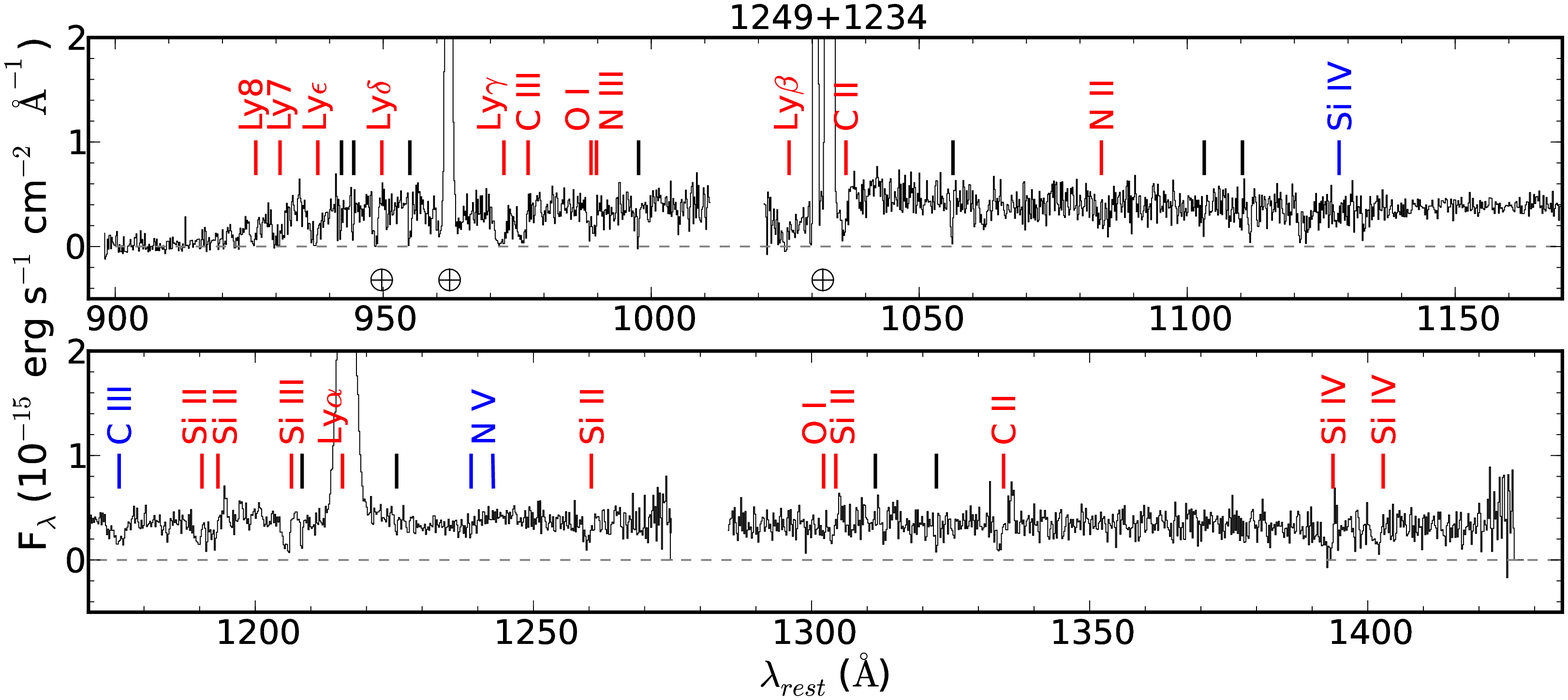} 
 \caption{Same as Figure \ref{0303_fullspec}, but for 1249+1249.   }  
 \end{center} 
\end{figure}

\begin{figure}
\begin{center}
  \includegraphics[scale=0.40, viewport=10 10 1000 430,clip] {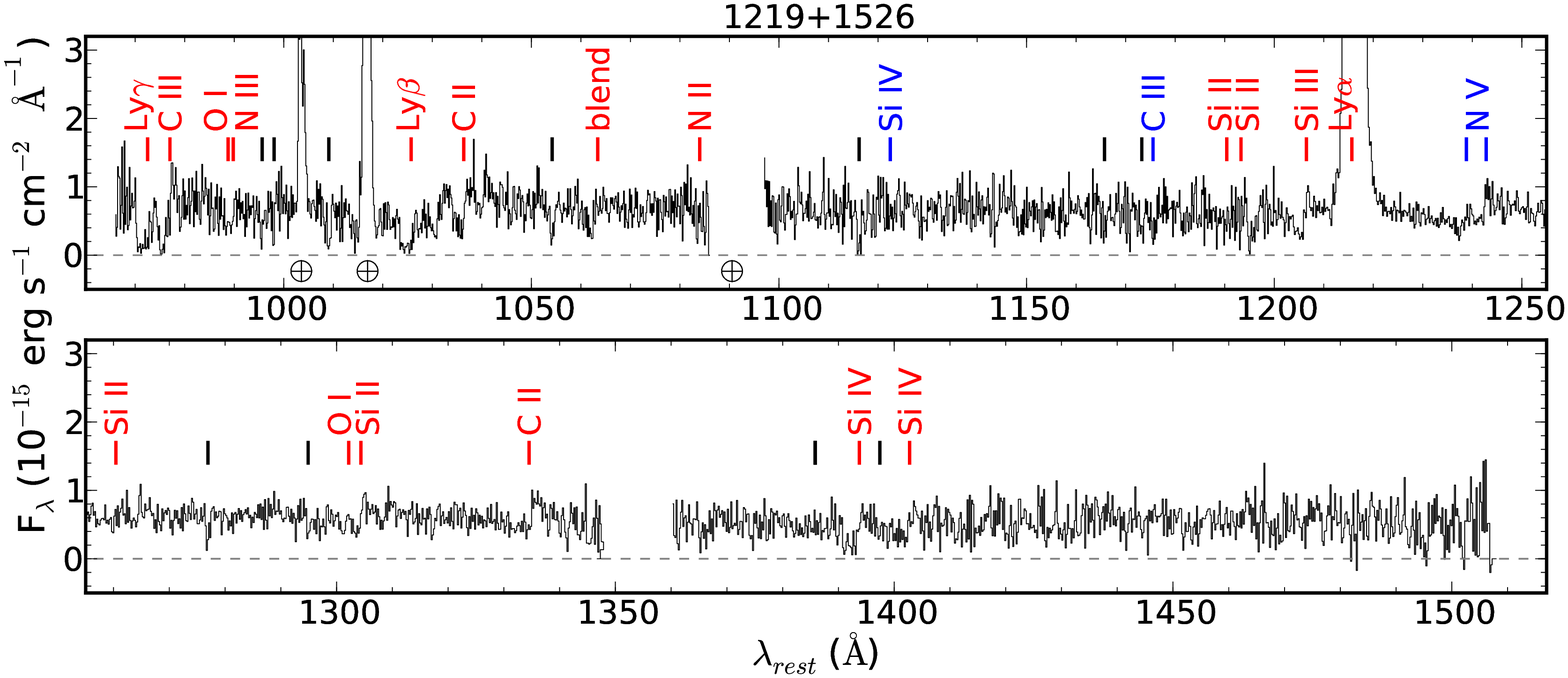} 
 \caption{Same as Figure \ref{0303_fullspec}, but for  1219+1526.   } 
 \end{center}
\end{figure}

\end{document}